\documentclass[final,onecolumn]{elsarticle}
\usepackage{caption}
\usepackage{amssymb}
\usepackage{graphicx}
\usepackage{subcaption}
\usepackage{mwe}
\usepackage{soul}
\usepackage{url}           %this package should fix any errors with URLs in refs.
\usepackage{dcolumn}       % needed for some tables
\usepackage{bm}            % for math
\usepackage{mathtools}     % for math
\usepackage{dsfont}        % for math
\usepackage{mathdots}      % for math
\usepackage{caption}
\usepackage{subcaption}
\DeclarePairedDelimiter\ket{\lvert}{\rangle}
\usepackage{amsmath}    % mathematical env.
\usepackage{pgfplots}
\pgfplotsset{compat=1.14}
\usepackage{tikz,pgfplots}                   %draw graphics
\usetikzlibrary{shapes,arrows,calc}
\usetikzlibrary{arrows, decorations.markings}
\usepackage{lipsum}

\journal{Computers \& Fluids}

\begin{document}

\begin{frontmatter}



\title{Two quantum algorithms for solving the one-dimensional advection-diffusion equation}

\author[label1]{Julia Ingelmann}

\affiliation[label1]{organization={Institute of Thermodynamics and Fluid Mechanics},
             addressline={Technische Universität Ilmenau, P.O.Box 100565},
             city={Ilmenau},
             postcode={D-98684},
             country={Germany}}

\author[label2]{Sachin S. Bharadwaj}

\affiliation[label2]{organization={Tandon School of Engineering},
             addressline={New York University},
             city={New York City},
             postcode={11201},
             state={NY},
             country={USA}}

\author[label1]{Philipp Pfeffer}

\author[label2,label3,label4,label5]{Katepalli R. Sreenivasan}

\affiliation[label3]{organization={Courant Institute of Mathematical Sciences},
             addressline={New York University},
             city={New York City},
             postcode={10012},
             state={NY},
             country={USA}}

\affiliation[label4]{organization={Department of Physics},
             addressline={New York University},
             city={New York City},
             postcode={10012},
             state={NY},
             country={USA}}

\affiliation[label5]{organization={Center for Space Science},
             addressline={New York University Abu Dhabi},
             city={Abu Dhabi},
             postcode={129188},
             country={United Arab Emirates}}

\author[label1,label2]{J\"org Schumacher}

\begin{abstract}
Two quantum algorithms are presented for the numerical solution of a linear one-dimensional advection-diffusion equation with periodic boundary conditions. Their accuracy and performance with increasing qubit number are compared point-by-point with each other. Specifically, we solve the linear partial differential equation with a Quantum Linear Systems Algorithms (QLSA) based on the Harrow--Hassidim--Lloyd method and a Variational Quantum Algorithm (VQA), for resolutions that can be encoded using up to 6 qubits, which corresponds to $N=64$ grid points on the unit interval. Both algorithms are of hybrid nature, i.e., they involve a combination of classical and quantum computing building blocks. The QLSA and VQA are solved as ideal statevector simulations using the in-house solver QFlowS and open-access Qiskit software, respectively. We discuss several aspects of both algorithms which are crucial for a successful performance in both cases. These are the sizes of an additional quantum register for the quantum phase estimation for the QLSA and the choice of the algorithm of the minimization of the cost function for the VQA. The latter algorithm is also implemented in the noisy Qiskit framework including measurement and decoherence circuit noise. We reflect the current limitations and suggest some possible routes of future research for the numerical simulation of classical fluid flows on a quantum computer.
\end{abstract}

%\includegraphics[width=0.9\textwidth]{graphical_abstract.png}

%%Research highlights (max. 85 characters incl. spacings)
%\item We compare two hybrid quantum-classical methods for 1d advection-diffusion equation.

%\item The scalability and computing time with the growing qubit number are studied.

%\item Performance of variational algorithm depends on parametric circuit and optimization.

%\item Performance of the linear systems algorithm relies on accurate eigenvalue estimation. 


%\begin{keyword}
%Quantum computing \sep Variational Quantum Algorithm \sep Quantum Linear Systems Algorithm



%\end{keyword}

\end{frontmatter}


\section{Introduction}
\label{sec:intro}
Quantum computing has the potential to open new ways to classify, generate, and process data \cite{Preskill2018,Deutsch2020} thus changing paradigms in many application fields, such as material science, renewable energy technology, and finance. The reason for the expected advantage over classical algorithms is the physical foundation of quantum computing. Quantum algorithms are capable of encoding information in superposition states and of combining several such quantum states into tensorial product states which span high-dimensional spaces. They can perform unitary transformations (quantum gates) on these product states in parallel rather than on individual bits, as done in classical computers. In this way, $n$ qubits---the smallest units of quantum information---span a $2^n$-dimensional Hilbert space. This parallelism is tightly connected to the possibility of entangling qubits, representing inseparable correlations between qubits, which is absent in classical bit registers \cite{Nielsen2010}. Already these two properties suggest faster solutions of problems with high computational complexity, as has been demonstrated for operations such as prime number factorization \cite{Shor1997}, data search \cite{Grover1997}, and data sampling \cite{Deng2023}; see ref. \cite{Choi2023} for a discussion. Still open is the question of whether similar advantages survive the application of quantum algorithms to solutions of nonlinear ordinary and partial differential equations.  

Fluid dynamics comprises many applications with high computational effort, for instance the modeling of flows over complex objects such as airplanes and the Direct Numerical Simulation (DNS) of turbulent flows \cite{Moin1998} that resolves all physically relevant flow scales from the system size down to those dominated by viscous and diffusive effects. The nonlinear partial differential equations (PDEs) relevant to us are the Navier-Stokes equations for the flow, and (simultaneously) the advection-diffusion equation for the transport of the scalar field such as a substance concentration and temperature. The numerical effort to resolve all these spatial scales increases as $N^3$ for the three-dimensional case, which varies at least as fast as $Re^{9/4}$. Here, $N$ the number of mesh points along one spatial direction and $Re$ is the flow Reynolds number that quantifies the vigor of the fluid turbulence. In many technological applications for which the geometry of the flow domain is complex, one requires in addition adaptive refinements of the computational meshes. Consequently, resource limits are reached quickly, even on the largest state-of-the-art supercomputers. The present solution to this problem is to model the small-scale part, e.g., in the form of Reynolds-averaged Navier-Stokes equation models or large eddy simulations. 

A further possible solution might be the transformation of classical fluid flow problems on a quantum computer to make use of the parallelism that originates from the quantum mechanical foundations. As one example, a single velocity component of a DNS of homogeneous isotropic turbulence in a periodic box with $N^3=8192^3\approx 5.5\times 10^{11}$ grid points \cite{Iyer2019,Buaria2019} could be encoded theoretically in less than 40 qubits, which should be eminently doable since the biggest quantum chip contains 433 qubits. This motivates our present work.

Several approaches have been suggested in the past years to study fluid flows on quantum computers. They include a transformation into a quantum computing-inspired tensor product framework with an effective mapping of the excited degrees of freedom of a three-dimensional turbulent flow \cite{Gourianov2022} or the mapping of specific classical flow problems to a Schrödinger-type quantum dynamics \cite{Meng2023,Jin2023,Succi2023}. They include also a surrogate modeling of thermally driven flows within quantum machine learning frameworks, such as hybrid quantum-classical reservoir computing \cite{Pfeffer2022,Pfeffer2023}. Implementations of mostly one-dimensional flow problems on a quantum computer in the form of pure or hybrid algorithms comprise quantum linear systems algorithms for steady pipe Poiseuille, plane Couette flows and Burgers equation \cite{Bharadwaj2020,Bharadwaj2023,Bharadwaj2023(2)}, quantum amplitude estimation for one-dimensional gas dynamics \cite{Gaitan2020}, Variational Quantum Algorithms for the one-dimensional nonlinear Burgers equation \cite{Lubasch2020,Pool2022}, advection-diffusion problems \cite{Demirdjian2022,Leong2022,Leong2023}, and quantum lattice Boltzmann methods \cite{Todorova2020,Budinski2021}. See also ref. \cite{Succi2023a} for a recent perspective.     

The potential of quantum computing algorithms for solving advection-diffu\-sion problems has been investigated in different ways recently. One approach is the decomposition of the PDE into finite differences such that the resulting system of linear equations can be solved. For sparse linear equation systems, the Harrow-Hassidim-Lloyd (HHL) algorithm can provide a exponential speed up in comparison to classical computation \cite{Harrow2009} under certain caveats \cite{Aaronson2015,Montanaro2016}. A further approach are variational methods. Different versions, such as the variational quantum imaginary time evolution \cite{Leong2023}, the Variational Quantum Linear Solver (VQLS) \cite{Demirdjian2022}, or the Variational Quantum Algorithm (VQA) \cite{Lubasch2020,Guseynov2023,Liu2023}, have been used, even for two-dimensional problems, such as the heat equation \cite{Liu2023}. 

The present work compares these two popular hybrid quantum-classical algorithms for a standard benchmark problem in fluid mechanics, which is the one-dimensional advection-diffusion equation with a constant advection velocity $U$, described by a {\em linear} partial differential equation. To this end, we will compare one-to-one a hybrid quantum-classical Variational Quantum Algorithm (VQA) with a Quantum Linear Systems Algorithm (QLSA). The purpose of the present study is to explore the scalability of both algorithms up to mesh grids which will be encoded in registers consisting of $n\le 6$ qubits giving resolutions of $N\le 64$ grid points. Furthermore, we identify the bottlenecks that exist in both cases for some of their main building blocks: for the VQA scheme, this turns out to be the classical optimization algorithm for the minimization of the cost function; for the QLSA it is the quantum phase estimation routine---an approximate method to find eigenvalues of a unitary matrix. Several classical optimization algorithms are therefore compared in the VQA case. Here, we also investigate the role of the depth of the parametric quantum circuit on the performance of the VQA algorithm and report the impact of measurements for data readout on the overall performance. In case of QLSA, the underlying hybrid algorithm presented here (which in itself preserves the speed-up \cite{Bharadwaj2023}) is customized carefully for the advection-diffusion problem. We analyse the algorithm's performance after prescribing specific strategies for accurate eigenvalue estimation. We also evaluate its dependence on the number of qubits, preconditioning and measurement. To keep our manuscript self-contained and accessible to the fluid dynamics community, we provide compact introductions to quantum computing as well as the two algorithms. Finally, we critically assess both algorithms for this simple fluid mechanical problem and thereby discuss possible limitations of quantum algorithms for (nonlinear) fluid flow problems in one \textit{red}{(and higher-dimensional)} cases. 

The article is structured as follows. First, the analytical solution for the one-dimensional advection-diffusion equation is obtained as the basis for the comparison of the quantum algorithms (Sec. \ref{sec:eq}). Second, the numerical scheme of the finite differences approach is given for forward and backward Euler stepping, which is the groundwork for the quantum algorithms considered here (Sec. \ref{sec:class}). Then, the quantum algorithms are introduced in detail (Sec. \ref{sec:algo}). The comparison of both quantum algorithms is shown in Sec. \ref{sec:comp} on aspects such as the time evolution of the concentration profiles, dependence on the number of qubits, dependence on parameter $T_0$ and the realisation on Noisy Intermediate Scale Quantum (NISQ) devices. The results are summarized and discussed in Sec. \ref{sec:discussion}.

\section{One-dimensional advection-diffusion equation}
\label{sec:eq}

We demonstrate and compare the performance of the two quantum computing algorithms considered here for the advection-diffusion equation given by
\begin{align}
    \label{eq:advection-diffusion-equation}
    \partial_t c = D \nabla^2 c - {\bm u}\cdot \nabla c,
\end{align}
where $c({\bm x},t)$ is the concentration field of the solvent, $D$ is the diffusion constant and ${\bm u}({\bm x},t)$ is the velocity vector field that advects the solvent. This equation describes the transport of the solvent, such as a dye or a cloud of tracer particles subject to diffusion and advection with the velocity field. In this paper, we consider the simplest case of a one-dimensional linear equation, which is given by
\begin{align}
    \label{eq:advection-diffusion-equation-1D}
    \partial_t c(x,t) = D \partial^2_x c(x,t) - U \partial_x c(x,t).
\end{align}
Here, the advection velocity $U$ in the $x$-direction is taken as a constant. The problem is discretized in space and time. For the spatial discretization, the interval $x\in [-L,L]$ is divided into $N$ segments of width $\Delta x=2L/N$. The time evolution is also discretized uniformly, such that $t= m \tau$, where $\tau$ is the time step. For the analytical solution, the wave-like ansatz $c(x,t)=\text{exp}\left( \omega t + i \lambda x\right)$ is chosen such that $\omega = - D \lambda^2  - i U \lambda$ follows from eq. \eqref{eq:advection-diffusion-equation-1D}. Hence, the concentration profile takes the form of
\begin{equation}
    \label{eq:ansatz-Euler identity} 
    c(x,t) = \left[a \cos{\lambda (x-Ut)} + b \sin{\lambda (x-Ut)}\right] \text{exp}(-D\lambda^2 t)\,.
\end{equation}
Periodic boundary conditions are imposed, such that $c(x=0,t) = c(x=N,t)$. Consequently, the wavenumber $\lambda = k \pi/L$ with $k\in \mathbb{N}$. Thus there follows the general solution to the problem in the form of a series expansion
\begin{align}
    \label{eq:solution}   
    c(x,t) &= \sum_{k=0}^{\infty} \left[a_k \cos{\left(\frac{k \pi}{L} (x-Ut)\right)} + b_k \sin{\left( \frac{k \pi}{L} (x-Ut)\right)}\right] \nonumber\\
    &\times \text{exp}\left(-D\left(\frac{k \pi}{L}\right)^2 t\right)
\end{align}
As initial condition the delta function is applied such that $c(x,0) = \delta(x)$. The delta function is standard and defined to be $\delta(x)=0$ for $x \neq 0$ and $\int_{-\infty}^\infty \delta (x) \text{d}x =1$. The initial condition specifies the expansion coefficients in the general solution as 
\begin{align}
    \label{eq:delta-Fourier}
    c(x,0) = \frac{a_0}{2} + \sum_{k=1}^{\infty} a_k \cos\left( \frac{k \pi}{L}x \right) + b_k \sin\left( \frac{k \pi}{L}x \right).
\end{align}
The Fourier coefficients are given by
\begin{align}
    a_0 &= \frac{1}{L} \int_{-L}^{L} \delta(x) \text{d}x = \frac{1}{L}, \\
    a_k &= \frac{1}{L} \int_{-L}^{L} \delta(x) \cos\left( \frac{k \pi}{L}x \right)\text{d}x = \frac{1}{L} \text{ and}\\
    b_k &= \frac{1}{L} \int_{-L}^{L} \delta(x) \sin \left( \frac{k \pi}{L}x \right)\text{d}x = 0,
\end{align}
such that the initial condition can written as
\begin{align}
    \label{eq:initial condition}
    c(x,0) = \frac{1}{2L} + \sum_{k=1}^{\infty} \frac{1}{L} \cos\left( \frac{k \pi}{L}x \right). 
\end{align}
Considering these coefficients, the analytical solution can be found to be
\begin{align}
    \label{eq:analytical solution}   
    c(x,t) &= \frac{1}{2L} + \sum_{k=1}^{\infty} \left[ \frac{1}{L} \cos{\left(\frac{k \pi}{L} (x-Ut)\right)} \right] \text{exp}\left(-D\left(\frac{k \pi}{L}\right)^2 t\right)\,.
\end{align}
Equation (\ref{eq:analytical solution}) describes a Gauss-shaped pulse that diffuses while moving to the right given that $U>0$. For the following, lengths are measured in units of the interval length $2L$. Times can be expressed in units of either the advection time $\tau_a=2L/U$ or the diffusive time $\tau_d=(2L)^2/D$. If not stated otherwise, we will use $\tau_a$. 

\section{Finite difference methods with Euler time stepping}
\label{sec:class}
The numerical solution of the advection-diffusion equation (\ref{eq:advection-diffusion-equation-1D}) can be obtained by a finite difference method (FDM). In the simplest case, these are Euler methods, either an explicit forward or an implicit backward Euler time step method. For this method, the partial differential equation is approximated by a system of algebraic discretization equations. Furthermore, the problem is discretized in space and time uniformly, such that $ x_i= x_0 + i\Delta x$ and $ t_m= t_0 + m\tau$ with $x_0=0$ and $t_0=0$. Indices $i=0,\dots, N-1$ and $m=0, \dots, M$. When the forward difference in time and the centered difference in space is taken, one gets the forward in time and centered in space (FCTS) method. It is of 1st order accuracy in time, of 2nd order accuracy in space, and given by 
\begin{align}
    \label{eq:explicit-scheme1}
    \frac{c_{i}^{m+1} - c_{i}^{m}}{\tau} = D \frac{c_{i+1}^{m} - 2c_{i}^{m}+c_{i-1}^{m}}{(\Delta x)^2} - U \frac{c_{i+1}^{m} - c_{i-1}^{m}}{2 \Delta x}\,,
\end{align}
and thus
\begin{equation*}
    \label{eq:explicit-scheme2}
    c_{i}^{m+1} = \left(\frac{D \tau}{(\Delta x)^2} - \frac{U\tau}{2 \Delta x} \right) c_{i+1}^{m} + \left( 1- 2 \frac{D \tau}{(\Delta x)^2} \right) c_{i}^{m} +\left( \frac{D \tau}{(\Delta x)^2} + \frac{U\tau}{2 \Delta x} \right) c_{i-1}^{m}\,.
\end{equation*}
We define the following abbreviations
\begin{equation}
r=\frac{D\tau}{(\Delta x)^2} \quad \mbox{and}\quad s = \frac{U\tau}{2\Delta x}\,,
\end{equation}
where $s$ is the parameter of the convective part and $r$ is the stability parameter. For this explicit scheme $r\leq 1/2$ should hold. The scheme can be expressed as a system of linear equations via %, $A {\bm c}^{m}={\bm c}^{m+1}$, where the coefficient matrix $A$ is given by
\begin{align} 
A {\bm c}^{m}&={\bm c}^{m+1} \\
    \label{eq:explicit-scheme3}
    \text{with}\ \  A &= \left[
    \begin{array}{cccccc}
        1-2r & r-s & 0 & \hdots & 0  & s+r \\
        s+r & 1-2r & r-s &  &  &  0 \\
        0 & s+r & 1-2r & r-s &  &  0 \\
        \vdots &  & \ddots & \ddots & \ddots & \vdots  \\
        0 &  &  & s+r & 1-2r &   r-s \\
        r-s & 0 & \hdots & 0 & s+r & 1-2r \\
    \end{array} \right].
\end{align}

\begin{figure*}
    \centering 
    \includegraphics[width=\textwidth]{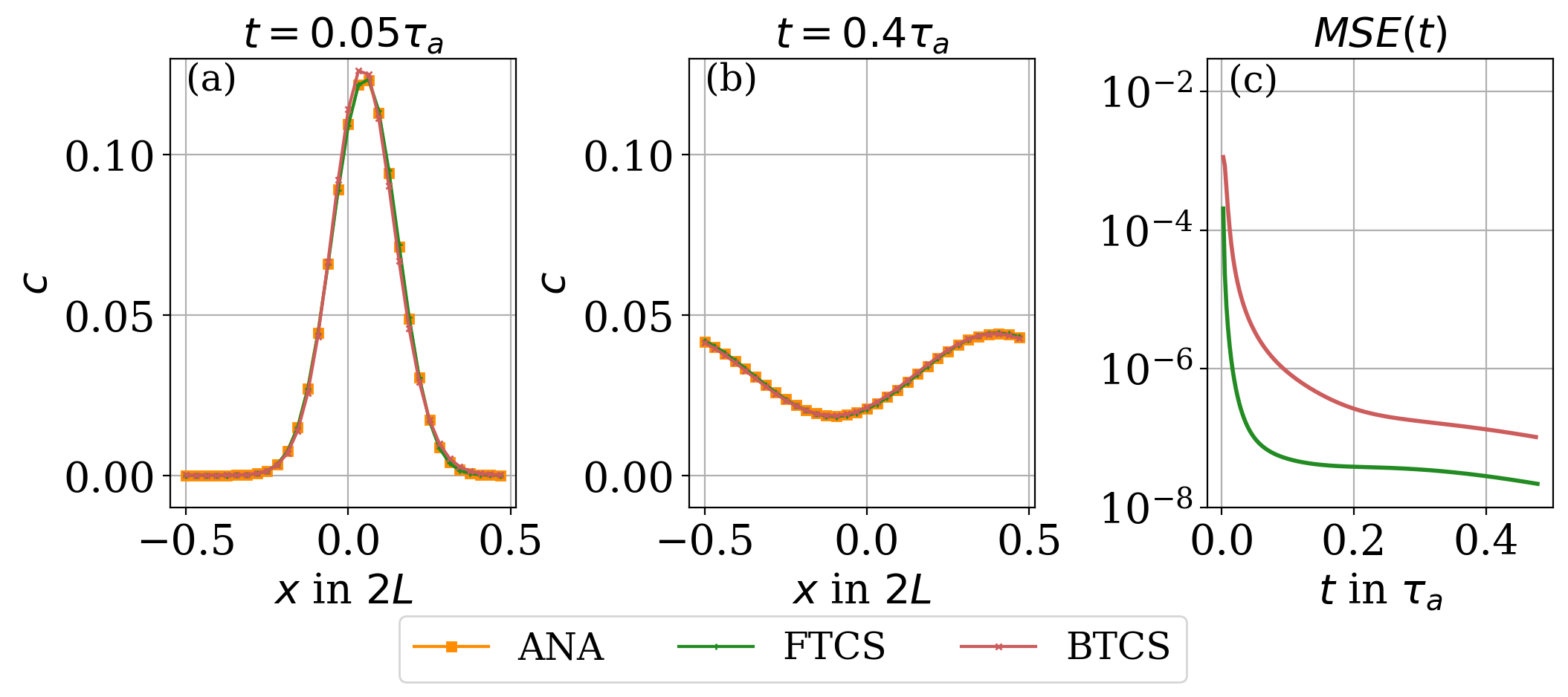}
    \caption{Time evolution of the concentration profiles of the analytical solution (ANA) and the results of the classical numerical methods, namely the explicit method (FTCS), the implicit method (BTCS) and the midpoint method (MP) for $N=32$, $D=1$, $U=10$. Panels (a) and (b) compare the concentration profiles of the methods at two time instants. The corresponding mean squared error (MSE) of the results of the classical numerical methods to the analytical solution is shown in panel (c).}
    \label{fig:Comp-Analyt_vs_FDA1} 
\end{figure*}

In case of the implicit backward Euler scheme (BTCS), the system of linear equation follows %$A {\bm c}^{m+1} ={\bm c}^{m}$,
such that the matrix $A$ has to be inverted to find the desired solution, since this method imposes the expression %. Thus
\begin{align}
    \label{eq:implicit-scheme1}
    \frac{c_{i}^{m+1} - c_{i}^{m}}{\tau} = D \frac{c_{i+1}^{m+1} - 2c_{i}^{m+1}+c_{i-1}^{m+1}}{(\Delta x)^2} - U \frac{c_{i+1}^{m+1} - c_{i-1}^{m+1}}{2 \Delta x}\,,
\end{align}
which can be reformulated to
\begin{equation*}
    \label{eq:implicit-scheme2}
    c_{i}^{m} = \left( -\frac{D \tau}{(\Delta x)^2} +\frac{U\tau}{2 \Delta x} \right) c_{i+1}^{m+1} + \left( 1+ 2 \frac{D \tau}{(\Delta x)^2} \right) c_{i}^{m+1} +\left(- \frac{D \tau}{(\Delta x)^2} - \frac{U\tau}{2 \Delta x} \right) c_{i-1}^{m+1}\,.
\end{equation*}
In other words, the scheme can be expressed as a system of linear equations via %Then the coefficient matrix $A$ is given by
\begin{align}
    A {\bm c}^{m+1} &={\bm c}^{m} \\
    \label{eq:implicit-scheme3}
    \text{with}\ \ A &= \left[
    \begin{array}{cccccc}
        1+2r & -r+s & 0 & \cdots & 0  & -r-s \\
        -r-s & 1+2r & -r+s &  &  &  0 \\
        0 & -r-s & 1+2r & -r+s &  &  0 \\
        \vdots &  & \ddots & \ddots & \ddots & \vdots  \\
        0 &  &  & -r-s & 1+2r &   -r+s \\
        -r+s & 0 & \cdots & 0 & -r-s & 1+2r \\
    \end{array} \right]
\end{align}
The comparison of the analytical solution (ANA) with those of FCTS and BCTS is shown in Fig. \ref{fig:Comp-Analyt_vs_FDA1}. The comparison is made via the mean squared error (MSE) defined as 
\begin{align}
    {\rm MSE}(t_m) = \frac{1}{N} \sum_{i=0}^{N-1}\left[c_i^{\text{ANA}}(t_m)-c_i^{\text{FDM}}(t_m) \right]^2\,,
    \label{eq:MSE0}
\end{align}
where $c_i=c(x_i,t)$. In Figs. \ref{fig:Comp-Analyt_vs_FDA1}(a) and \ref{fig:Comp-Analyt_vs_FDA1}(b), it can be seen that the numerical methods approximate the analytical solution sufficiently well. This could be different when nonlinear equations have to be solved with VQA.

%--------------------------------------------------------
%--------------------------------------------------------

\section{Quantum algorithms}
\label{sec:algo} 
This section describes both quantum algorithms, namely the VQA and the QLSA. The quantum part of the VQA is implemented in the quantum simulation environment Qiskit \cite{Qiskit}. The QLSA is done with QFlowS, a C++ based simulation package \cite{Bharadwaj2023}. For the direct comparison of both algorithms, an ideal statevector simulation will be used. In the following, we will briefly introduce the basics of both quantum algorithms. The building block of both algorithms are the qubits, the smallest information units in a quantum algorithm. While a single classical bit can take two discrete values only, namely $\{0,1\}$, a qubit is a superposition of the two basis states of the  Hilbert space $\mathbb{C}^2$
\begin{equation}
|q_1\rangle=\alpha_1|0\rangle+\alpha_2|1\rangle=\alpha_1
\left(
\begin{array}{c}
1 \\ 0\\
\end{array}\right)
+\alpha_2
\left(
\begin{array}{c}
0 \\ 1\\
\end{array}\right)\,, \label{Eq:B1}
\end{equation}
with $\alpha_1, \alpha_2\in \mathbb{C}$ and $\|q_1\|_2=\sqrt{|\alpha_1|^2+|\alpha_2|^2}=1$ and basis vectors $|0\rangle$ and $|1\rangle$ in Dirac's notation \cite{Nielsen2010}. It can be combined into an $n$-qubit system, also denoted as an $n$-qubit quantum register, by successive tensor products of qubits. An unentangled two-qubit state vector is the tensor product of two single-qubit vectors, 
\begin{equation}
|q_1\rangle\otimes |q_1^{\prime}\rangle \in \mathbb{C}^2\otimes\mathbb{C}^2\,.
\end{equation}
The basis of this tensor product space is given by 4 vectors, usually formulated in integer or binary bit-string notation: $| j_1\rangle=|0\rangle\otimes |0\rangle= |00\rangle$, $|j_2\rangle=|01\rangle$, $|j_3\rangle=|10\rangle$, and $|j_4\rangle=|11\rangle$. The $n$-qubit quantum state $|c(t)\rangle$ at a time $t$ is consequently defined in a $2^n$-dimensional tensor product Hilbert space ${\cal H}=(\mathbb{C}^2)^{\otimes n}$ and given by
\begin{equation}
   \ket{c(t)} = \sum\limits_{k=1}^{2^n} c_k(t) \ket{j_k} \hspace*{1em} \text{with} \hspace*{1em} \sum\limits_{k=1}^{2^n} |c_k(t)|^2 = 1\,.
\label{nqubit}   
\end{equation}
In other words, when connecting this formalism to the present flow problem, the discretization of the concentration profile $c(x,t)$ on $N=2^n$ grid points at time $t$ is obtained by an $n$-qubit quantum state vector. In eq.~\eqref{nqubit}, the quantum state vector is normalized to 1. Technically, the square magnitude of each coefficient represents the probability of measuring the respective basis state. Thereby, they naturally have to sum up to 1. For a classical concentration profile this does not have to be the case; see subsection \ref{subsec:vqa}. The time evolution of the state vector in quantum algorithms is established by unitary transformations or operators $\hat U$ with $\hat U^{-1}=\hat U^{\dagger}$, realized on a quantum computer by a sequence of quantum gates. These gates can be viewed as rotation operators on quantum state vectors that can also generate entanglement between qubit states \cite{Nielsen2010}. 

Note that the accuracy of the quantum algorithms depends on the accuracy of the numerical input data described in Sec.~3. For the comparison of the quantum algorithms, the analytical solution, which we derived in Sec. \ref{sec:eq}, is also considered.

\subsection{Quantum Linear Systems Algorithm (QLSA)}
\label{subsec:qlsa}
\begin{figure}[htpb!]
    \centering{
    \subfloat[]{
    \includegraphics[trim={0.0cm 0.0cm 0cm 0.0cm},clip=true,scale=0.85]{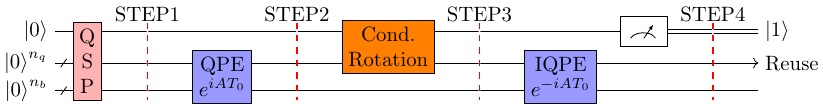}}\\
    \vspace{1cm}
    \subfloat[]{\includegraphics[trim={0.0cm 0.0cm 0cm 0.0cm},clip=true,scale=0.25]{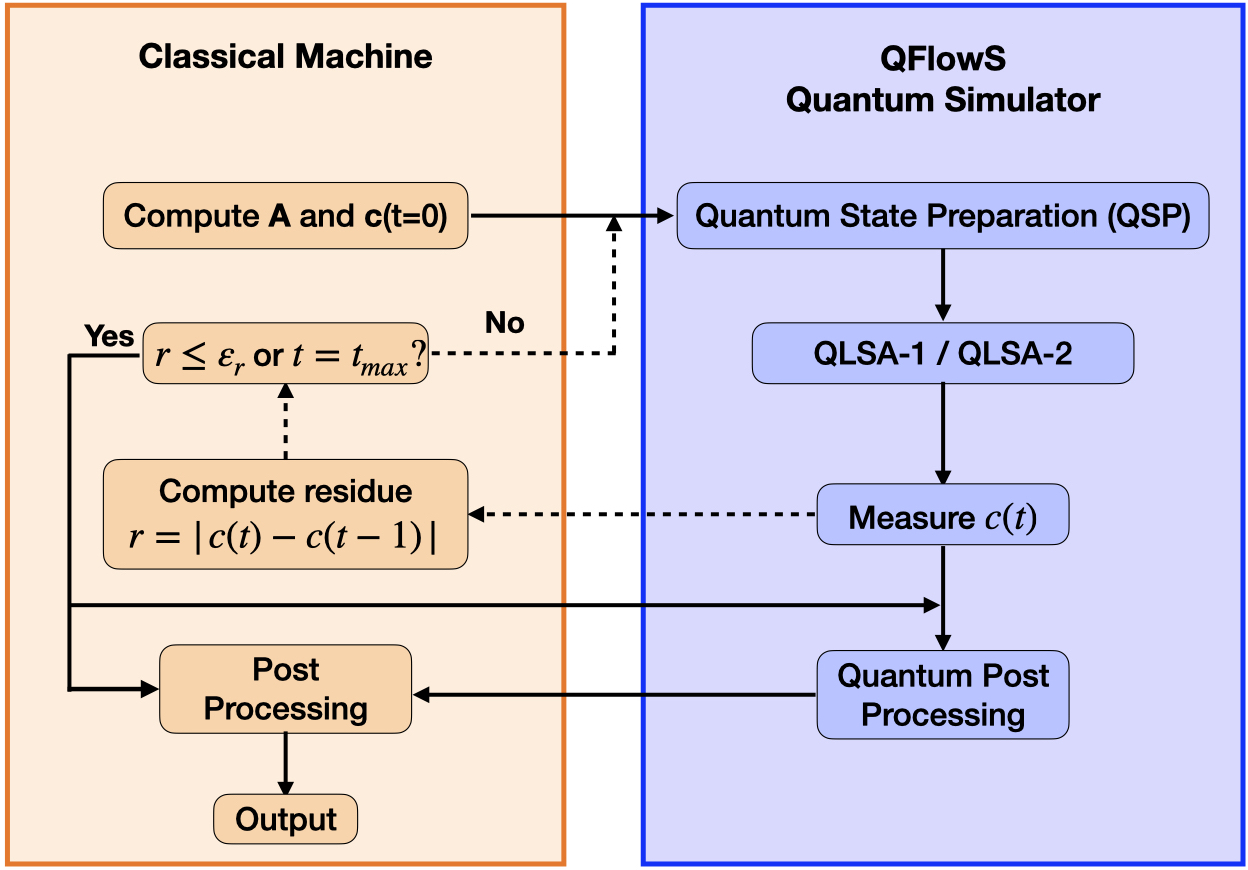}}}
    
    \caption{(a) A sketch of the quantum circuit depicting the QLSA-1 architecture. The horizontal lines in the diagram stand either for individual qubits or collections of several qubits, so-called quantum registers. Boxes across qubits stand for parametric circuits, collections of multiple unitary transformations on a single or multiple qubits. These are from left to right Quantum State Preparation (QSP), Quantum Phase Estimation (QPE), Conditional Rotation and inverse Quantum Phase Estimation (IQPE).  (b) A sketch depicting the hybrid workflow of the QLSA algorithm. The classical preparation of the matrix elements and pre-conditioning is done classically. Then, on the quantum simulator the quantum states are prepared and processed with quantum state preparation (QSP) and QLSA, respectively. The solutions are measured or post processed at the quantum level and read into the classical machine. The residue and convergence checks are made classically and the solutions are post-processed when the iterations have finished or converged.}
    \label{fig:qlsa flowchart}
\end{figure}
Quantum algorithms which solve a linear system of equations of the form, $A\mathbf{x}=\mathbf{b}$, belong to the category of Quantum Linear Systems Algorithm (QLSA). All such algorithms \cite{Harrow2009,Childs2017,Childs2021} (excluding variational methods, which will be described subsequently in Sec. \ref{subsec:vqa}) can be broadly categorized into two approaches which compute a quantum-numerical approximation to $A^{-1}\mathbf{b}$ (BTCS) or $A\mathbf{x}$ (FTCS). The approach presented here, which we call QLSA-1, is a modified version of the original HHL algorithm \cite{Harrow2009}. Here, we compute the eigenvalues ($\sigma_{j}$) of the matrix $A$ and thereby approximate the solution $A^{-1}\mathbf{b}$. The central computational issue here is to identify the eigenvalues of $A$ and the following evaluation of their inverse. An alternative algorithm we call QLSA-2 proceeds by approximating the action of the matrix $A$ (or $A^{-1}$) as purely a matrix-vector multiplication operation, implemented by decomposing the matrix into a Linear Combination of Unitary (LCU) quantum gates \cite{Childs2017}, acting on a suitably prepared quantum state. The central goal in that case is to find the best unitary basis to produce a probabilistic implementation of the matrix. Both methods have been implemented on QFlowS in \cite{Bharadwaj2023} to solve laminar Poiseuille and Couette flows. The solution $c(t)$ can be obtained either iteratively at every time-step or by one-shot QLSA algorithms that would offer higher quantum advantage \cite{Liu2021,Bharadwaj2023}. However, the latter strategy can be computationally expensive to simulate large system sizes over long integration times. For our present purposes, we present results for QLSA-1 using the former approach.

The QLSA-1 algorithm is implemented as a full gate-level circuit simulation with at most single qubit or (double) controlled NOT gates \cite{Nielsen2010} to integrate eq.~\eqref{eq:advection-diffusion-equation-1D} using the BTCS method. The outline of the algorithm's work flow and its circuit is shown in Fig. \ref{fig:qlsa flowchart}. It comprises the steps or quantum sub-routines briefly outlined below (whose details can be found in \cite{Bharadwaj2023}). The herein flow problem has the matrix A of eq. \eqref{eq:implicit-scheme3}, which is not Hermitian if advection is present, namely $U\neq0 \ (\Rightarrow s\neq0)$. Since the algorithm admits only Hermitian matrices, the matrix $A$ is first extended to an Hermitian classically as
    \begin{equation}
        \tilde{A} = \begin{pmatrix}
            0 & A\\
            A^{\dag} & 0
        \end{pmatrix}.
        \label{eq: hermitian}
    \end{equation}
The implementation involves the following steps.

Step 1 - Quantum State Preparation (QSP): The concentration field at every time step $m$ is loaded onto an $(n+1)$-qubit ($=n_{b}$ from here on) state proportional $\tilde{\mathbf{c}}^{m} =[\mathbf{c}^{m};0]\,\|\mathbf{c}^{m}\|_2^{-1}$ to make it compatible with eq. \eqref{eq: hermitian} (and therefore one expects the solution state in the form $\mathbf{x} = [0;\mathbf{c}^{m+1}]$). As will be described shortly, the algorithm also requires an additional $n_{q}+1$ ancillary qubits. The latter are helper qubits or in short ancillas. All these qubits that are initially set to basis state $|0\rangle$, are then initialized using either the functional form type state preparation or the sparse-state preparation oracle $\hat U_{\rm QSP}$ (see Sec.~3 of SI Appendix in~\cite{Bharadwaj2023}),
    \begin{equation}
       \vert \psi_\text{STEP1}\rangle =\ \hat U_{\rm QSP}\vert0\rangle_{n_{b}+n_{q}+1} = \ \vert \tilde{c}(t)\rangle_{n_{b}}\otimes\vert0\rangle_{n_{q}}\otimes\vert0\rangle .
    \end{equation}
    
Step 2 - Quantum Phase Estimation (QPE): 
    Given a linear operator $U$, if $e^{i\pi\sigma}$ is an eigenvalue, the QPE essentially estimates the phase angle $\sigma$ as a binary representation $n_{q}$-bit $|\varphi_{1}\varphi_{2}\cdots\varphi_{n_{q}}\rangle$, $\forall \varphi_{k}\in\{0,1\}$. Using this algorithm, an $n_{q}$-bit binary approximation to the eigenvalues $\tilde{\sigma}_{j}$ of $\tilde{A}$ is computed. For this purpose, we first rescale the matrix by a suitable value so that its eigenvalues lie in a range that is \textit{optimal} for the algorithm's performance \cite{Bharadwaj2023,Childs2021}, and, in addition, is a subset of $[-0.5,0.5]$, to obtain the matrix $\bar{A}$. To now invoke QPE, this matrix is exponentiated as $e^{i\bar{A}T_{0}}$ to form a linear unitary operator, where $T_{0}$ is the Hamiltonian simulation time \cite{Harrow2009,Nielsen2010}. This parameter can be regarded as a scaling parameter that rescales the eigenvalues of $\bar{A}$ such that the eigenvalues $\bar{\sigma}_{j}$ can be represented nearly exactly using an $n_{q}$-bit binary state with minimal truncation error. The matrix $\bar{A}$ can be expanded in the eigenbasis $\vert v_{j}\rangle\langle v_{j}\vert$ such that  
    \begin{equation}
        e^{i\bar{A}T_{0}} :=  \sum\limits_{j=1}^{2^{n_{b}}} e^{i\bar{\sigma}_{j}T_{0}}\vert {v_{j}}\rangle\langle {v_{j}}\vert.
    \label{eq:QPE0}    
    \end{equation} Following this, the QPE then produces the state proportional to
    \begin{equation}
        \vert \psi_\text{STEP2}\rangle =  \sum\limits_{j=1}^{2^{n_{b}}}\hat{\mathbf{c}}_{j}^{m}\vert v_j\rangle_{n_{b}}\otimes\vert\bar{\sigma}_{j0}\rangle_{n_{q}}\otimes \vert0\rangle,
        \label{eq: QPE binary}
    \end{equation}
    where $\bar{\sigma}_{j0} = \bar{\sigma}_{j}T_{0}$ are the binary represented eigenvalues of A rescaled by $T_0$ while $\hat{\mathbf{c}}_{j}^{m}$ are the coefficients of the normalized $\tilde{\mathbf{c}}^{m}$ generated by rotating into the basis of A's eigenvectors $\vert v_j \rangle$.
    
 Step 3 - Conditional Rotation: Here we apply a relative rotation operator on the last ancilla qubit, conditioned on $\bar{\sigma}_{j}$ to compute the inverse $1/\bar{\sigma}$,
    \begin{equation}  \vert \psi_\text{STEP3}\rangle=\sum\limits_{j=1}^{2^{n_{b}}} \hat{\mathbf{c}}_{j}^{m}\vert v_j\rangle_{n_{b}}\otimes\vert\bar{\sigma}_{j0}\rangle_{n_{q}} \otimes\Bigg(\sqrt{1-\frac{K^{2}}{\bar{\sigma}_{j0}^{2}}}\vert 0\rangle + \frac{K}{\bar{\sigma}_{j0}}\vert 1\rangle \Bigg)
    \end{equation}
    where $K$ is a suitably chosen normalization constant.\\    
Step 4: Finally, we perform the inverse QPE (IQPE) operation to reset $n_{q}$ to $|0\rangle$, and follow it up by a measurement of the last ancilla qubit in the computational basis, producing a state proportional to
\begin{equation}
    {|\mathbf{c}^{m+1}\rangle}\sim R \times \sqrt{\dfrac{1}{\sum\limits_{j=0}^{2^{n_{b}-1}}|b_{j}|^{2}/|\bar{\sigma}_{j0}|^{2}}}\, \sum\limits_{j=1}^{2^{n_{b}}}\frac{\hat{\textbf{c}}^m_{j}}{\bar{\sigma}_{j0}} |v_{j}\rangle_{n_{b}}\otimes \vert 0 \rangle_{n_q} \otimes \vert 1 \rangle \,,
    \label{eq: QLSA solution}
\end{equation}
where $R$ is the corresponding rescaling constant to extract the solution appropriately. The solution can now either be read into classical formats by sampling every component of the wavefunction from performing multiple runs of the circuit (so-called shots), or the state can also be post-processed within the quantum simulator to estimate linear and nonlinear functions of the concentration field as shown in \cite{Bharadwaj2023}. This would help conserve a certain degree of quantum advantage. In any case, the results are then finally assimilated in the classical device for post processing and output. 

\subsection{Variational Quantum Algorithm (VQA)}
\label{subsec:vqa}
\begin{figure}
    \centering 
    \includegraphics[width=0.95\textwidth]{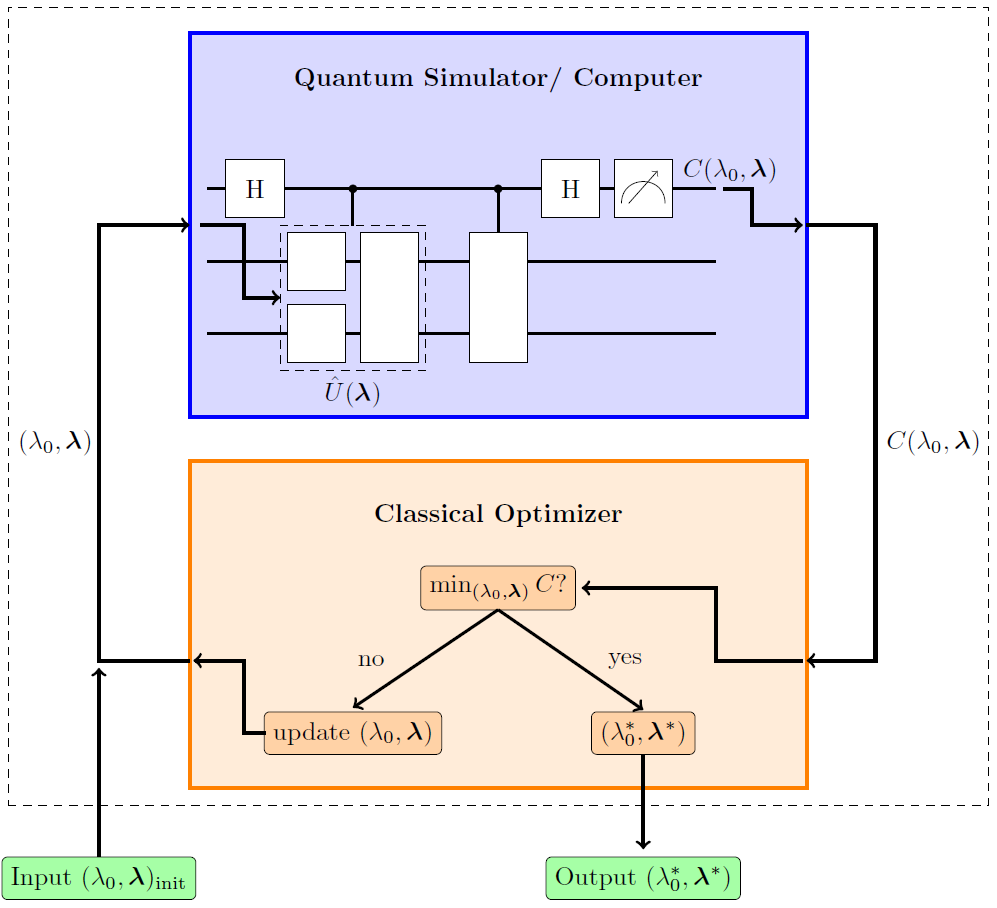}
        \caption{Principal sketch of the Variational Quantum Algorithm (VQA) which interacts between a quantum computational cost evaluation and a classical supervised parameter update. Here, $(\lambda_0,{\bm \lambda})$ is the parameter vector which is optimized, $C(\lambda_0,{\bm \lambda})$ is the cost function, $\hat U({\bm \lambda})$ is the quantum ansatz function and $(\lambda_0^*, {\bm \lambda}^*)$ is the optimal parameter vector, which corresponds with the minimum of the cost function, hence with the solution of the considered problem. The output of the quantum simulator results from measurements of the quantum Hadamard test circuit which is indicated in the quantum simulator block.}
        \label{fig:VQA-concept2}
\end{figure}

The variational quantum algorithm (VQA) is a hybrid quantum-classical algorithm where a parameterized cost function $C$ is minimized by an optimizer \cite{Cerezo2021}. While the cost function is evaluated by a quantum circuit composed of single- and two-qubit gates, the optimization is performed classically. This defines the hybrid character of the algorithm. The general principle of the VQA is shown in Fig.~\ref{fig:VQA-concept2}. The initial parameter set $(\lambda_0, {\bm \lambda})_{\text{init}}$, which consists of the normalization factor $\lambda_0$ and the angles of the single-qubit unitary rotation gates ${\bm \lambda} = (\lambda_1, \lambda_2, \dots )$, is the input to the algorithm. Then, a cost function $C(\lambda_0, {\bm \lambda})$, which is parameterized with $(\lambda_0, {\bm \lambda})$, is evaluated on a quantum device. For our approach, a classical device adds results of multiple quantum circuits together to generate the final cost. The minimum of the cost function corresponds to the solution of the considered problem. This solution is modeled by a quantum ansatz function $\hat U({\bm \lambda})$ which is initialized with the parameter set ${\bm \lambda}$. Measurements of the quantum circuits evaluate the costs. These costs are minimized with an classical optimizer. The optimal parameter set $(\lambda_0^*,{\bm \lambda}^*)$ initializes the ansatz function such that the solution of the given problem can be observed \cite{Cerezo2021}.

%--------------------------------------------------------
%-------------------------------------------------------------

Note that VQA can also straightforwardly be applied to nonlinear equations \cite{Lubasch2020,Pool2022}. In order to derive the cost function for the present advection-diffusion problem, the discrete concentration profile is transformed in vector notation such that eq.~(\ref{eq:advection-diffusion-equation-1D}) can be written explicitly as in Euler type FTCS methods by 
\begin{align}
    \label{eq:Euler-method}
    \vert c(t+\tau) \rangle = (\mathds{1}+\tau \hat{O}) \vert c(t) \rangle,
\end{align}
with the linear operator $\hat{O}=D \partial^2_x - U \partial_x$ and the identity operator $\mathds{1}$. Then, the corresponding cost function $C$ can be found as 
\begin{align}
    \label{eq:Cost function}
    C(\vert c(t+\tau) \rangle) = \| \vert c(t+\tau) \rangle - (\mathds{1}+\tau \hat{O}) \vert c(t) \rangle \|_2 ^2,   
\end{align}
where the minimum of the cost function $C$ corresponds to the solution $\vert c(t+\tau) \rangle$. Following Lubasch et al.~\cite{Lubasch2020}, we define 
\begin{align}
    \label{eq:Definition-Quantum-Ansatz1}
    \vert c(t+\tau) \rangle &= \lambda_0 \vert \psi({\bm \lambda})\rangle = \lambda_0 \hat{U}({\bm \lambda}) \vert 0 \rangle, \\
    \label{eq:Definition-Quantum-Ansatz2}
    \vert c(t) \rangle &= \Tilde{\lambda}_0 \vert\Tilde{\psi} \rangle = \Tilde{\lambda}_0 \hat{\Tilde{U}}\vert 0 \rangle,
\end{align}
where ${\bm \lambda}$ is the parameter vector which initializes the quantum ansatz for the solution $\vert c(t+\tau) \rangle$. The quantum states $\vert \psi \rangle$ are normalized, such that $\| \psi \|_2^2=1$, while for the concentration profile holds 
\begin{equation}
\int_{-L}^L c(x,t) dx = \mbox{const} \quad \Rightarrow \sum_{k=1}^{2^n} c_k =1\,.
\end{equation}
The constant is 1 for the present case due to $c(x,0)=\delta(x)$. In order to fulfill both constraints, normalization parameters, $\lambda_0$ and $\tilde{\lambda}_0$ are introduced. %\footnote
{In the present work we will rescale our solution to an $L_2$-norm of 1 to be directly comparable to the QLSA case.} Thus eq.~\eqref{eq:Cost function} results in
\begin{align}
    \label{eq:Cost function(lambda)}
    C(\lambda_0, {\bm \lambda} ) = \| \lambda_0 \vert \psi ({\bm \lambda}) \rangle - \Tilde{\lambda}_0 (\mathds{1}+\tau \hat{O}) \vert \Tilde{\psi} \rangle \|_2^2.
\end{align}
The norm is evaluated by the scalar product and gives
\begin{align}
    \label{eq:Cost function- scalar product}
    C( \lambda_0,{\bm \lambda} ) &= \lambda_0^2 \underbrace{\langle \psi ({\bm \lambda}) \vert \psi ({\bm \lambda}) \rangle }_{=1}- 2 \lambda_0 \Tilde{\lambda}_0 \langle \psi ({\bm \lambda}) \vert ( \mathds{1} + \tau \hat{O}) \Tilde{\psi} \rangle \\ \notag
    &+ \underbrace{\Tilde{\lambda}_0^2 \langle \Tilde{\psi}| ( \mathds{1} + \tau \hat{O})^{\dagger} ( \mathds{1} + \tau \hat{O})  
    |\Tilde{\psi} \rangle}_{=\text{const.}}
\end{align}
The last term is constant for each time step because it depends only on $\vert \Tilde{\psi} \rangle$ and $\tilde{\lambda}_0^2$, which are fixed from the previous time step. A further decomposition of the scalar product leads to  
\begin{align}
    \label{eq:Cost function- scalar product2}
    C(\lambda_0, {\bm \lambda}) &= \lambda_0^2 - 2 \lambda_0 \Tilde{\lambda}_0 \left[ \langle \psi ({\bm \lambda}) \vert \Tilde{\psi} \rangle  +  \tau  \langle \psi ({\bm \lambda}) \vert \hat{O} \Tilde{\psi} \rangle\right] \nonumber\\ 
    &+ \Tilde{\lambda}_0^2 \left[ 1 + 2 \tau \langle \Tilde{\psi} \vert \hat{O} \Tilde{\psi} \rangle + \tau^2 \langle \Tilde{\psi} \vert \hat{O}^{\dagger}\hat{O} \Tilde{\psi} \rangle \right]\,. 
\end{align}
The linear operator $\hat{O}$ consists of 2 terms, the diffusion term and the advection term. Rather than implementing these terms directly, a 2nd-order finite difference discretization of both operators is used, which is in line with the discussion in Sec.~\ref{sec:class}. For this, the unitary shift operators $\psi_{i+1} = \hat{S}_+ \psi_i$ and $\psi_{i-1} = \hat{S}_- \psi_i$ are defined; see \ref{sec:Cost function-Appendix} for details. Note that thereby, the periodic boundary conditions are imposed. With the definition of these shift operators and $2l/\Delta x = N$ one gets
\begin{align}
    \label{eq:O-pure diffusion}
    \hat{O}  &= D N^2(\hat{S}_+ - 2\mathds{1} + \hat{S}_-) - U \frac{N}{2}(\hat{S}_+ - \hat{S}_-) \nonumber\\ 
    &= (\underbrace{D N^2- U \frac{N}{2}}_{=\alpha}) \hat{S}_+ - \underbrace{2DN^2}_{=\beta}\mathds{1} + (\underbrace{D N^2 + U \frac{N}{2}}_{=\gamma}) \hat{S}_-\,.
\end{align}
Consequently, the cost function can be written as a further decomposition of the scalar products, leading to the following summary of the cost function:
\begin{align}
    \label{eq:Cost function- adv-diffusion-0}
    C(\lambda_0, {\bm \lambda} )  = \lambda_0^2 &- 2 \lambda_0 \Tilde{\lambda}_0 \bigg[ (1-\tau \beta) \langle \psi ({\bm \lambda}) \vert \Tilde{\psi} \rangle + \tau \alpha \langle \psi ({\bm \lambda}) \vert \hat S_+ \Tilde{\psi} \rangle + \tau \gamma \langle \psi ({\bm \lambda}) \vert \hat S_- \Tilde{\psi} \rangle \bigg] \nonumber\\
    &+ \Tilde{\lambda}_0^2 \bigg[ 1 + 2 \tau \alpha \langle \Tilde{\psi} \vert  \hat S_+ \Tilde{\psi} \rangle - 2 \tau \beta + 2 \tau \gamma \underbrace{\langle \Tilde{\psi} \vert  \hat S_- \Tilde{\psi} \rangle}_{=\langle \Tilde{\psi} \vert  \hat S_+ \Tilde{\psi} \rangle}  \bigg] \nonumber \\ 
    &+ \Tilde{\lambda}_0^2 \tau^2 \bigg[ \alpha^2 \underbrace{\langle \hat S_+ \Tilde{\psi} \vert  \hat S_+ \Tilde{\psi} \rangle}_{=1} - \alpha \beta \underbrace{\langle \hat S_+ \Tilde{\psi} \vert \Tilde{\psi} \rangle}_{=\langle \Tilde{\psi} \vert  \hat S_+ \Tilde{\psi} \rangle} + \alpha \gamma \underbrace{\langle \hat S_+ \Tilde{\psi} \vert  \hat S_- \Tilde{\psi} \rangle}_{=\langle \Tilde{\psi} \vert  \hat S_{++} \Tilde{\psi} \rangle} \bigg] \nonumber\\ 
    &+ \Tilde{\lambda}_0^2 \tau^2 \bigg[- \beta \alpha \langle \Tilde{\psi} \vert  \hat S_+ \Tilde{\psi} \rangle +\beta^2  \underbrace{\langle \Tilde{\psi} \vert  \Tilde{\psi} \rangle}_{=\mathds{1}} - \beta \gamma \underbrace{\langle \Tilde{\psi} \vert  \hat S_- \Tilde{\psi} \rangle}_{=\langle \Tilde{\psi} \vert  \hat S_{+} \Tilde{\psi} \rangle}\bigg] \nonumber\\
    &+ \Tilde{\lambda}_0^2 \tau^2 \bigg[ \gamma \alpha \underbrace{\langle \hat S_- \Tilde{\psi} \vert  \hat S_+ \Tilde{\psi} \rangle}_{=\langle \Tilde{\psi} \vert  \hat S_{++} \Tilde{\psi} \rangle} - \gamma \beta \underbrace{\langle \hat S_- \Tilde{\psi} \vert  \Tilde{\psi} \rangle}_{=\langle \Tilde{\psi} \vert  \hat S_{+} \Tilde{\psi} \rangle} + \gamma^2 \underbrace{\langle \hat S_- \Tilde{\psi} \vert  \hat S_- \Tilde{\psi} \rangle}_{=1} \bigg]\,.
\end{align}
We use that $\hat{S}_+=\hat{S}_-^{-1}=\hat{S}_-^{\dagger}$ and $\hat{S}_-=\hat{S}_+^{\dagger}$. This leads to the final cost function
\begin{align}
    \label{eq:Cost function- adv-diffusion}
    C(\lambda_0, {\bm \lambda} )  &= \lambda_0^2 - 2 \lambda_0 \Tilde{\lambda}_0 \bigg[ (1-\tau \beta) \underbrace{\langle \psi ({\bm \lambda}) \vert \Tilde{\psi} \rangle}_{=C_{\mathds{1}}} + \tau \alpha \underbrace{\langle \psi ({\bm \lambda}) \vert \hat S_+ \Tilde{\psi} \rangle}_{=C_{S_+}} + \tau \gamma \underbrace{\langle \psi ({\bm \lambda}) \vert \hat S_- \Tilde{\psi} \rangle}_{=C_{S_-}} \bigg] \nonumber\\ 
    &+ \Tilde{\lambda}_0^2 \bigg[ 1 - 2 \tau \beta + 2 \tau (\alpha + \gamma) \underbrace{\langle \Tilde{\psi} \vert \hat S_+ \Tilde{\psi} \rangle}_{=\Tilde{C}_{S_+}} \bigg] \nonumber \\
    &+ \Tilde{\lambda}_0^2 \tau^2 \bigg [ \alpha^2 + \beta^2 + \gamma^2 - 2 \beta (\alpha + \gamma) \underbrace{\langle \Tilde{\psi} \vert \hat S_+ \Tilde{\psi} \rangle}_{=\Tilde{C}_{S_+}} + 2 \alpha \gamma \underbrace{\langle \Tilde{\psi} \vert \hat S_{++} \Tilde{\psi} \rangle}_{=\Tilde{C}_{S_{++}}} \bigg],
\end{align}
where $C_\mathds{1}$ expresses the contribution of the identity part and $C_{S_{+/-}}$ the contribution of the shift parts. The contributions $\Tilde{C}_{S_{+}}$ and $\Tilde{C}_{S_{++}}$ to the cost function depend on the solution of the previous time step only, and are hence constants. Note that these different terms are evaluated separately and summed classically to give the cost function. This means that one re-prepares the parametrized state a few times every integration time step.

\begin{figure*}
     \begin{subfigure}[b]{1\textwidth}

        \begin{tikzpicture}[object/.style={thin,double,<->}]
        \draw (0,0) -- (11.75,0); % -- node[left=1pt,fill=white] {$\sin \alpha$} (-2,0); ;
        \draw (0,-1) -- (11.75,-1) ;
        %\draw (0,-1.7) -- (8,-1.7) ;
        \draw (0,-2.4) -- (11.75,-2.4) ;
        \node at (0,0) [rectangle,draw=white ,fill=white] {$q_0$};
        \node at (0,-1) [rectangle,draw=white ,fill=white] {$q_1$};
        \node at (0,-1.7) [rectangle,draw=white ,fill=white] {$\vdots$};
        \node at (0,-2.4) [rectangle,draw=white ,fill=white] {$q_{n}$};
        %\draw[black, thick, fill=white] (0.5,-0.4) rectangle (1.2,0.4) node[anchor = west] {H} (-1,-1);
        \draw (0.8,0) node[minimum height=0.8cm,minimum width=0.8cm, fill=white,draw] {$H$};
        \draw (10.2,0) node[minimum height=0.8cm,minimum width=0.8cm, fill=white,draw] {$H$};
        \draw (2,0) -- (2,-0.75) ;
        \filldraw[black] (2,0) circle (2pt);
        \draw (2,-1.7) node[minimum height=2cm,minimum width=1.4cm, fill=white,draw] {$\hat{U}({\bm \lambda})$};
        
        %Shift -
        \node at (4.1,0.8) [rectangle,draw=white ,fill=white] {$\hat{S}_-$};
        \draw (4.1,-1.2) node[minimum height=3.5cm,minimum width=2.4cm, fill= none,draw][dashed] {};
        \draw (3.3,-1.0) circle [radius=7pt];
        \draw (4.8,-2.4) circle [radius=7pt];
        \node at (4.05,-1.6) [rectangle,draw=white ,fill=white] {$\ddots$};
        \draw (3.3,-1.26) -- (3.3,0) ;
        \filldraw[black] (3.3,0) circle (2pt);
        \draw (4.8,-2.66) -- (4.8,0) ;
        \filldraw[black] (4.8,0) circle (2pt);
        \filldraw[black] (4.8,-1.0) circle (2pt);
        
        %Shift +
        \node at (6.9,0.8) [rectangle,draw=white ,fill=white] {$\hat{S}_+$};
        \draw (6.9,-1.2) node[minimum height=3.5cm,minimum width=2.4cm, fill= none,draw][dashed] {};
        \draw (7.7,-1.0) circle [radius=7pt];
        \draw (6.1,-2.4) circle [radius=7pt];
        \node at (6.9,-1.6) [rectangle,draw=white ,fill=white] {$\iddots$};
        \draw (7.7,-1.26) -- (7.7,0) ;
        \filldraw[black] (7.7,0) circle (2pt);
        \draw (6.1,-2.66) -- (6.1,0) ;
        \filldraw[black] (6.1,0) circle (2pt);
        \filldraw[black] (6.1,-1.0) circle (2pt);
        
        \draw (9,0) -- (9,-0.75) ;
        \filldraw[black] (9,0) circle (2pt);
        \draw (9,-1.7) node[minimum height=2cm,minimum width=1.4cm, fill=white,draw] {$\hat{\Tilde{U}}^\dagger$};
        \draw (11.2,0) node[minimum height=0.8cm,minimum width=0.8cm, fill=white,draw] {};
        \draw (11.0,-0.20) .. controls (11.0,0.1) and (11.4,0.1) .. (11.4,-0.20);
        \draw [->] (11.15,-0.20) -- (11.4,0.25) ;
        \end{tikzpicture}
        %\caption{Quantum circuit which evaluates $\Re \langle \psi (\vec{\lambda}) \vert \Tilde{\psi} \rangle$.}
        %\label{fig:VQA_id-qc1}
     \end{subfigure}
        \caption{Quantum circuits for the evaluation of the main cost contributions $C_{\mathds{1}}$, $C_{S_+}$ and $C_{S_-}$. For the evaluation of $C_{\mathds{1}}$, the both shift blocks are neglected, while the evaluation of $C_{S_{-/+}}$ requires the $\hat S_{-/+}$ block. Qubit $q_0$ is an ancillary (in short ancilla) qubit which collects the information for the measurement to the right.}
        \label{fig:Quantum circuits1}
\end{figure*}
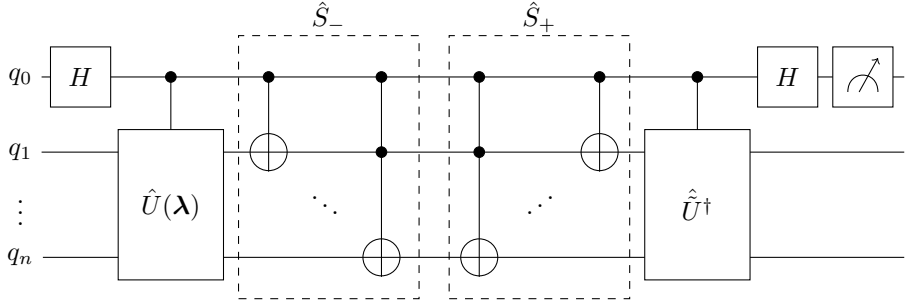

The cost function is evaluated by an adaption of a fundamental quantum circuit, the so-called Hadamard test. In general, the Hadamard test provides an expectation value $\Re \langle \psi \vert \hat U\vert \psi \rangle$ for any variable $\hat U\vert \psi \rangle$ (see \ref{sec:Hadamard-test}). The measurement on the ancilla qubit $q_0$ delivers a measure for the manipulation on the lower qubits $q_1$ to $q_{n-1}$. This measurement is performed such that $p_0-p_1$ is evaluated, where $p_0$ and $p_1$  is the probability to measure the standard basis eigenstates $\vert 0 \rangle$ and $\vert 1 \rangle$ at the ancilla qubit $q_0$, respectively.  In order to evaluate $C_\mathds{1}$ which is $\langle \psi ({\bm \lambda}) \vert \Tilde{\psi} \rangle$, the parameterized quantum ansatz for the solution $\hat U({\bm \lambda})$ and the inverse of previous time step $\tilde{U}^\dagger$ are implemented as controlled gates. If $\hat U({\bm \lambda})$ initializes a state which is completely removed by $\Tilde{U}^\dagger$, the probability to measure the $\vert 0\rangle $ state would be $p_0=1$ because just the Hadamard gates by themselves, cancel their effects causing no net rotation in total. For the evaluation of the $C_{S+/-}$ which is $\langle \psi ({\bm \lambda}) \vert \hat S_{+/-} \Tilde{\psi} \rangle$, the shift operation is added by implementing controlled NOT gates (CNOT) and Toffoli gates which are organized in a particular way. For the $C_{S+}$ case, the CNOT and Toffoli gates are organized in reverse order compared to $C_{S-}$. The structure is shown in Fig.~\ref{fig:Quantum circuits1}. The evaluation of $\Tilde{C}_{S_+}$ and $\Tilde{C}_{S_{++}}$ requires the implementation of $\Hat{\Tilde{U}}$ instead of $\Hat{U}({\bm \lambda})$. In order to realize the double shift operation for $\Tilde{C}_{S_{++}}$, the $\hat S_+$ block can be either implemented twice, or more efficiently, the processing structure starts one qubit lower such that $q_1$ is not affected by the shift operator.

%-------------------------------------------------------------

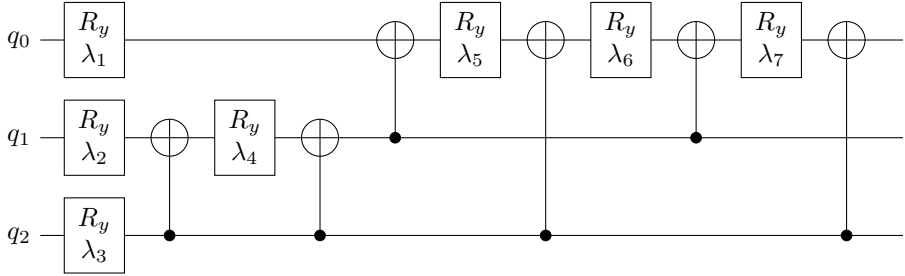
\begin{figure}
     \begin{subfigure}[b]{1\textwidth}

        \begin{tikzpicture}[object/.style={thin,double,<->}]
        \draw (0,0) -- (11.75,0);
        \draw (0,-1.3) -- (11.75,-1.3);
        \draw (0,-2.6) -- (11.75,-2.6);
        \node at (0,0) [rectangle,draw=white ,fill=white] {$q_0$};
        \node at (0,-1.3) [rectangle,draw=white ,fill=white] {$q_1$};
        \node at (0,-2.6) [rectangle,draw=white ,fill=white] {$q_2$};
        \draw (1,0) node[minimum height=1cm,minimum width=0.8cm, fill=white,draw] {};
        \node at (1,-0.2) [rectangle,draw=none ,fill=none] {$\lambda_1$};
        \node at (1,0.2) [rectangle,draw=none ,fill=none] {$R_y$};
        \draw (1,-1.3) node[minimum height=1cm,minimum width=0.8cm, fill=white,draw] {};
        \node at (1,-1.5) [rectangle,draw=none ,fill=none] {$\lambda_2$};
        \node at (1,-1.1) [rectangle,draw=none ,fill=none] {$R_y$};
        \draw (1,-2.6) node[minimum height=1cm,minimum width=0.8cm, fill=white,draw] {};
        \node at (1,-2.8) [rectangle,draw=none ,fill=none] {$\lambda_3$};
        \node at (1,-2.4) [rectangle,draw=none ,fill=none] {$R_y$};
        \filldraw[black] (2,-2.6) circle (2pt);
        \draw (2,-1.3) circle [radius=7pt];
        \draw (2,-1.05) -- (2,-2.6);
        \draw (3,-1.3) node[minimum height=1cm,minimum width=0.8cm, fill=white,draw] {};
        \node at (3,-1.5) [rectangle,draw=none ,fill=none] {$\lambda_4$};
        \node at (3,-1.1) [rectangle,draw=none ,fill=none] {$R_y$};
        \filldraw[black] (4,-2.6) circle (2pt);
        \draw (4,-1.3) circle [radius=7pt];
        \draw (4,-1.05) -- (4,-2.6);
        \filldraw[black] (5,-1.3) circle (2pt);
        \draw (5,0) circle [radius=7pt];
        \draw (5,0.25) -- (5,-1.3);
        \draw (6,0) node[minimum height=1cm,minimum width=0.8cm, fill=white,draw] {};
        \node at (6,-0.2) [rectangle,draw=none ,fill=none] {$\lambda_5$};
        \node at (6,0.2) [rectangle,draw=none ,fill=none] {$R_y$};
        \filldraw[black] (7,-2.6) circle (2pt);
        \draw (7,0) circle [radius=7pt];
        \draw (7,0.25) -- (7,-2.6);
        \draw (8,0) node[minimum height=1cm,minimum width=0.8cm, fill=white,draw] {};
        \node at (8,-0.2) [rectangle,draw=none ,fill=none] {$\lambda_6$};
        \node at (8,0.2) [rectangle,draw=none ,fill=none] {$R_y$};
        \filldraw[black] (9,-1.3) circle (2pt);
        \draw (9,0) circle [radius=7pt];
        \draw (9,0.25) -- (9,-1.3);
        \draw (10,0) node[minimum height=1cm,minimum width=0.8cm, fill=white,draw] {};
        \node at (10,-0.2) [rectangle,draw=none ,fill=none] {$\lambda_7$};
        \node at (10,0.2) [rectangle,draw=none ,fill=none] {$R_y$};
        \filldraw[black] (11,-2.6) circle (2pt);
        \draw (11,0) circle [radius=7pt];
        \draw (11,0.25) -- (11,-2.6);
        %\draw (9.7,-1.7) node[minimum height=2cm,minimum width=1.4cm, fill=white,draw] {$\hat{\Tilde{U}}^\dagger$};
        %\draw (12.2,0) node[minimum height=0.8cm,minimum width=0.8cm, fill=white,draw] {};
        %\draw [->] (12.15,-0.20) -- (12.4,0.25) ;
        \end{tikzpicture}
     \end{subfigure}
        \caption{Example of a universal quantum ansatz function $\hat U({\bm \lambda})$ for a qubit amount of $n=3$ with parameterized $R_y(\lambda_i)$ and CNOT gates.}
        \label{fig:Quantum ansatz}
\end{figure}

The quantum ansatz function can be designed problem-specific or generic without any knowledge about the form of the solution. In this work, a quantum ansatz with an universal function is used which is shown in Fig. \ref{fig:Quantum ansatz}. The ansatz is defined by a special structure of $R_y$ rotation gates and CNOT gates. The $R_y$ gates are parameterized with the parameter set and perform rotations by $\lambda_i/2$ about the y axis of the qubit. CNOT gates negate the state of the target qubit whenever the control qubit is in state $\vert 1\rangle$. This ansatz allows to implement all possible quantum states. The trade-off is, that the solution can always be found, but the optimization has as many parameters to tune as reasonably possible. The usage of other ansatz functions showed no improvement in performance, as discussed in Sec. \ref{subsec:vqa-cost func}. However, the inherent disadvantage of the considered universal ansatz function is the circuit depth which would diminish a possible quantum advantage. This ansatz requires $2^n-1$ parameterized gates and thus, $2^n$ parameters (one additional parameter $\lambda_0$ for normalization purpose) need to be optimized which leads to a high computational effort in circuit execution and optimization. 

%-------------------------------------------------------------

The Nelder-Mead algorithm \cite{Nelder1965} is chosen as the classical optimization algorithm. This algorithm is designed to solve unconstrained optimization problems by a geometric method. For this, the function values of the cost function are evaluated at some points. These points define the so-called simplex. For a two-dimensional data space, a simplex would correspond with a triangle. In the optimization process, the simplex is transformed by reflections and expansions or contractions of the sides of the triangle. 

We also tested other classical optimization algorithms and found that the the Nelder-Mead algorithm is most suitable for the present problem in low-dimensional data spaces. Thus, it is used mainly. The comparison of our results with the other popular classical optimization algorithms is reported in detail in \ref{sec:Appendix-Classical optimization methods}.

%-------------------------------------------------------------
\begin{figure}
    \includegraphics[scale=.6]{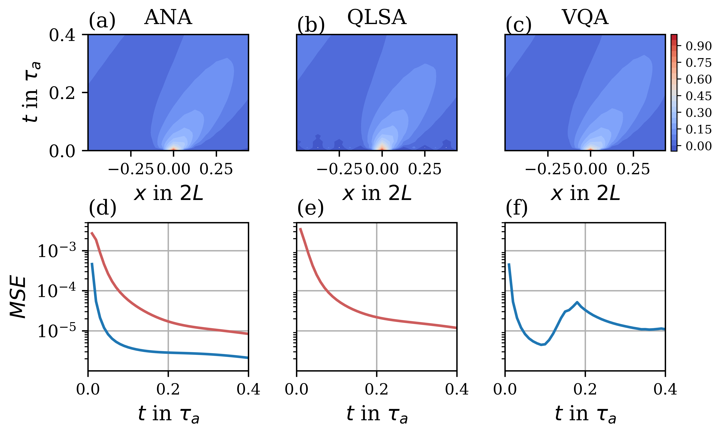}
    \caption{Comparison of the analytical solution to the results of the quantum algorithms. The results are shown as contour plots in $x$ direction over the time $t$ for (a) the analytical solution (ANA), (b) the QLSA results and (c) the VQA results. The mean squared error (MSE) is evaluated to determine the deviation to the analytical solution. For this, the MSE is evaluated with respect (d) to the classical backward (red) and forward (blue) Euler method, (e) to the QLSA results and (f) to the VQA results. Times are given in units of $\tau_a = 2L/U$ and length in units of $2L$. The $x$-axis was resolved with 4 qubits in both cases.}
    \label{fig:Comparison-Analyt-VQA-QLSA}
\end{figure}
%----------------------------------------------------------------

\section{Comparison of quantum algorithms}
\label{sec:comp}

\subsection{Time evolution of concentration profile}
\label{subsec:time-evolution}
In this section, the time evolution of the concentration profiles of the QLSA and the VQA is shown and compared to the analytical solution, cf. eq.~\eqref{eq:analytical solution}. For this comparison, the 4-qubit case with $N=16$ is chosen. The parameters are time step width $\tau=0.001$ s, diffusion constant $D=1$ m$^2$/s, unit length $2L=1$ m, and constant advection velocity $U=10$ m/s. The characteristic time scales of the problem are the advection and the diffusion times. They are given by $\tau_a=0.1$ s and $\tau_d=1$ s. From now on, we proceed with the dimensionless form. Characteristic scales are combined in the dimensionless P\'{e}clet number which is given by 
\begin{equation}
{\rm Pe}=\frac{2UL}{D}=\frac{\tau_d}{\tau_a}=10\,.
\end{equation}
Figure \ref{fig:Comparison-Analyt-VQA-QLSA} provides a first impression of the dynamical evolution of the concentration profile in the form of a contour plot. We provide the analytical solution together with those obtained from QLSA and VQA. The bottom row shows the time evolution of the corresponding mean squared errors which will be detailed below.

\begin{figure*}
    \centering 
    \includegraphics[width=\textwidth]{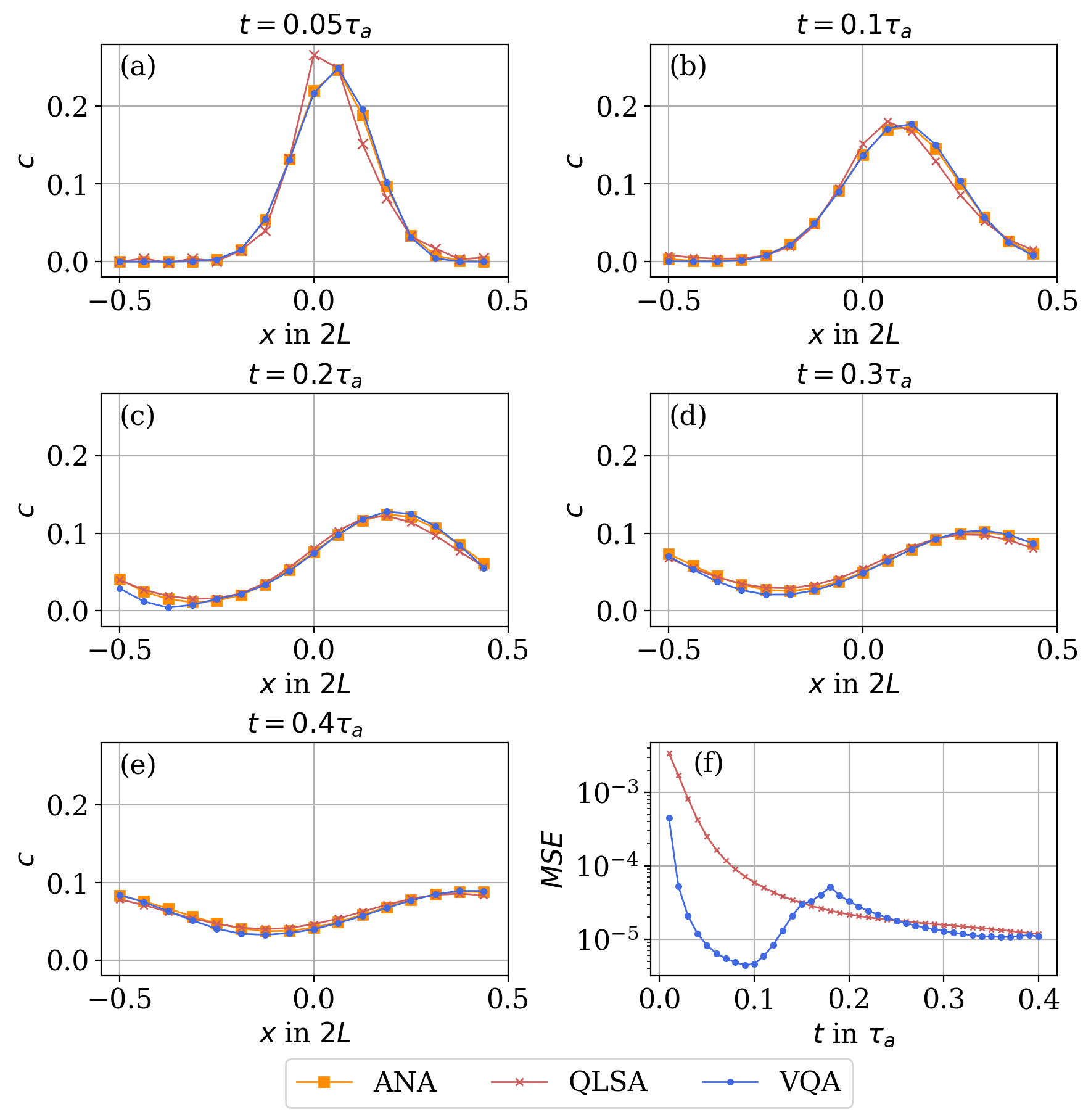}
    \caption{Comparison of the concentration profiles of the analytical solution (ANA) and the results of the VQA and the QLSA for $N=16$ and a time step width $\tau=0.01 \tau_a$. (a)--(e) illustrate the time evolution of the concentration profiles for increasing time. In panel (f), the corresponding MSE is shown where the results of the quantum methods are compared to the analytical solution. The VQA algorithm used 4 qubits for the spatial discretization plus 1 ancilla qubit for the Hadamard tests which gives $n_{\rm tot}=5$. The QLSA algorithm used 4 qubits for the spatial discretization plus 1 ancilla qubit for converting the matrix $A$ into a hermitian matrix. A further ancilla qubit is used together with 8 additional qubits for the register of the QPE. Thus, a total of $n_{\rm tot}=14$ qubits are used in QLSA.}
    \label{fig:Comp-Analyt_vs_VQA-QLSA}
\end{figure*}

The corresponding concentration profiles are plotted in Fig.~\ref{fig:Comp-Analyt_vs_VQA-QLSA} where the time interval is approximately $ 1/30 \tau_d$ or $1/3 \tau_a$. The concentration profiles of the VQA reproduce the advection and diffusion process as expected but the accuracy is limited by the Euler method used in the cost function considered; see subsection \ref{subsec:vqa}. Especially for the early time steps, the performance of the VQA is excellent (see Figs. \ref{fig:Comp-Analyt_vs_VQA-QLSA}a and \ref{fig:Comp-Analyt_vs_VQA-QLSA}b). In the course of the time evolution, the advection-diffusion process starts to depart slightly in comparison to the analytical solution: see Figs. \ref{fig:Comp-Analyt_vs_VQA-QLSA}(c)-(e). 

This behaviour can also be seen in the time evolution of the mean squared error (MSE) in Fig.~\ref{fig:Comp-Analyt_vs_VQA-QLSA}(f) where the VQA result is compared to the analytical solution. The MSE is given by, cf. eq. \eqref{eq:MSE0}, 
\begin{equation}
{\rm MSE}(t_m)=\frac{1}{N}\sum_{i=1}^N \left[c_i^m-c_i^{\rm ANA}(t_m)\right]^2\,.
\label{eq:MSE}
\end{equation}
The MSE curve decreases first, but starts to increase for $t\gtrsim 0.1$ $L/U$. The reason for this behaviour can be assigned to the phase when the largest fraction of the concentration profile crosses the periodic boundary for the first time. This is connected with stronger changes \textcolor{blue}{in} the parameterized vector components $(\lambda_0,{\bm \lambda})$. In more detail, comparing the iterative update of the parameter set, for a time step where the maximum concentration is away from the periodic boundaries, the components $(\lambda_0,{\bm \lambda})$ instead change rapidly when the boundary is crossed by the bulk of distribution. This is a specific property of the geometric Nelder-Mead algorithm. We continued with this algorithm since it still gave the lowest MSE amplitudes for the present problem, boundary conditions, and qubit number range. See also Appendix C for more discussion. With progressing time evolution, the problem fades out and the MSE curve decreases again. This non-monotoneous behaviour was observed for system sizes $N\geq 16$; here the number of optimization parameters was always similar to the number of grid points $N$. 

The concentration profiles computed with the QLSA using the implicit Euler method (BTCS) also capture the physics of the advection-diffusion process very well, both qualitatively and quantitatively. It should be reiterated here that the error in the QLSA solutions (or from any algorithm for that matter) with respect to the analytical solution is bounded from below by the error of the classical solutions from the same underlying numerical scheme, in this case the implicit Euler method. With this in mind, we can now observe from Figs.~\ref{fig:Comp-Analyt_vs_VQA-QLSA}(a) and (b) that the QLSA, in contrast to the VQA, deviates from the analytical solution only during the initial few time steps (for small $t$). However, this is natural since the classical implicit Euler solution also deviates almost exactly in the same manner the QLSA solution does. In fact, the QLSA performs excellently when compared to the classical BTCS solution alone, which we shall discuss more closely in the next subsection. This behaviour is anticipated from Fig.~\ref{fig:Comp-Analyt_vs_FDA1}(c) where, for the problem under discussion, the MSE of the BTCS is in general higher than the FTCS scheme which forms the basis to the VQA solutions. Proceeding further, the QLSA performs progressively better for increasing $t$, as can be seen in Figs.~\ref{fig:Comp-Analyt_vs_VQA-QLSA}(c)-(e), when it begins to closely follow the analytical solution. This is quantified by observing the monotonic decay in the MSE of QLSA with evolving time, as shown in Fig.~\ref{fig:Comp-Analyt_vs_VQA-QLSA}(f). 

In contrast to the VQA, the performance of QLSA improves with increasing system size as one would naturally expect from higher degrees of resolution. On the other hand, similar to the blockades posed by the parametric optimization in the VQA, the accuracy of QLSA critically depends on (a) large enough registers $n_{q}$ for the Quantum Phase Estimation and (b) the right choice of $T_{0}$, see eq.~\eqref{eq:QPE0} in Step 2. For instance, though the MSE of QLSA asymptotes closely with the MSE of VQA for large $t$ in Fig.~\ref{fig:Comp-Analyt_vs_VQA-QLSA} (f), these MSE values of the QLSA can, in general, be further lowered by providing a higher $n_{q}$, without any increase in the finite difference resolution. We investigate all such dependencies more closely in the following sections. 

\begin{figure*}
    \centering 
    \includegraphics[width=\textwidth]{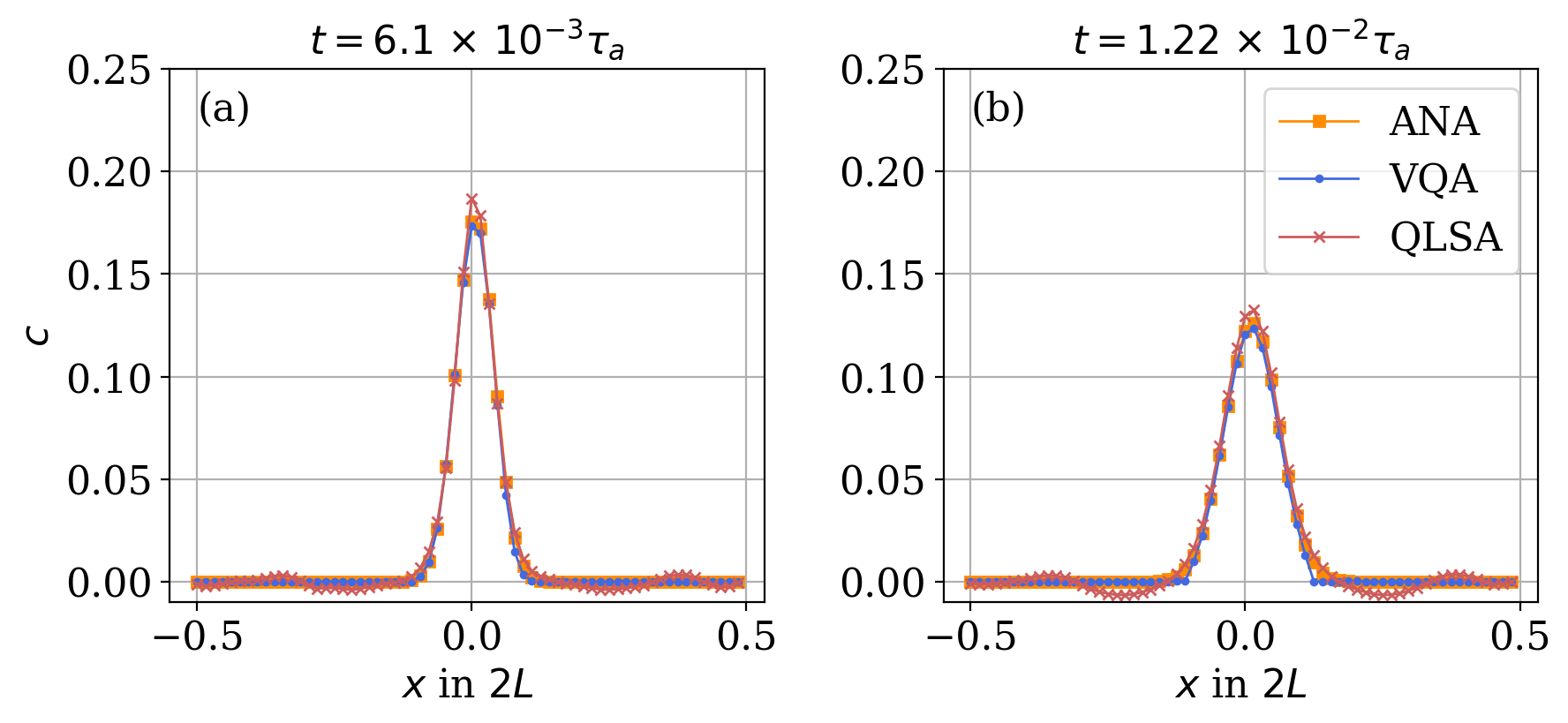}
    \caption{Concentration profiles for $N=64$ of the analytical solution (ANA), the VQA and QLSA results for (a) $t=6.1 \times 10^{-3} \tau_a$ and  (b) $t=1.22 \times 10^{-2}\tau_a$ where the time step width is $\tau=6.1\times 10^{-4} \tau_a$. The VQA algorithm used 6 qubits for the spatial discretization plus 1 ancilla qubit which gives $n_{\rm tot}=7$. The QLSA algorithm used 6 qubits for the spatial discretization plus 1 ancilla qubit for the converting the matrix $A$ into a hermitian matrix. A further ancilla qubit is used together with 6 additional qubits for the register of the QPE. This results to $n_{\rm tot}=14$ qubits for QLSA.}
    \label{fig:Comp-Analyt_vs_VQA-QLSA-N64}
\end{figure*}

The maximum number of grid points is $N=64$ which corresponds to 6 qubits. In this case, $64$ parameters have to be optimized, as described in \ref{sec:Appendix-Classical optimization methods} for a detailed explanation of the classical optimization. The corresponding concentration profiles are shown in Fig.~\ref{fig:Comp-Analyt_vs_VQA-QLSA-N64}. Within the short time interval considered, the advection-diffusion dynamics can be reproduced very well by the VQA. 

\subsection{Dependence on the number of qubits}

With the number of qubits the resolution of the spatial discretization $N$ increases exponentially as $N=2^n$. Apart from the spatial resolution $N$, the total number of qubits $n_{\rm tot}$ in both algorithms is as follows:
\begin{align}
n_{\rm tot}^{\rm VQA}&=\log_2 (N)+1 \,,\\
n_{\rm tot}^{\rm QLSA}&=\log_2 (N) +2 + n_q\,. 
\label{eq:n QLSA}
\end{align}
    In case of QLSA, an additional register with $n_{q}$ qubits is required which corresponds to the QPE qubits.\footnote{In case of QLSA-2, the need for $n_{q}$ can be eliminated given the absence of QPE.} They determine the accuracy of the eigenvalue estimation. The specific dependence on $n_{q}$ will be dealt with in subsection \ref{subsec:qlsa eigenvalue}. The discussion in this section focuses on the dependence on the number of qubits associated with resolution alone. 

For this investigation, the diffusion constant and the velocity are fixed to $D=1$ and $U=10$ and the cases for $N=8,16,32$ grid points are analyzed. In case of the VQA, the time constant $\tau$ is adapted for each discretization. The cost function (see eq.~\eqref{eq:Cost function- adv-diffusion}), which is derived in Sec.~\ref{subsec:vqa}, includes the prefactor $(1-2DN^2 \tau)$. In order to include the term with this prefactor, the time step width $\tau<1/2DN^2$ besides the Courant-Friedrichs-Lewy (CFL) condition, $\tau < 1/(NU)$. Thus, the time steps are $\tau=4\times 10^{-3}$ for $N=8$, $\tau=10^{-3}$ for $N=16$, and $\tau=2.4\times 10^{-4}$ for $N=32$. If we take a CFL number of 0.5, the time steps are smaller by about a factor of 1.5, 3, and 7 than would be classically possible for the first-order scheme. 

\begin{figure*}[htp!]
    \centering 
    \includegraphics[width=\textwidth]{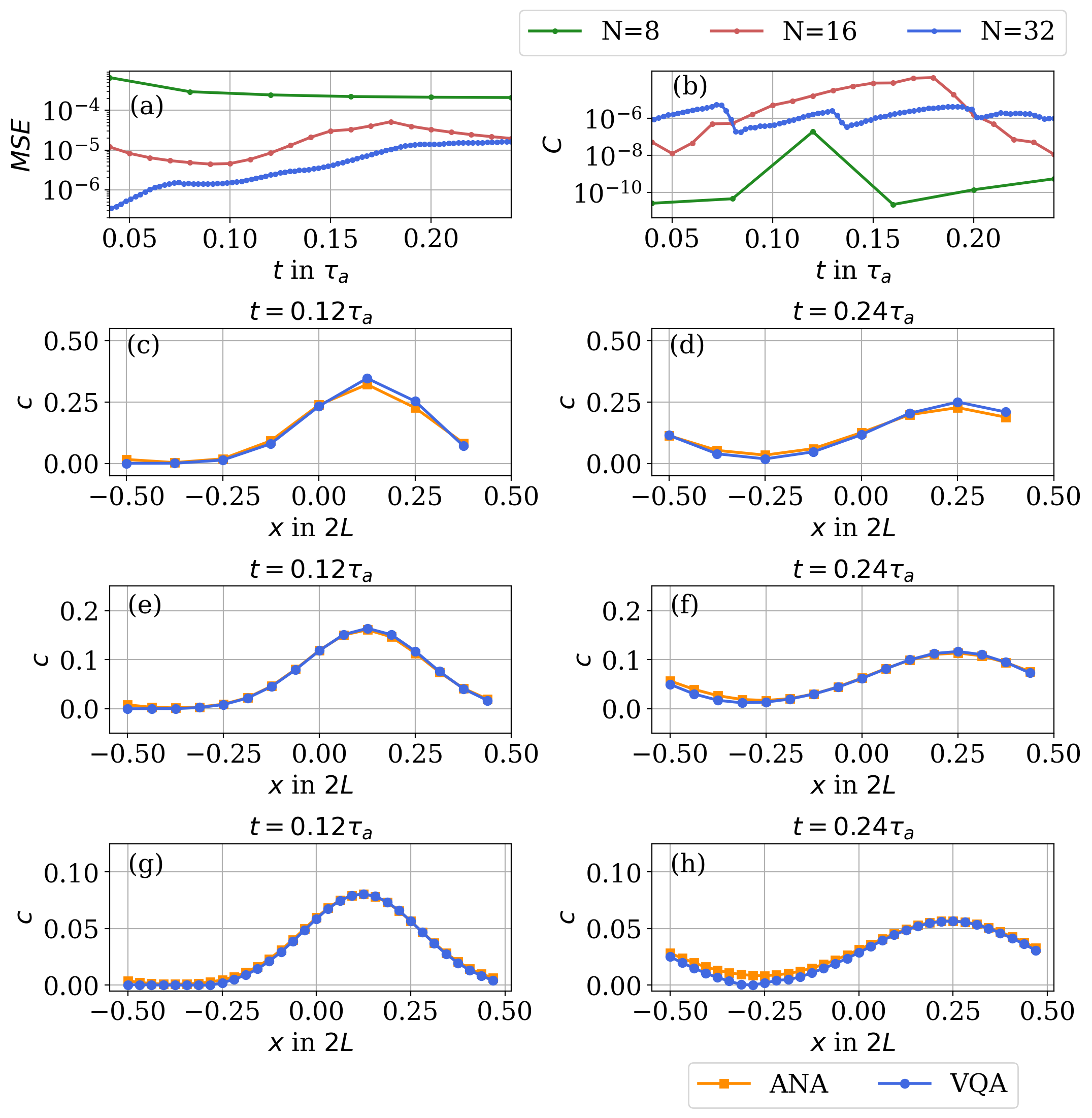}
    \caption{Comparison of VQA results for $N=8,16,32$ to the analytical solution (ANA) with (a) the Mean Squared Error (MSE) over time and (b) the cost function over time. Panels (c) and (d) show the concentration profiles for $N=8$, panels (e) and (f) for $N=16$ and panels (g) and (h) for $N=32$ where $t=0.12\tau_a$ and $t=0.24\tau_a$ are considered for each discretization case.}
    \label{fig:QAmount}
\end{figure*}

For fair comparison, the same set of system parameters is prescribed for the QLSA simulations as well. In the classical finite difference method, which is the basis for the cost function of the VQA, a finer resolution results in a decreased error. For the VQA, this means a finer resolution increases the number of states, hence an increase of the qubit amount. For a time $t \leq 0.12 \tau_a$, it can be shown that the error decreases for cases with a higher number of qubits. The evaluation of the MSE over the time $t=0.04-0.24 \tau_a$ is shown in Fig. \ref{fig:QAmount}(a). It can be seen that a larger number of qubits lead to smaller errors. However, the error for $N=8$ decreases while the curves for $N=16$ and $32$ show a rapid increase. The reason for these inaccuracies in the concentration profiles, which lead to the increased MSE, is the crossing of the periodic boundary of the bulk of the concentration profile as discussed above. This is also shown in Fig. \ref{fig:QAmount}(b). Here, it can be seen that the case for $N=8$ reaches lower cost than the cases for $N=16$, $32$. Thus, it can be assumed that the global minimum in the higher-dimensional parameter space of the optimization is harder to find and hence, the concentration profiles differ slightly from the analytical solution. This can be seen in Figs.~\ref{fig:QAmount}(c)--(h). It can also be seen that problems in reproducing the analytical solution appear especially after crossing the boundary. 

In the case of QLSA, we compare its performance with respect to both the analytical solution and the classical BTCS solution. It should be noted here that, in order to study the effect of resolution alone, we assign a fixed, sufficient number of $n_{q} = \log_{2}(N)+2$ qubits for each case for QPE module of the algorithm. This corresponds to the minimum number of qubits required for any given $N$, such that the solutions are of comparable performance and $n_{q}$ affects every case somewhat similarly, although for increasing resolutions one would need far larger $n_{q}$ registers and therefore lack of which would still bear a weak effect. For $N=8$, $16$, and $32$, one would therefore need to assign a total of $n_{\rm tot}=2\log_{2}(N)+3$ qubits, that is, $n_{\rm tot}=10$, $12$ and $14$ qubits, respectively. 

\begin{figure*}[htp!]
    \centering 
    \includegraphics[width=\textwidth]{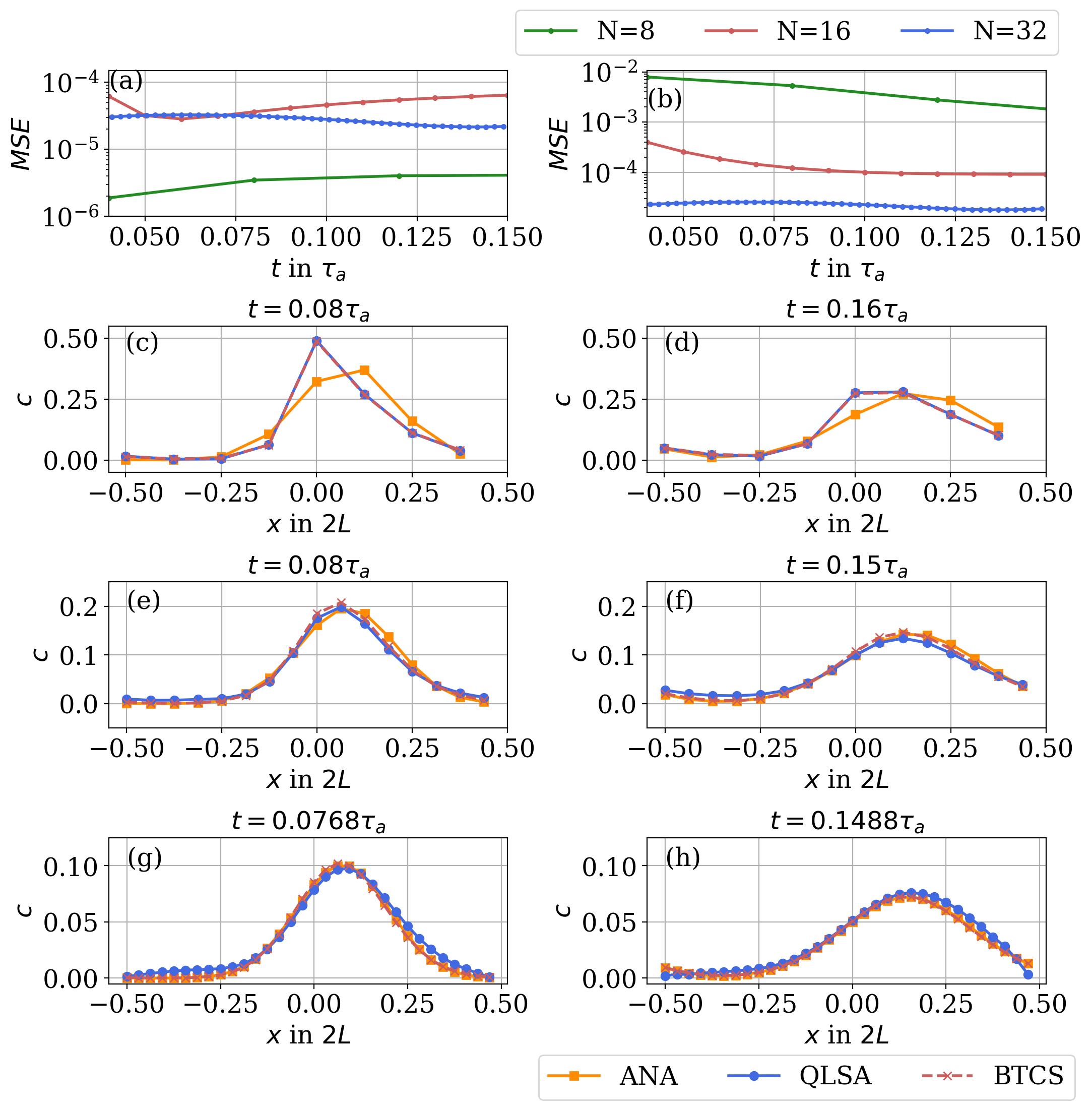}
    \caption{Comparison of QLSA results for $N=8,16,32$ and their MSE over time with respect to (a) implicit time stepping (BTCS) classical solution and (b) the analytical solution (ANA). Panels (c) and (d) show the concentration profiles for $N=8$ at times $t=0.08\tau_a$ and $t=0.16\tau_a$, panels (e) and (f) for $N=16$ at times $t=0.08\tau_a$ and $t=0.15\tau_a$ and panels (g) and (h) for $N=32$ at times $t=0.0768\tau_a$ and $t=0.1488\tau_a$, respectively.}
    \label{fig:QAmount-QLSA}
\end{figure*}

More generally, one needs to assign at least
$n_{q} = \max\{\log_{2}(N)+1,\log_{2}(\kappa)\}$, where $\kappa$ is the condition number of $\tilde{A}$ (see eq.~\eqref{eq: hermitian}). That said, the effect of increasing resolution has a clear consequence of lowering the MSE with respect to the analytical solution for all $t$, which is shown in Fig.~\ref{fig:QAmount-QLSA}(b). This can also be qualitatively observed from Figs.~\ref{fig:QAmount-QLSA}(c)--(h). It can also be seen that in all those figures that QLSA follows the BTCS very closely; however, when quantified by computing the MSE with respect to classical BTCS solution as shown in Fig.~\ref{fig:QAmount-QLSA}(a), we observe a rather non-trivial trend with increasing resolution. Firstly, this MSE in this case is overall lower than the MSE in Fig.~\ref{fig:QAmount-QLSA} (b), suggesting that QLSA solutions are performing extremely well in closely reproducing the classical BTCS, which forms the basis of QLSA discretization. However, one can see that, overall, the $N=8$ case performs the best followed by $N=32$ and $N=16$ which, more or less, have close time evolution trends. This behavior is purely an artifact of the $n_{q}$ assigned in each case. The $n_{q}$ provided for increasing $N$ is progressively inadequate to foster an accurate eigenvalue estimation. 

This can be seen more pronounced in Figs. \ref{fig:Comp-Analyt_vs_VQA-QLSA-N64}(a)-(b) which shows the $N=64$ case for small $t$. The QLSA solution though effectively reproduces the analytical solution barring a modest quantitative error which is seen as spurious oscillations around zero. This quantitative deviation can again be attributed to two factors -- (1) Inadequate $n_{q}$ (which also causes sign flips around zero for small values of the solution field) which in turn causes improper sign handling and evaluation of negative eigenvalues. The seemingly small and negative concentration values, are not essentially non-physical, but are just values with the wrong sign, which once measured can readily be flipped to positive values classically. However, we still show this to highlight that the sign handling quantum subroutines also suffer with insufficient $n_{q}$. (3) The expected errors of solutions from BTCS based schemes in the initial few time steps.  The consequence of such insufficient resource allocation and its remedy is further detailed in subsection \ref{subsec:qlsa eigenvalue}. In essence, we can summarize from the above that the performance and accuracy of QLSA when compared to the analytical solution clearly increases with increasing resolution, when provided with adequate algorithmic resources performs very well as already seen in Figs. (\ref{fig:Comparison-Analyt-VQA-QLSA}) and (\ref{fig:QAmount-QLSA}).

\subsection{Accurate eigenvalue estimation with QLSA}\label{subsec:qlsa eigenvalue}
\begin{figure*}[htpb!]
    \centering{
    \subfloat{
    \includegraphics[trim={0.1cm 0.0cm 1cm 0cm},clip=true,scale=0.45]{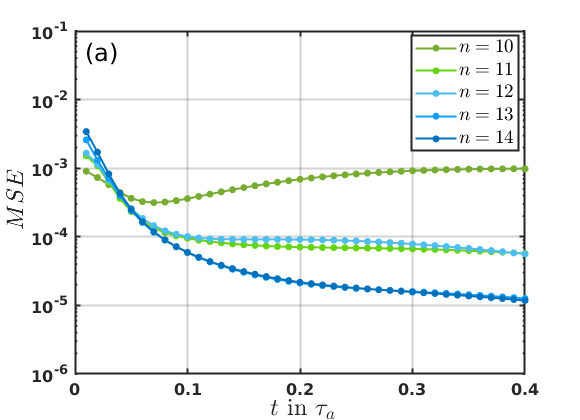}}
    \subfloat{\includegraphics[trim={0.1cm 0.0cm 1cm 0.0cm},clip=true,scale=0.45]{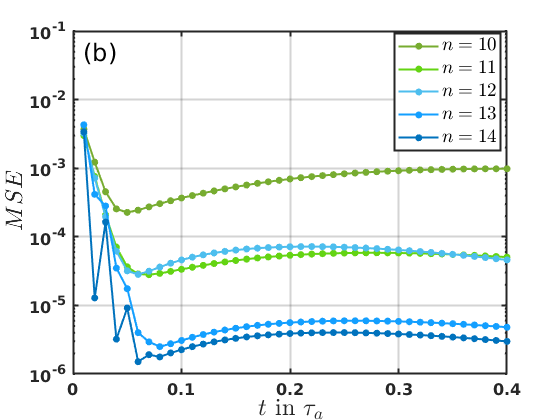}}
    
    \subfloat{\includegraphics[trim={0.1cm 0.0cm 1cm 0.0cm},clip=true,scale=0.45]{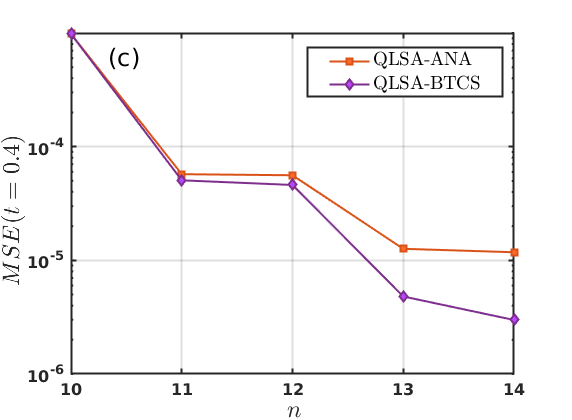}}
    \subfloat{\includegraphics[trim={0.1cm 0.0cm 0cm 0.0cm},clip=true,scale=0.45]{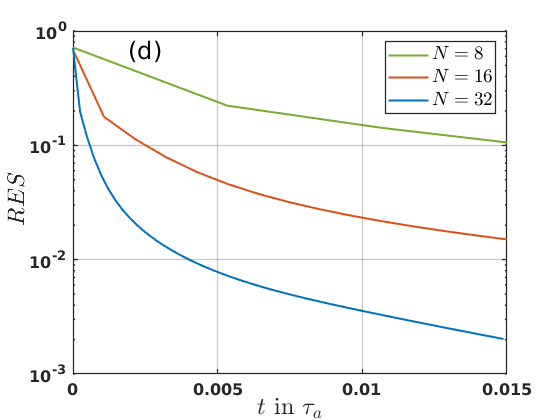}}%{QLSA/Residue_QLSA.png}}
    
    }

    \subfloat{\includegraphics[trim={0.5cm 0cm 0.5cm 0.0cm},clip=true,scale=0.5]{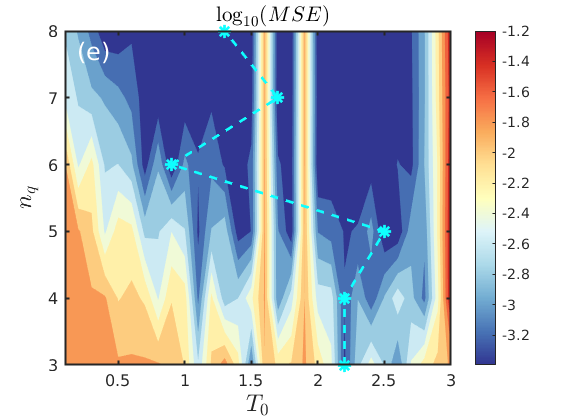}}
       
  \caption{(a) and (b) depict the evolution in time of the MSE from QLSA computed with respect to the analytical and BTCS solutions respectively, plotted for increasing number of qubits. (c) compares the MSE at $t=0.4$ (in $\tau_a$) with respect to analytical and BTCS solutions as function of $n$. (d) shows the time-decay of residue for $N=8,16$ and $32$. (e) Shows the contour of MSE of QLSA with respect to analytical solution as function of both $n_{q}$ and $T_{0}$. The dotted line (cyan) plots the trajectory of minimum MSE for every combination of $(T_{0},n_{q}$). The color bar shows MSE in logarithmic scale (of base 10).}
    \label{fig:QLSA QPE MSE}
\end{figure*}

QLSA, specifically the QLSA-1 discussed in this work, relies heavily on accurate estimation by QPE of the eigenvalues of the matrices under discussion. The errors in this module occur from two primary sources:

(1) \textit{Numerical truncation:} We recall from expression~\eqref{eq: QPE binary}, that the process of estimating eigenvalues in the QPE module requires an intermediate encoding of those values into a binary format using $n_{q}$ qubits. For a given value, an insufficient $n_{q}$ will naturally cause truncation errors of the order $O(2^{-(n_{q}+1)})$. However, given an $n_{q}$, the eigenvalues can always be scaled by choosing an appropriate $T_{0}$ such that $\sigma_{j}T_{0}$ can be represented with the required accuracy. Unfortunately, since the eigenvalues are unknown a priori, the choice of both $n_{q}$ and $T_{0}$ becomes elusive. Figure \ref{fig:QLSA QPE MSE}(e) depicts the intricate connection between the two quantities and their effect on the MSE. A similar contour can be made for the fidelity of the solution as well. The fidelity $F$ would be given by 
\begin{equation}
F= \dfrac{\sum_{i=1}^N\vert c_i^{\rm QLSA}  c_i^{\rm BTCS}\vert}{\| c^{\rm QLSA}\|_2\,\| c^{\rm BTCS}\|_2}\,.
\end{equation}
$F$ is a measure of overlap instead of the difference. However, as noted in \cite{Bharadwaj2023}, it would provide only a rough indication of the QLSA performance. So, for brevity of discussion, we limit ourselves to the computing of MSE only. Though some recommendations for the choice of $T_{0}$ can be made by bounding the minimum and maximum eigenvalues, with functions of either the condition number $\kappa$ or the trace of the matrix, they would still be rough estimates. 

The optimal choice of $T_{0}$ would be such that (a) all eigenvalues are almost exactly represented, and (b) the MSE (with respect to analytical solution) of the concentration field, should neither diverge nor oscillate with time, and decrease with increasing $n_{q}$. This estimation of $T_{0}$ is described in \cite{Bharadwaj2023}. In summary, it requires one to first compute the behavior of condition number $\kappa$ with increasing system sizes. If accurate estimation of $\kappa$ turns out to be expensive, they can also be estimated by tight, theoretical upper bounds (which, of course, would give less accurate results). With this relation in hand, the system of equations is then solved with QLSA for a smaller range of system sizes ($N,t$), $n_{q}$ and $T_{0}\in[0,2\kappa(1-2^{-n_{q, max}})]$. From these results an MSE is computed with respect to classical or analytical solution (available in this case) for every combination of $n_{q}$ and $T_{0}$ as shown in Fig.~\ref{fig:QLSA QPE MSE}(e), for the $N=8$ case integrated up to $t=0.1\tau_a$. The MSE could be of either the entire concentration field (as is the case here), or of a function of the concentration field, such as the scalar dissipation computed using the by Quantum Post Processing (QPP) protocol \cite{Bharadwaj2023}, as denoted in Fig.~\ref{fig:qlsa flowchart}(b). Computing the latter is more efficient and speed-up preserving since it avoids measuring the entire field -- which is a $\mathcal{O}(N)$ operation, and also minimizes the measurement errors associated with it.

Proceeding further, the trajectory of the minimum MSE is traced for every $n_{q}$ and $T_{0}$ as shown in~Fig.~\ref{fig:QLSA QPE MSE}(e) (cyan dotted line) to find a $T^{*}_{0}$ for which most eigenvalues are accurately represented with $n_{q}$ qubits. Finally, using the previously computed $\kappa-N$ relation, a new relation between $N$ and $T^{*}_{0}$ is determined (power-law like behavior), with which one can predict with nominal confidence a $T^{*}_{0}$ for all large system sizes. Note that, for a given problem, this exercise needs to be performed only once and larger system sizes can thereafter be solved with minimal classical precomputing. With the right choice of $T_{0}$ we now solve the system for $N=16$, $\tau=0.001$ with increasing $n_{q}$---and therefore $n$ is given by eq. (\ref{eq:n QLSA}). In this case, $n_{q} \in [4,8]$ and thus $n \in [10,14]$. The MSE is computed with respect to analytical and BTCS solution as shown in Figs.~\ref{fig:QLSA QPE MSE}(a)--(c). Three observations are possible: 

(i) The overall magnitudes of MSE between QLSA and BTCS is lower than with QLSA and the analytical solution. This is expected since QLSA is based on the BTCS scheme and thus follows the classical solution closely. 

(ii) The MSE seems to exhibit a non-monotonic trend with time. The error is initially high as expected when estimating a delta peak, consistent with Fig.~\ref{fig:Comp-Analyt_vs_FDA1}(c). It decreases as the field tends to become more uniform due to diffusion. On the other hand, the initial few time steps pose a beneficial setting to QLSA, since many values of $c(t)$ are close to 0, and therefore errors in eigenvalue estimation are somewhat diminished (except sign issues, which can be corrected easily). 

As the concentration field becomes more uniform and mixed, inaccuracies in $T_{0}$ estimation manifest as a slight increase in MSE. It has to be emphasized here that these errors and their trends also have contributions from errors due to finite differences that are of the order $O(\tau,(\Delta x)^{2})$. Another reason is the following, from eq. \eqref{eq: QLSA solution} the solution is of the form $b_{j}/\sigma_{j}$. So the maximum error stems from the smallest eigenvalue. If the $b_{j}$ associated with the smallest eigenvalue is negligible, then the error from that is also relatively smaller. However, as the concentration peak is advected in space and more $b_{j}$ become finite, the error from the smallest eigenvalue becomes magnified. Further as the field tends to diffuse, again the inaccuracy in eigenvalue estimation is smaller. However, the final large $t$ value of MSE is bounded by $2^{-n_{q}}$. To accurately estimate very small values of $c(t)$ would require a larger $n_{q}$. 

(iii) Finally, the effect of $n_{q}$ is clear from Figs. \ref{fig:QLSA QPE MSE} (a)  and (b). Increasing $n_{q}$ tends to lower MSE in general, but it is in step-like fashion as shown in Fig.~\ref{fig:QLSA QPE MSE}(c) plotted for final time $t=0.4\tau_a$. This is because increasing $n_{q}$ in small steps (of $\mathcal{O}(1)$) does not lower the least count significantly (in $\log_{10}$ or $\log_{e}$). We also plot the residue, given by $RES=\vert c(t)-c(t-\tau)\vert$, as a function of $t$ as shown in Fig.~\ref{fig:QLSA QPE MSE} (d). The monotonic decay in residue symbolizes two aspects: (a) The numerical method and choice of $\tau$ and $\Delta x$ produces stable non-diverging solutions. (b) When the residue falls below a threshold (which can be set arbitrarily small), a steady-state of the solution is reached. The overall residue also decreases with increasing refinement as expected. 

(2) \textit{Sign flips:} Finally, a finer observation is that of the somewhat erratic behavior in MSE in the initial few time steps as shown Fig.~\ref{fig:QLSA QPE MSE} (b) for the $n_{\rm tot}=14$ case. This is because of improper handling of very small negative eigenvalues. The eigenvalues are generally mirrored about 0, to lie in the range [-0.5,0.5]. If the eigenvalues get too close to 0 or when improperly scaled with $T_{0}$, the signs might flip causing a rough MSE profile. This also manifests physically in the concentration field, as depicted by the tiny dark peaks for $t$ close to 0 in Fig.~\ref{fig:Comparison-Analyt-VQA-QLSA}(b). This can be minimized by adding another sign control qubit or by increasing $n_{q}$ and estimating $T_{0}$ more accurately.

\subsection{Comparison of different ansatz functions for VQA}
\label{subsec:vqa-cost func}

\begin{figure}
     \begin{subfigure}[b]{1\textwidth}
        \begin{tikzpicture}[object/.style={thin,double,<->}]
        \node at (2.5,1) [rectangle,draw=white ,fill=white] {(a) TN1};
        \draw (0,0) -- (5,0);
        \draw (0,-1) -- (5,-1);
        \draw (0,-2) -- (5,-2);
        \draw (0,-3) -- (5,-3);
        \node at (0,0) [rectangle,draw=white ,fill=white] {$q_0$};
        \node at (0,-1) [rectangle,draw=white ,fill=white] {$q_1$};
        \node at (0,-2) [rectangle,draw=white ,fill=white] {$q_2$};
        \node at (0,-3) [rectangle,draw=white ,fill=white] {$q_3$};
        \draw (1.2,0) node[minimum height=0.8cm,minimum width=0.8cm, fill=white,draw] {};
        \node at (1.2,-0.2) [rectangle,draw=none ,fill=none] {$\lambda_1$};
        \node at (1.2,0.2) [rectangle,draw=none ,fill=none] {$R_y$};
        \filldraw[black] (2,0) circle (2pt);
        \draw (2,-1) circle [radius=7pt];
        \draw (2,0) -- (2,-1.265);
        \draw (1.5,-0.4) node[minimum height=1.8cm,minimum width=1.8cm, fill=none,draw] {};
        \draw (1.5,-2.4) node[minimum height=1.8cm,minimum width=1.8cm, fill=white,draw] {$\lambda_2$};
        \draw (3.5,-1.4) node[minimum height=1.8cm,minimum width=1.8cm, fill=white,draw] {$\lambda_3$};
        \draw (2.5,-1.4)  node[minimum height=4cm,minimum width=4cm, fill=none,draw] [dashed] {};
        \node at (2,-3.8) [rectangle,draw=white ,fill=none] {{\LARGE$\circlearrowright$}};
        \node at (9.5,1) [rectangle,draw=white ,fill=white] {(b) TN2};
        \draw (6,0) -- (11.8,0);
        \draw (6,-1) -- (11.8,-1);
        \draw (6,-2) -- (11.8,-2);
        \draw (6,-3) -- (11.8,-3);
        \node at (6,0) [rectangle,draw=white ,fill=white] {$q_0$};
        \node at (6,-1) [rectangle,draw=white ,fill=white] {$q_1$};
        \node at (6,-2) [rectangle,draw=white ,fill=white] {$q_2$};
        \node at (6,-3) [rectangle,draw=white ,fill=white] {$q_3$};
        \draw (6.8,0) node[minimum height=0.8cm,minimum width=0.8cm, fill=white,draw] {};
        \node at (6.8,-0.2) [rectangle,draw=none ,fill=none] {$\lambda_{1}$};
        \node at (6.8,0.2) [rectangle,draw=none ,fill=none] {$R_y$};
        \draw (6.8,-1) node[minimum height=0.8cm,minimum width=0.8cm, fill=white,draw] {};
        \node at (6.8,-1.2) [rectangle,draw=none ,fill=none] {$\lambda_{2}$};
        \node at (6.8,-0.8) [rectangle,draw=none ,fill=none] {$R_y$};
        \draw (6.8,-2) node[minimum height=0.8cm,minimum width=0.8cm, fill=white,draw] {};
        \node at (6.8,-2.2) [rectangle,draw=none ,fill=none] {$\lambda_{3}$};
        \node at (6.8,-1.8) [rectangle,draw=none ,fill=none] {$R_y$};
        \draw (6.8,-3) node[minimum height=0.8cm,minimum width=0.8cm, fill=white,draw] {};
        \node at (6.8,-3.2) [rectangle,draw=none ,fill=none] {$\lambda_4$};
        \node at (6.8,-2.8) [rectangle,draw=none ,fill=none] {$R_y$};
        \draw (9.5,-1.4)  node[minimum height=4cm,minimum width=4cm, fill=none,draw] [dashed] {};
    
        \draw (8.5,-0.4) node[minimum height=1.8cm,minimum width=1.8cm, fill=none,draw] {};
        \draw (8.2,0) node[minimum height=0.8cm,minimum width=0.8cm, fill=white,draw] {};
        \node at (8.2,-0.2) [rectangle,draw=none ,fill=none] {$\lambda_5$};
        \node at (8.2,0.2) [rectangle,draw=none ,fill=none] {$R_y$};
        \filldraw[black] (9,0) circle (2pt);
        \draw (9,-1) circle [radius=7pt];
        \draw (9,0) -- (9,-1.265);
        \draw (8.5,-2.4) node[minimum height=1.8cm,minimum width=1.8cm, fill=white,draw] {$\lambda_6$};
        \draw (10.5,-1.4) node[minimum height=1.8cm,minimum width=1.8cm, fill=white,draw] {$\lambda_7$};
        \node at (9.5,-3.8) [rectangle,draw=white ,fill=none] {{\LARGE$\circlearrowright$}};
        \end{tikzpicture}
     \end{subfigure}
        \caption{Example of tensor networks (TN) as generic quantum ansatz functions $\hat U({\bm \lambda})$ for $n=4$ qubits with parameterized $R_y(\lambda_i)$ and CNOT gates. The parameters are continuously indexed for repeated blocks ($\circlearrowright$). (a) Full generic ansatz TN1 and (b) Ansatz TN2 which includes an additional row of $R_y$ gates.}
        \label{fig:Quantum ansatz - generic}
\end{figure}
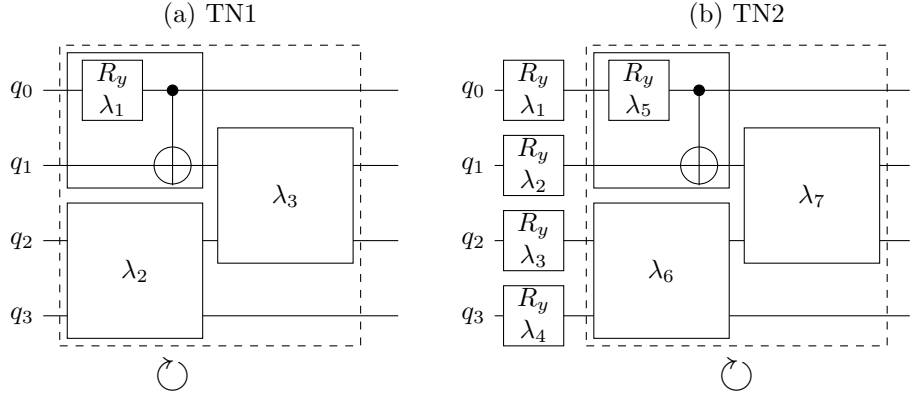

The quantum ansatz $\hat U({\bm \lambda})$ has to meet several requirements. The ansatz should be able to construct the unknown solution for the next time step in the given problem. Secondly, the amount of rotation and entanglement gates should be reduced to a minimum in order to get an efficient and shallow parametric circuit $\hat U({\bm \lambda})$. The efficiency of the ansatz is a key factor in determining whether a quantum advantage can be achieved at all or not. The currently applied universal ansatz (see Fig. \ref{fig:Quantum ansatz}) contains $2^n-1$ parameterized gates to generate the next $n$-qubit state. However with ${\cal O}(N)$ parameterized gates, one cannot obtain a quantum advantage. In order to improve the efficiency of the algorithm, further ansatz functions are now tested. To this end, we analyse the performance of shallow tensor networks (TNs) where $R_y$ and CNOT gates are structured in a staggered way, see e.g. ref. \cite{Barratt2021}. These TNs are visualized in Fig. \ref{fig:Quantum ansatz - generic}. The TN1 ansatz shows a generic structure where the marked code block can be repeated as often as desired in order to build $\hat U({\bm \lambda})$ with a different number of gates. In TN2, a row of $R_y$ gates is added which is inspired by the universal ansatz and should enable an easier generation.

The quality of an ansatz is quantified by a Hadamard test-like quantum circuit which is shown in Fig. \ref{fig:Quantum circuits1}. In the $\hat{U}({\bm \lambda})$ code block, the parameterized ansatz is initialized. Then, the inverse of the wanted solution which is given as the classical FTCS solution of a certain time step $t$ is initialized. The ancilla qubit is measured to obtain the probability of state $\vert 1 \rangle$. In this way we determine the match (or overlap) of the generated with the desired state. In detail, if the probability for the state $\vert 1 \rangle$ is zero, the ansatz $\hat{U}({\bm \lambda})$ generates a state which is perfectly uncomputed by the inverse of the wanted FTCS solution. In other words, the ansatz allows to reconstruct the considered quantum state exactly. If the probability  for the state $\vert 1 \rangle$ is greater than zero, we obtain a measure for the deviation between both states. This measure is called identity costs $C_{\text{id}}$.

\begin{figure*}
    \centering 
    \includegraphics[width=\textwidth]{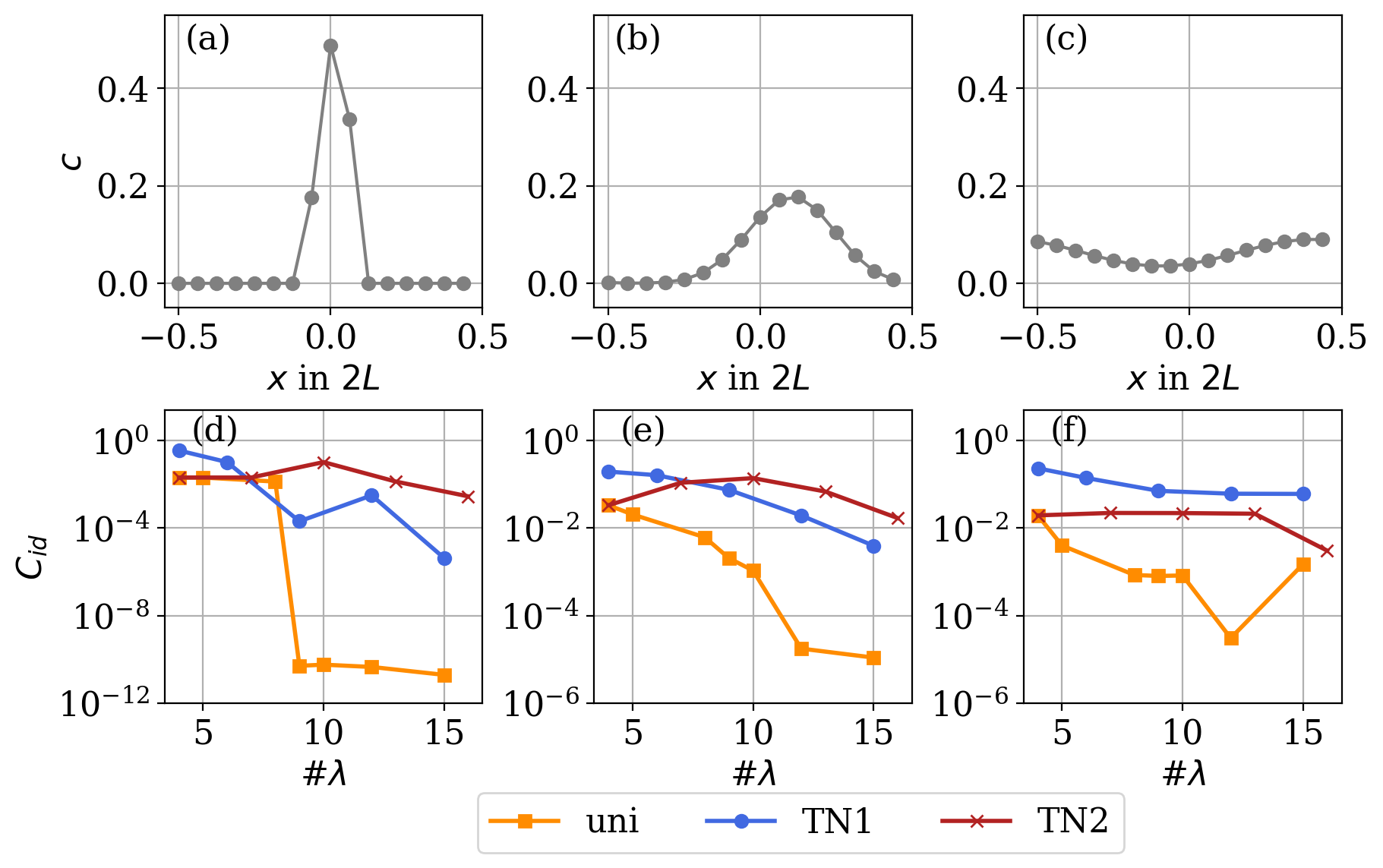}
    \caption{Comparison of the performance of the different investigated ansatz structures for $n=4$ qubits. The upper panel shows the concentration profiles of the FTCS solution for the time steps (a) $t=0.01\tau_a$, (b) $t=0.1\tau_a$ and (c) $t=0.4\tau_a$ with $\tau_a=2L/U$. In the lower panel, the corresponding identity costs $C_{\text{id}}$ of the investigated ansatz structures are shown for a varying number of parameters $\#\lambda$. }
    \label{fig:Ansatz-Comp-4qu}
\end{figure*}

In this work, the considered ansatz structures are evaluated for quantum registers of size $n=4$ and 6. For this, three concentration profiles are chosen which capture the significant shapes of the advection-diffusion problem. The corresponding identity costs $C_{\text{id}}$ are evaluated for these chosen concentration profiles and for a varying number of parameterized $R_y$ gates. In Fig.~\ref{fig:Ansatz-Comp-4qu}, the results for the $n=4$ qubits are shown. We can observe that the universal ansatz leads to low identity costs of $\approx 10^{-11}$--$10^{-5}$, such that this ansatz is suitable to construct the wanted concentration profile, but the number of used parameterized gates is ${\cal O}(2^n)$. The considered tensor networks (TN1, TN2) lead to increased identity costs of $\approx 10^{-4}$--$10^{-2}$. The identity costs decrease slowly for a higher number of parameterized gates, but the costs cannot reach the level of the universal ansatz. Thus, one observes that the investigated TNs are less suitable as an ansatz function for the considered advection-diffusion problem. 

\begin{figure*}
    \centering 
    \includegraphics[width=\textwidth]{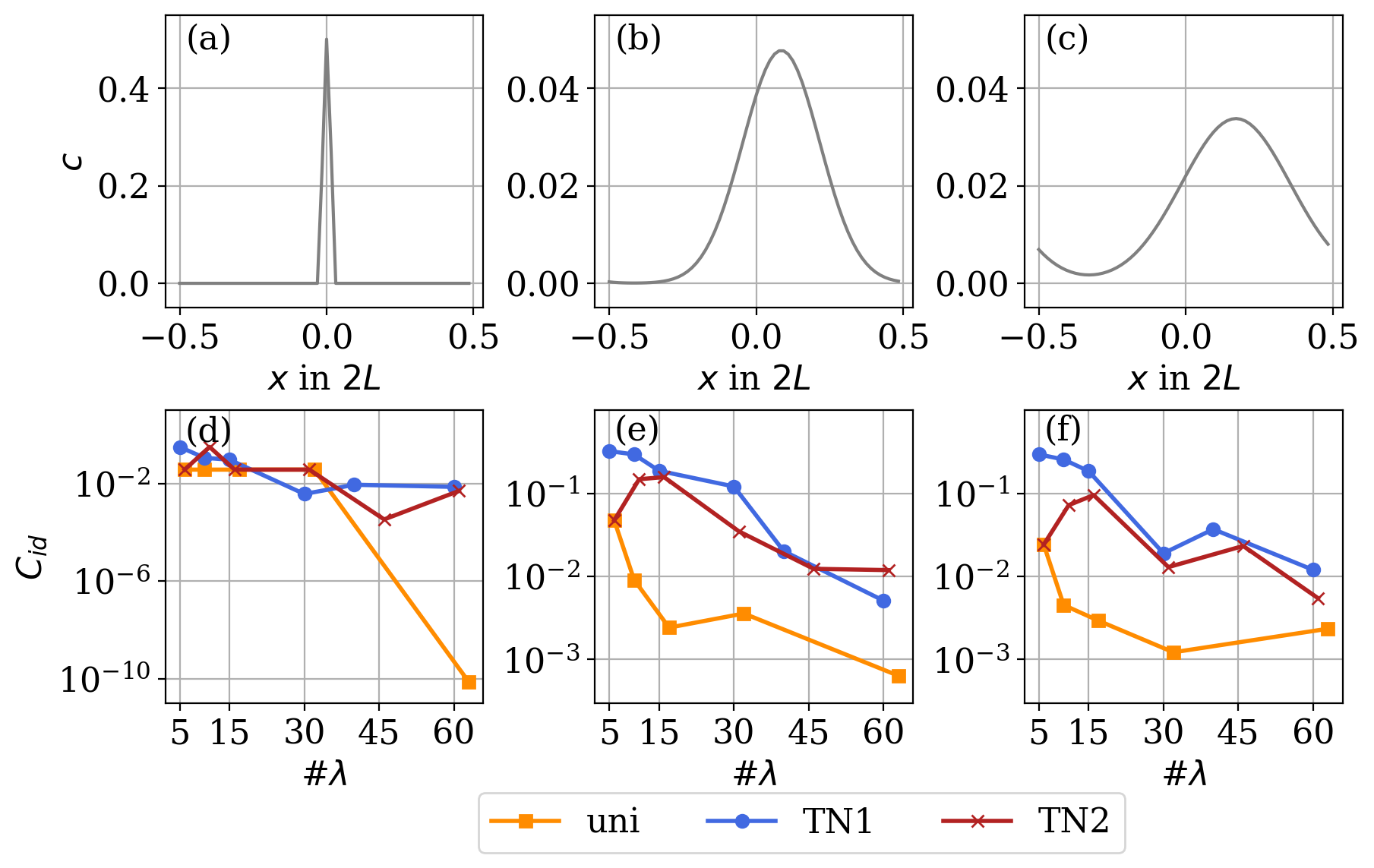}
    \caption{Comparison of the performance of the different investigated ansatz structures for $n=6$ qubits. The upper panel shows again the concentration profiles of the FTCS solution for the time steps (a) $t=6.1 \times 10^{-4}\tau_a$, (b) $t=5.54 \times 10^{-2}\tau_a$ and (c) $t=1.7 \times 10^{-1} \tau_a$. In the lower panel, the corresponding identity costs $C_{\text{id}}$ of the investigated ansatz structures are shown for a varying number of parameters $\#\lambda$. We do not show the individual grid points in (a)--(c) which sum to $N=64$.}
    \label{fig:Ansatz-Comp-6qu}
\end{figure*}

Interestingly, even if the number of gates is similar to the universal ansatz, the tensor networks cannot reproduce the given concentration profiles well. The evaluation of both investigated TNs results in similar identity costs, whereby TN1 tends to be more suitable for sharp Gaussian shaped concentration profiles and TN2 seems to be more appropriate for concentration profiles which are further decayed. Similar results are obtained for the $n=6$ qubit case (see Fig. \ref{fig:Ansatz-Comp-6qu}). The identity costs for TN1 and TN2 differ marginally. The evaluation of the universal ansatz results in significantly lower costs for No. of $\lambda_i\to N$. Furthermore, a reduction of the parameter space is not possible, particularly for the reconstruction of the concentration profiles in early times a high amount of parameterized gates is necessary (see Fig. \ref{fig:Ansatz-Comp-6qu}(d)). 

To conclude, the investigated TN structures for $n=4$ and 6 qubits could not achieve the required accuracy for the state vector generation. Thus, we proceed with the universal ansatz for this investigation. It is also worth noting here that, in contrast to the ansatz used here, the general quantum state preparation (QSP) on the other hand (as used in QLSA), prepares a state exactly, however preparing arbitrary states requires $O(N)$ depth as well. Nevertheless efficient state preparation algorithms exist that prepare states which either have functional forms (such as Gaussian-like or wave-like forms as encountered in the current problem under discussion) or when they are sparse states \cite{Bharadwaj2023}.

\subsection{Qiskit implementation of VQA with circuit noise}
\label{subsec:NISQ-VQA}
In this section, the VQA implementation is performed with a noisy quantum back end, which is close to a real noisy intermediate scale quantum device. The difference from the ideal state vector (SV) simulation is that it includes a measurement process with sampling noise and error rates of the gates, such as bit flips or phase flips. For this, the QASM simulator environment in Qiskit is used which is a noisy quantum circuit back end. A single measurement of an $n$-qubit quantum state on a quantum computer is a random projection on one of the $2^n$ eigenstates with respect to an observable. This observable is the $Z$ matrix on each qubit. In order to obtain the full quantum state vector, such measurements have to be repeated many times to sample all eigenstates sufficiently well. These repetitions of identically prepared quantum simulations of each integration time step of the advection-diffusion equation are termed shots. In this investigation, the number of shots is fixed to $N_S=2^{20}$. This sampling error of the shots decreases with $1/\sqrt{N_S}$.

Real quantum computers are never perfectly isolated from the environment; thus many different types of decoherence errors appear at each of the individual gates. They are smaller for single qubit gates than for entanglement gates. In the simulation software Qiskit, the decoherence noise model implemented is such that customized quantum errors can be set. Thereby, the probabilities for the appearance of quantum gate errors ($p_{\text{gate}}$), errors in measurement ($p_{\text{meas}}$) and in resetting ($p_{\text{reset}}$) of qubits are defined. We have done a study for the case with $N=8$ and compared the results to the corresponding ideal SV simulations reported earlier. To this end, the noise model is implemented as follows. We choose the probabilities $p_{\text{gate}} = 0.008$, $p_{\text{meas}} = 0.03$, and $p_{\text{reset}} = 0.0003$ that a gate error, a measurement error, and a qubit reset error appear in the course of the quantum simulation. 

Furthermore, the evaluation of the cost function is simplified to reduce the appearance of decoherence noise in the quantum circuits. As discussed in subsection \ref{subsec:vqa}, the costs are calculated on the basis of eq.~\eqref{eq:Cost function- scalar product}. When the last term is dropped, which is always a constant term, the number of quantum circuits for the evaluation of the overlap terms can be reduced from 5 to 3. Thus, the minimum of $C(\lambda_0, {\bm \lambda})$ is technically no longer at zero, but at a negative constant value.

\begin{figure*}
    \centering
    \includegraphics[width=\textwidth]{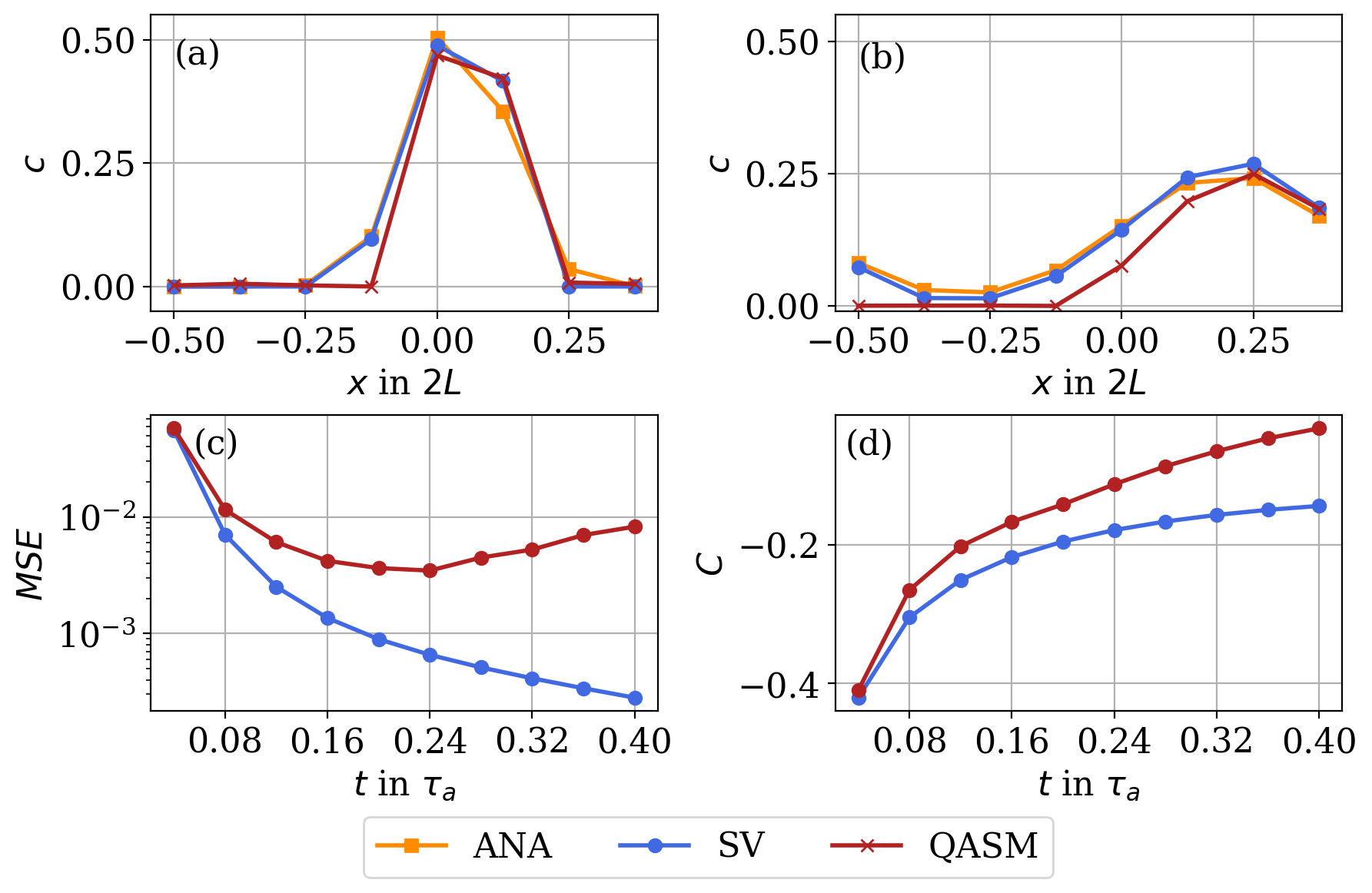}
    \caption{Comparison between the shot based simulator (QASM) with statistical noise where shot number is $2^{20}$ and the ideal state vector (SV) simulator results and the analytical solution (ANA) with the concentration profiles for (a) $t=0.04\tau_a$ and (b) $t=0.2\tau_a$ for the the advection-diffusion case. In (c), the mean squared errors (MSE) of the different simulation methods are compared from which the deviation from the analytical solution is evaluated; in (d) the corresponding cost functions are shown.}  
    \label{fig:Comp-SV-QASM}   
\end{figure*}

The direct comparison with the ideal state vector simulation is shown in Figs.~\ref{fig:Comp-SV-QASM}(a)--(b). It can be seen that the concentration profile with the QASM simulator can reproduce advection and diffusion, but the profile differs slightly from the those of the ideal simulation and the analytical solution. The MSE is evaluated again as a measure of the deviation from the analytical solution and shown in Fig.~\ref{fig:Comp-SV-QASM}(c). As expected, the MSE for the QASM simulation case is higher than that of the ideal simulation. Furthermore, it increases with respect to time. This can be explained by the error propagation from the previous step, which is included in this iterative framework. The cost function of the QASM simulation case is increased in comparison with the state vector simulation case, see Fig.~\ref{fig:Comp-SV-QASM}(d), because the additional noise complicates the optimization procedure. 

\section{Final discussion and outlook}
\label{sec:discussion}

The goal of the present work has been to present a one-to-one comparison of two quantum algorithms to simulate a simple linear flow problem numerically. In the present work we considered a time-dependent linear and one-dimensional advection-diffusion problem on the unit interval with periodic boundary conditions. The time evolution of this fluid mechanical problem is not unitary and hence requires specific steps to be taken in both algorithms. A passive scalar or concentration profile $c(x,t)$ is advected by a constant velocity $U>0$ and subject to a constant molecular diffusion $D$. The P\'{e}clet number is $Pe=10$.  

The two algorithms chosen were a Quantum Linear Systems Algorithm (QLSA) and a Variational Quantum Algorithm (VQA), both of which are hybrid quantum-classical in nature. We have investigated their performance on computational grids varying between $N=8$ and 64, which correspond to 3 and 6 qubits, respectively. We were able to show that both algorithms perform well for the numerical solution of the fluid mechanical problem with a first order time integration scheme, using either a backward or a forward Euler integration scheme. The accuracy of the time evolution in both quantum algorithms, i.e., the forward Euler (FTCS) for VQA and backward Euler (BTCS) for QLSA is bounded from below by the round-off errors of the corresponding classical integration schemes. Accuracy was quantified by a mean squared error (MSE). 

We have shown that both algorithms involved detailed pre-conditioning with respect to specific aspects; this was the major part of this work and could, in our view, be of interest to other users of these specific quantum algorithms. In QLSA, the central point that required comprehensive investigations is related to the approximate determination of eigenvalues of the unitary matrix $\hat U(t)=\exp(i\tilde{A}t)$ in the Quantum Phase Estimation (QPE) stage. It is demonstrated that the number of additional qubits $n_q$ needed for this task and appropriate pre-conditioning is key to the accuracy of the QLSA method. In case of the VQA, the classical optimization algorithm to determine the minimum of the cost function $C(\lambda_0,{\bm \lambda})$ turns out to be the bottleneck. In the present work, we found that the geometric Nelder-Mead algorithm gave the best results, despite a non-monotonic time evolution of the MSE; see also Appendix C. This result holds for the present benchmark task and the chosen boundary conditions. For example, Dirichlet boundary conditions of the concentration profile at the wall might eliminate this behaviour. 

Our results suggest immediate directions for future research for both algorithms. For QLSA, the ongoing and upcoming work focuses on developing algorithms, which are mainly based on the concept of Linear Combination of Unitaries (LCU) \cite{Childs2017,Bharadwaj2023} (QLSA-2) and eliminate the need for QPE. This ameliorates the higher circuit depths and gate count encountered in QLSA-1, making it more suitable for implementation on NISQ devices. Extending these tools to solve nonlinear flow problems by new embedding techniques such as Homotopy Analysis forms a major part of the future work \cite{Bharadwaj2023(2)}. In the case of VQA, surrogate algorithms for the global minimum search of the cost function have been suggested recently \cite{Shaffer2023}. Finding minima in high-dimensional parameter spaces is a general problem of quantum algorithms. This includes quantum machine learning where barren plateaus limit the efficiency of implementation \cite{Cerezo2021}. The strength of the VQA might become better visible for nonlinear problems already attempted, e.g. in Lubasch et al.~\cite{Lubasch2020}. These problems will, however, require higher-order time integration scheme to avoid numerical instabilities, e.g. when time-dependent nonlinear Schrödinger equations have to be solved. These equations describe also nonlinear wave phenomena in fluid mechanics \cite{Dudley2019}.

In the end, we wish to provide a few more general comments on the subject as a whole. The numerical implementations of the classical fluid flow problems as a quantum algorithm have so far not gone beyond the proof-of-concept level. We discuss mostly one-dimensional linear and nonlinear problems while the realistic flows are two- and three-dimensional. Many studies, including most of the existing ones, are implemented in ideal quantum simulation frameworks, thus avoiding the decoherence problems of real and noisy quantum devices that are state-of-the art today. These considerations appear to hinder demonstration of true quantum advantage. In case of the variational methods, most algorithms do not come with theoretical guarantees of quantum advantage (or complexity). The advantage is contingent on problem-specific implementation of the ansatz as well as the parametrization and the optimization methods. On the other hand, QLSA algorithms come with theoretical guarantees of quantum complexity and advantage. However, these algorithms tend to be very sensitive to parameters such as sparsity $S$ of the linear systems matrix $\tilde A$, its condition number $\kappa$, or the choice of unitary bases in case of methods of LCU, making it hard to project their performance on real quantum devices. In both approaches, one also needs to account for the number of shots needed to sample the final quantum state. Therefore careful implementation of Quantum Amplitude Amplification \cite{Brassard2002} is necessary such that one obtains the solution while maintaining quantum advantage.

A desired quantum advantage will most possibly require us to rethink the solution of classical flow problems even more as a quantum mechanical problem. This might be obtained by transforming a nonlinear problem, which is numerically formulated in a finite-dimensional space (for example by a Galerkin method), to a corresponding linear problem in a much higher-dimensional (theoretically infinite-dimensional) Hilbert space. In the latter, the encoding capacity of quantum algorithms would fully unfold. One possible pathway in this respect can be provided by the quantum mechanical implementation of Carleman embeddings \cite{Liu2021,Engel2021}, the Koopman operator formalism \cite{Joseph2020,Giannakis2022,Lin2022} or the Homotopy analysis method. Apart from these, the quantum volume\footnote{Representing the combined measure of the size of quantum circuits (qubits and depth) that can be reliably used \cite{Cross2019}.} of the current and near-term quantum devices is an important consideration while designing algorithms in the hope of harnessing any quantum advantage. Future investigations will probably show us if these routes are indeed successful, and leave us with new scenarios in this research field.       
\\

\noindent \textbf{Acknowledgements}\\
We wish to thank Georgy Zinchenko for insightful discussions. The work of J.I. is funded by the European Union (ERC, MesoComp, 101052786). Views and opinions expressed are however those of the author(s) only and do not necessarily reflect those of the European Union or the European Research Council. Neither the European Union nor the granting authority can be held responsible for them. P.P. is supported by the project no. P2018-02-001 "DeepTurb -- Deep Learning in and of Turbulence" of the Carl Zeiss Foundation, Germany. 

\appendix
\section{Shift operations for cost function $C(\lambda_0,{\bm \lambda})$}
\label{sec:Cost function-Appendix}

\begin{figure}
        \begin{subfigure}[b]{0.45\textwidth}
 
        \begin{tikzpicture}[object/.style={thin,double,<->}]
        \draw (0,0) -- (4,0); % -- node[left=1pt,fill=white] {$\sin \alpha$} (-2,0); ;
        \draw (0,-1) -- (4,-1) ;
        \node at (0,0) [rectangle,draw=white ,fill=white] {$q_1 = \vert 0 \rangle$};
        \node at (0,-1) [rectangle,draw=white ,fill=white] {$q_2=\vert 1 \rangle$};
        \draw[dashed] (0.75,0.5) -- (0.75,-1.5) ;
        \node at (0.75,0.5) [rectangle,draw=white ,fill=white] {$A$};
        %\draw[black, thick, fill=white] (0.5,-0.4) rectangle (1.2,0.4) node[anchor = west] {H} (-1,-1);
        \draw (1.5,0) node[minimum height=0.8cm,minimum width=0.8cm, fill=white,draw] {$X$};
        \draw[dashed] (2.25,0.5) -- (2.25,-1.5) ;
        \node at (2.29,0.5) [rectangle,draw=white ,fill=white] {$B^\prime$};
        \draw (3,-1.0) circle [radius=7pt];
        \draw (3.0,0) -- (3.0,-1.25) ;
        \draw[dashed] (3.75,0.5) -- (3.75,-1.5) ;
        \node at (3.75,0.5) [rectangle,draw=white ,fill=white] {$B$};
        \filldraw[black] (3.0,0) circle (2pt);
        %\node at (4.8,0) [rectangle,draw=white ,fill=white] {$q_1 = \vert 1 \rangle$};
       % \node at (4.8,-1) [rectangle,draw=white ,fill=white] {$q_2=\vert 0 \rangle$};
        \end{tikzpicture}
        \caption{$\hat S_-$ operation.}
        \label{fig:Definiton S-}
     \end{subfigure}
     \hfill
     \begin{subfigure}[b]{0.45\textwidth}
        \begin{tikzpicture}[object/.style={thin,double,<->}]
        \draw (0,0) -- (4,0); % -- node[left=1pt,fill=white] {$\sin \alpha$} (-2,0); ;
        \draw (0,-1) -- (4,-1) ;
        \node at (0,0) [rectangle,draw=white ,fill=white] {$q_1 = \vert0 \rangle$};
        \node at (0,-1) [rectangle,draw=white ,fill=white] {$q_2=\vert 1 \rangle$};
        \draw[dashed] (0.75,0.5) -- (0.75,-1.5) ;
        \node at (0.75,0.5) [rectangle,draw=white ,fill=white] {$A$};
        %\draw[black, thick, fill=white] (0.5,-0.4) rectangle (1.2,0.4) node[anchor = west] {H} (-1,-1);
        \draw (3,0) node[minimum height=0.8cm,minimum width=0.8cm, fill=white,draw] {$X$};
        \draw[dashed] (2.25,0.5) -- (2.25,-1.5) ;
        \node at (2.21,0.5) [rectangle,draw=white ,fill=white] {$C^\prime$};
        \draw (1.5,-1.0) circle [radius=7pt];
        \draw (1.5,0) -- (1.5,-1.25) ;
        \draw[dashed] (3.75,0.5) -- (3.75,-1.5) ;
        \node at (3.75,0.5) [rectangle,draw=white ,fill=white] {$C$};
        \filldraw[black] (1.5,0) circle (2pt);
        %\node at (4.8,0) [rectangle,draw=white ,fill=white] {$q_1 = \vert 1 \rangle$};
        %\node at (4.8,-1) [rectangle,draw=white ,fill=white] {$q_2=\vert 1 \rangle$};     
        \end{tikzpicture}
        \caption{$\hat S_+$ operation.}
        \label{fig:Definiton S+}
     \end{subfigure}
        \caption{Definition of the shift operations for two qubits with $X$ and CNOT gates.}
        \label{fig:Definition Shift}
\end{figure}
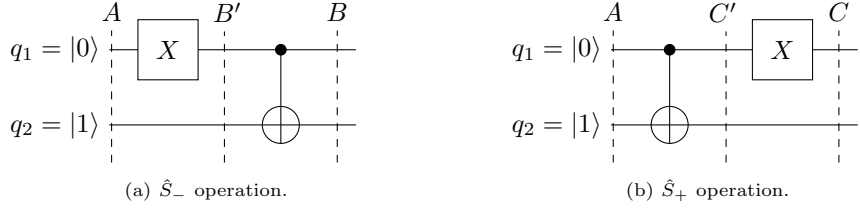

In the following, the shift operations are explained by a two-qubit example. These operations have been used in the evaluation of the cost function in the VQA. For this, a quantum register with the qubits $q_1$ and $q_2$ is considered to be in the initial state $\vert 0 1 \rangle$. For a shift to the left which is defined as $\hat S_-$ operation, an $X$ gate is implemented on the first qubit and afterwards, a controlled NOT gate (CNOT) acts on the register as it is shown in Fig. \ref{fig:Definiton S-}. Consequently, the register is in the following state:
\begin{align}
    \vert 1 0 \rangle \xlongrightarrow[]{B^\prime} \vert 1 1 \rangle \xlongrightarrow[]{B} \vert 0 1 \rangle
    \label{eq:Shift 2}
\end{align}
For the $\hat S_+$ operation, the gates are organized reversely as it is shown in Fig. \ref{fig:Definiton S+}. Then, the following states can be found:
\begin{align}
    \vert 1 0 \rangle \xlongrightarrow[]{C^\prime} \vert 1 0 \rangle \xlongrightarrow[]{C} \vert 1 1 \rangle
    \label{eq:Shift 3}
\end{align}
In the case of the considered cost function (see \ref{subsec:vqa}), the shift operations are applied to fixed quantum states $\vert \Tilde{\psi} \rangle$ within a Hadamard test which is analogous to the evaluation of a scalar product in classical computation. Now we want to show by an example that an application of a $\hat S_+$ operations ($\langle \Tilde{\psi} \vert \hat S_+ \Tilde{\psi} \rangle$) equals the application of any single shift operation for this special case. For this, the two-qubit quantum state is $\vert \Tilde{\psi} \rangle= (a,b,c,d)^T$. Then follows,
\begin{align}
    \langle \Tilde{\psi} \vert \hat S_+ \Tilde{\psi} \rangle &= (a,b,c,d) \cdot (d,a,b,c)^T = ad +ba +cb +dc \\
    \langle \Tilde{\psi} \vert \hat S_- \Tilde{\psi} \rangle &= (a,b,c,d) \cdot (b,c,d,a)^T = ab + bc + cd +da \\
    \langle \hat S_+ \Tilde{\psi} \vert \Tilde{\psi} \rangle &= (d,a,b,c) \cdot (a,b,c,d)^T = da +ab +bc +cd \\
    \langle \hat S_- \Tilde{\psi} \vert \Tilde{\psi} \rangle &= (b,c,d,a) \cdot (a,b,c,d)^T = ba + cb +dc + ad 
    \label{eq:Shift 4}
\end{align}
It can be observed that $\langle \Tilde{\psi} \vert \hat S_+ \Tilde{\psi} \rangle = \langle \Tilde{\psi} \vert \hat S_- \Tilde{\psi} \rangle =  \langle \hat S_+ \Tilde{\psi} \vert \Tilde{\psi} \rangle =  \langle \hat S_- \Tilde{\psi} \vert \Tilde{\psi} \rangle$. Furthermore, these shift operations can be applied to both factors of the scalar product. If the same shift operation is applied to both scalar product entries, e.g., $\langle \hat S_+ \Tilde{\psi} \vert \hat S_+ \Tilde{\psi} \rangle$, the identity will be computed because 
\begin{align}
    \langle \hat S_{(+/-)} \Tilde{\psi} \vert \hat S_{(+/-)} \Tilde{\psi} \rangle = \langle \Tilde{\psi} \vert \Tilde{\psi} \rangle = \sum_i \Tilde{\psi}_i^2 = 1.
    \label{eq:Shift 5}
\end{align}
Furthermore, $\langle \hat S_- \Tilde{\psi} \vert \hat S_+ \Tilde{\psi} \rangle$ can be rewritten to $\langle \Tilde{\psi} \vert \hat S_+ \hat S_+ \Tilde{\psi} \rangle = \langle \Tilde{\psi} \vert \hat S_{++} \Tilde{\psi} \rangle$,
\begin{align}
    \langle \hat S_- \Tilde{\psi} \vert \hat S_+ \Tilde{\psi} \rangle &= (b,c,d,a) \cdot (d,a,b,c)^T = bd + ca +db + ac \\
    \langle \Tilde{\psi} \vert \hat S_{++} \Tilde{\psi} \rangle &= (a,b,c,d) \cdot (c,d,a,b)^T = ac + bd +ca +db
    \label{eq:Shift 6}
\end{align}
Analogously, it holds $\langle \Tilde{\psi} \vert \hat S_{--} \Tilde{\psi} \rangle = \langle \Tilde{\psi} \vert \hat S_{++} \Tilde{\psi} \rangle$.

\section{The Hadamard test}
\label{sec:Hadamard-test}
\begin{figure*}
    \centering
    \begin{tikzpicture}[object/.style={thin,double,<->}]
        \draw (3,0) -- (10,0); % -- node[left=1pt,fill=white] {$\sin \alpha$} (-2,0); ;
        \draw (3,-1) -- (10,-1) ;
        \node at (3,0) [rectangle,draw=white ,fill=white] {$q_0 = \vert 0 \rangle$};
        \node at (3,-1) [rectangle,draw=white ,fill=white] {$q_1=\vert \varphi \rangle$};
        \draw[dashed] (3.75,0.5) -- (3.75,-1.5) ;
        \node at (3.75,0.5) [rectangle,draw=white ,fill=white] {$A$};
        %\draw[black, thick, fill=white] (0.5,-0.4) rectangle (1.2,0.4) node[anchor = west] {H} (-1,-1);
        \draw (4.5,0) node[minimum height=0.8cm,minimum width=0.8cm, fill=white,draw] {$H$};
        \draw (6,-1) node[minimum height=0.8cm,minimum width=0.8cm, fill=white,draw] {$\hat U$};
        \draw (7.5,0) node[minimum height=0.8cm,minimum width=0.8cm, fill=white,draw] {$H$};
        \draw[dashed] (5.25,0.5) -- (5.25,-1.5) ;
        \node at (5.25,0.5) [rectangle,draw=white ,fill=white] {$B$};
        \draw (6.0,0) -- (6.0,-0.6) ;
        \draw[dashed] (6.75,0.5) -- (6.75,-1.5) ;
        \node at (6.75,0.5) [rectangle,draw=white ,fill=white] {$C$};
        \filldraw[black] (6.0,0) circle (2pt);
        \draw[dashed] (8.25,0.5) -- (8.25,-1.5) ;
        \node at (8.25,0.5) [rectangle,draw=white ,fill=white] {$D$};
        \draw (9,0) node[minimum height=0.8cm,minimum width=0.8cm, fill=white,draw] {};
        \draw (8.8,-0.20) .. controls (8.8,0.1) and (9.2,0.1) .. (9.2,-0.20);
        \draw [->] (9,-0.20) -- (9.2,0.25) ;
        \end{tikzpicture}
    \caption{Two qubit quantum circuit which defines the general Hadamard test. Two Hadamard gates are applied on the ancilla qubit $q_0$ and a controlled unitary transformation $\hat U$ is applied between these Hadamard gates. The measurement of the ancilla qubit returns the expectation value $\Re \langle\varphi \vert \hat U \vert \varphi \rangle$ for the variable $\hat U\vert \varphi\rangle$. This Hadamard test principle is used in the VQA to evaluate the cost function.}
    \label{fig:Hadamard-test}
\end{figure*}
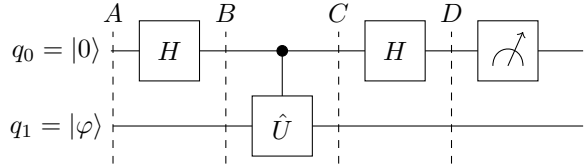

In general, the Hadamard test is a method which allows to find the expectation value $\Re \langle \varphi \vert \hat U \vert \varphi \rangle$. For this, a unitary gate $\hat U$ acts on the qubit $q_1$ which is in the state $\vert \varphi \rangle$. The corresponding quantum circuit is shown in Fig. \ref{fig:Hadamard-test}. First, the quantum circuit is in the state
\begin{align}
\label{eq:Hadamard-test-1}
    \vert q_1 q_0 \rangle_A = \vert \varphi \rangle \otimes \vert 0 \rangle.
\end{align}
The first Hadamard gate acts on the zeroth qubit such that
\begin{align}
\label{eq:Hadamard-test-2}
    \vert q_1 q_0 \rangle_B &= \vert \varphi \rangle \otimes \frac{1}{\sqrt{2}} \left( \vert 0 \rangle + \vert1 \rangle\right)
    = \frac{1}{\sqrt{2}} \left( \vert \varphi \rangle \otimes\vert 0 \rangle + \vert \varphi \rangle \otimes\vert1 \rangle\right).
\end{align}
The unitary gate is implemented on the qubit $q_1$ and is controlled to the ancilla qubit $q_0$ such that follows:
\begin{align}
\label{eq:Hadamard-test-3}
    \vert q_1 q_0 \rangle_C &= \frac{1}{\sqrt{2}} \left( \vert \varphi \rangle \otimes\vert 0 \rangle + \hat U\vert \varphi \rangle \otimes\vert1 \rangle\right).
\end{align}
Considering the second Hadamard gate, the state of the quantum circuit changes to the following one: 
\begin{align}
\label{eq:Hadamard-test-4}
    \vert q_1 q_0 \rangle_D &= \frac{1}{2} \left( \vert \varphi \rangle \otimes (\vert 0 \rangle +\vert 1 \rangle)  + \hat{U}\vert \varphi \rangle \otimes (\vert 0 \rangle -\vert1 \rangle\right) \\ \notag
    &= \frac{1}{2} \left( (\mathds{1} + \hat U) \vert \varphi \rangle \otimes\vert 0 \rangle + (\mathds{1} - \hat U)\vert \varphi \rangle \otimes\vert1 \rangle\right) 
\end{align}
The measurement is performed in the standard $Z$ basis such that 
\begin{align}
\label{eq:Hadamard-test-5}
    p_0-p_1 &= \frac{1}{4} \langle \varphi \vert  (\mathds{1} + \hat U)^\dagger (\mathds{1} + \hat U) \vert 
    \varphi\rangle- \frac{1}{4} \langle \varphi \vert  (\mathds{1} - \hat U)^\dagger (\mathds{1} - \hat U) \vert \varphi\rangle \\ \notag
    &= \frac{1}{2} \langle \varphi \vert  \hat U^\dagger + \hat U \vert \varphi\rangle 
    = \Re \langle \varphi \vert  \hat U \vert \varphi\rangle \,.
\end{align}

\section{Classical optimization methods for the VQA}
\label{sec:Appendix-Classical optimization methods}
In this appendix, the classical optimization methods for the VQA are introduced and compared. The optimization in the considered advection-diffusion problem is challenging due to different aspects. First, the number of parameters which are optimized and hence, the complexity of the optimization problem, scales with the qubit number. Consequently, a fine discretization with a large number of qubits and the chosen ansatz function lead to a high-dimensional parameter space and a complex-shaped cost function for the classical optimization. Secondly, vanishing gradients which drive the search in a local minimum, also known as barren plateaus \cite{Uvarov2021}, complicate the search for the global minimum. 

\begin{figure*}
    \includegraphics[width=\textwidth]{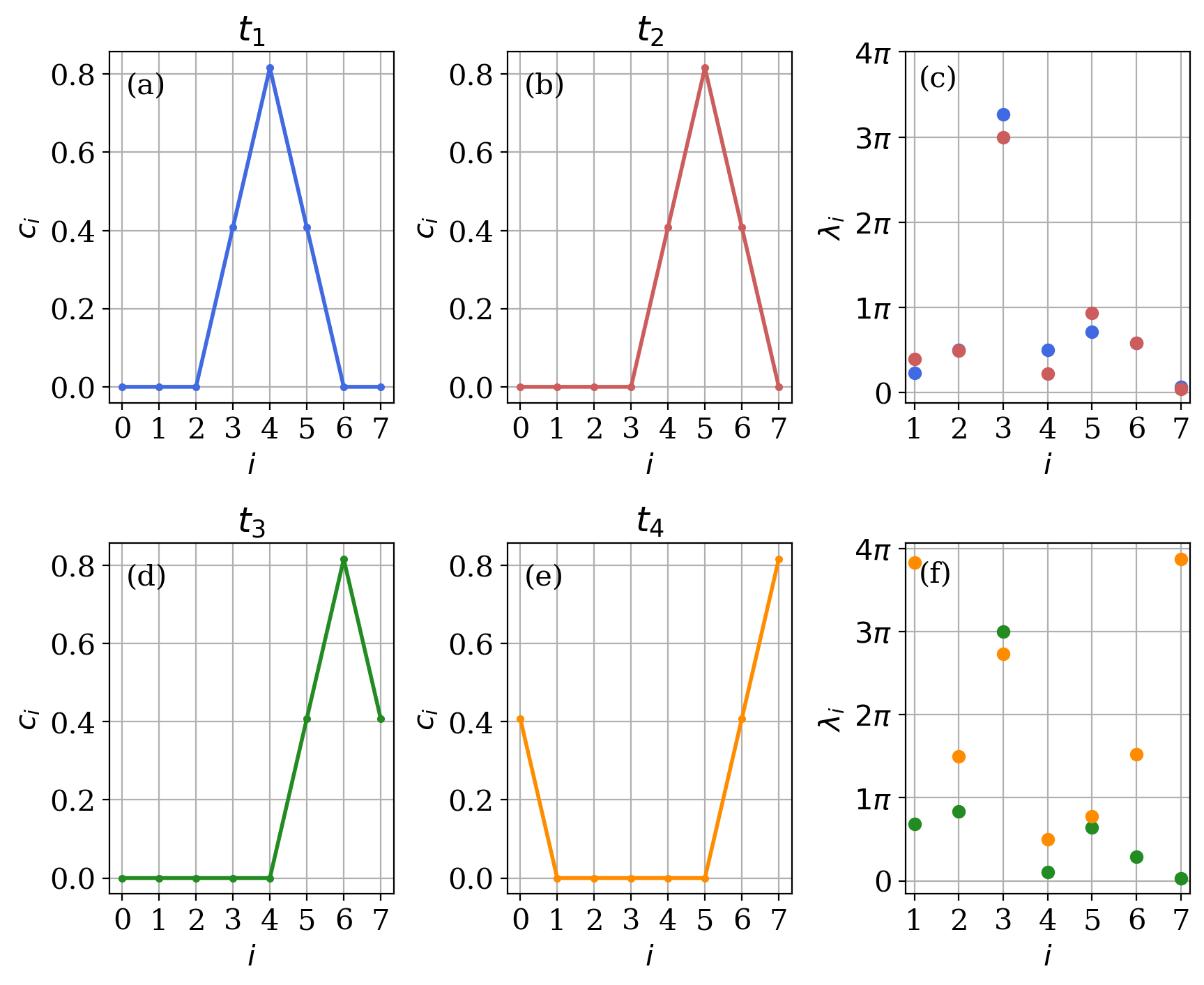}
    \caption{Comparison of the changing parameter set ${\bm \lambda}$ for an advection crossing the periodic boundary by the example of a simple advection of a hat signal for $N=8$. For advection far from the boundaries, the concentration profile for a time (a) $t_1$ and (b) $t_2$ is shown with (c) the corresponding distribution of the parameter set ${\bm \lambda} = (\lambda_1, \dots, \lambda_7)$. The concentration profiles for a time (d) $t_3$ and (e) $t_4$ show the case where the periodic boundaries are crossed and the corresponding distribution of the parameters ${\bm \lambda}$ can be found in panel (f). The comparison between the panels (c) and (f) show that crossing the boundary layer results in larger changes of the parameter vector which can be problematic for the optimization routine.}
    \label{fig:Lambda-Change}
\end{figure*}

Furthermore, the imposed periodic boundary conditions can induce rapidly changing parameter sets at the boundaries. This aspect is visualized by the simple example of a triangle function which is moving by one cell per time step in Fig. \ref{fig:Lambda-Change}. The movement far away from the boundaries results in small changes of the parameter set $(\lambda_0, {\bm \lambda})$. If the periodic boundary is crossed and entries at the other boundary appear, the state vector which models the concentration profile changes strongly and hence, the corresponding parameter set ${\bm \lambda}$ shows major modifications. This aspect is specific to geometric optimization algorithms. Lastly, the existence of noise in the evaluation of quantum circuits contributes to the challenges of the classical optimization. In this work, ideal simulations were considered and hence, the impact of noise is neglected in the selection of the optimization algorithm. For the comparison of the classical optimization algorithms, the VQA is applied to the one-dimensional advection-diffusion equation for the case with $N=16$. The parameters are $D=1$, $u=10$, and $\tau= 0.001$ and the computation is performed for a total time $T=30\tau$. 

The {\em Nelder-Mead algorithm} (NM) \cite{Nelder1965} or downhill-simplex algorithm is designed to solve classical unconstrained optimization problems without any gradient approximation. The algorithm only uses the function values at some points which construct the simplex in the hyperplane. This simplex is transformed by geometric operations such as reflection, expansion, contraction and shrinking. The Nelder-Mead algorithm uses a geometric method in order to find the minimum of the given cost function. The application of the Nelder-Mead algorithm in the considered advection-diffusion problem showed that this algorithm is robust for small search spaces ($N\leq 16$). Moreover, the mean computation time per time step is small in comparison to other methods. However, problems appear if the qubit amount is increased or there are entries at the elements at the boundary. Reasonable for this might be the fact that the parameter set changes rapidly at the periodic boundary in comparison to the changes far away from the boundary, see again Fig. \ref{fig:Lambda-Change}. Consequently, other optimization algorithms need to be considered for an increased qubit amount. 

The {\em Broyden-Fletcher-Goldfarb-Shanno algorithm} (BFGS) applies a quasi-Newton method for solving unconstrained, nonlinear optimization problems. Thereby, the Hessian matrix of the cost function is approximated by the evaluation of the gradients (or the approximated gradients) in order to find the descent direction in the hyperparameter landscape. In this work, an adaption of the BFGS algorithm is used. The {\em Limited memory-BFGS algorithm for bound constraints} (L-BFGS-B) \cite{Liu1989} uses a limited amount of computer memory which makes the algorithm suitable for large search spaces. Furthermore, it can handle bound constraints. In this investigation, the L-BFGS-B algorithm cannot find the global minimum such that the mean MSE is approximately $10^{-4}$. A possible reason for this is the disadvantageous initial parameter set. In order to improve the performance, the L-BFGS-B method is combined with other preceding optimization methods which aim at finding an appropriate region for further optimization. 

\begin{figure*}
    \includegraphics[width=\textwidth]{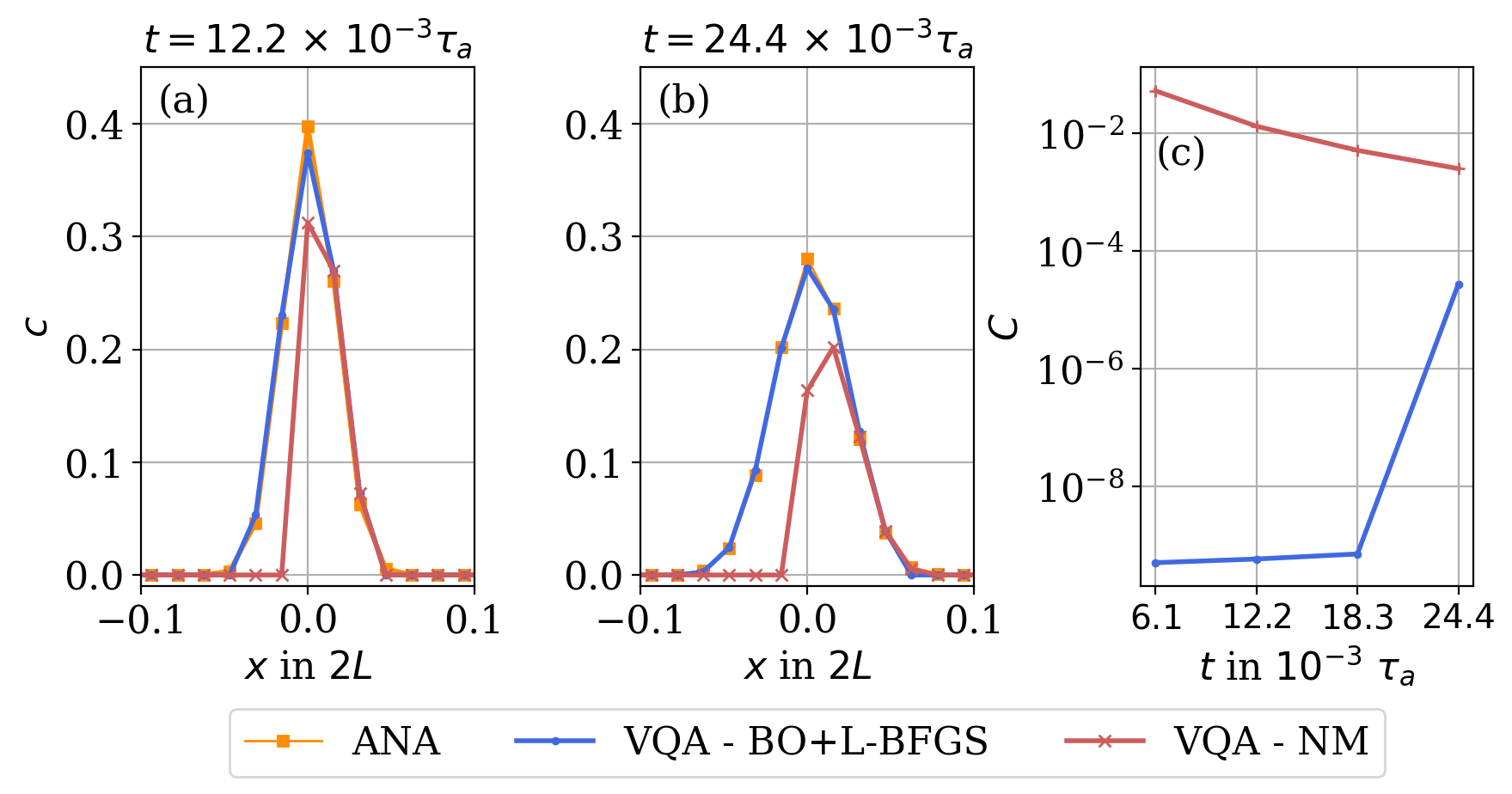}
    \caption{Comparison of the concentration profiles of the analytical solution (ANA) and the VQA results with the Nelder-Mead optimization (NM) and the combined Bayesian and Broyden-Fletcher-Goldfarb-Shanno algorithm optimization (BO+BFGS-L) for $N=64$. In (a) the concentration profile of the  for $t=2\tau$ is shown and in (b) for $t=4\tau$. Here, the parameters are chosen to be $D=1$, $U=10$, $\tau=6.1\times 10^{-5}$. In (c) the corresponding cost function is shown.}
    \label{fig:Comp-Optimizer-N64}
\end{figure*}

The first one is the combination of the {\em Bayesian optimization} and {\em L-BFGS-B algorithm} (BO+L-BFGS). Bayesian optimization \cite{Mockus1978} is suitable for optimization problems where the costs are given as black box functions, are expensive to evaluate or include noise. This method approximates the cost function by a Gaussian process regression based on previous observations. An acquisition function determines the next samples for the observations whereby random exploration steps can be added in order to include a wide range of observations for the fitting of the cost landscape. This aims at finding a promising region for further optimization with the L-BFGS-B algorithm. With this preceding Bayesian optimization, the results of L-BFGS-B algorithm could be improved such that the mean MSE is approximate $10^{-5}$, but the computation time is increased. However, the application of this combination of methods is reasonable if the system size is increased. For this, the test case was expanded to $N=64$. Thereby, the Nelder-Mead optimization cannot find the global minimum of the cost function ($C\approx 10^{-2}$) and hence, the optimization fails. In contrast, the combination of Bayesian optimization and L-BFGS-B algorithm shows small costs ($C\approx 10^{-10},\dots,10^{-5}$) and good results in accuracy. This is shown qualitatively with the concentration profiles in Fig. \ref{fig:Comp-Optimizer-N64}a and \ref{fig:Comp-Optimizer-N64}b and with the comparison of the cost functions (Fig. \ref{fig:Comp-Optimizer-N64}c).

Secondly, the {\em Adaptive moments algorithm} (Adam) \cite{Kingma2014} is combined with the L-BFGS-B algorithm. The Adam algorithm uses a gradient-based method to determine the descent direction in the hyperparameter landscape. It includes an adaptive learning rate and momentum for each update step of the parameter which improves the performance in cases of sparse gradients and non-stationary problems. Furthermore, it is suitable for the optimization of large parameter sets. In this investigation, the combination of Adam and L-BFGS-B algorithm can process the test case with an accuracy $\approx 10^{-5}$, but the computational effort is too high to use this method for increased system sizes. Reasonable for this high computation time is the large amount of required iteration steps which all include the calculation of the gradients. 

The {\em Simultaneous Perturbation Stochastic Approximation (SPSA)} \cite{Spall1998} is an optimization algorithm which uses a stochastic method to approximate the gradient of the cost function. Thereby, the cost function is evaluated twice with completely perturbed parameter sets. The parameters are chosen randomly using a zero-mean distribution. This algorithm is robust to noise. In this work, the SPSA optimization could not find the minimum such that the costs were found to be $\approx 10^{-2}$ which results in high mean squared errors ($\approx 10^{-3}$). 

\begin{figure*}
    \centering
    \includegraphics[width=\textwidth]{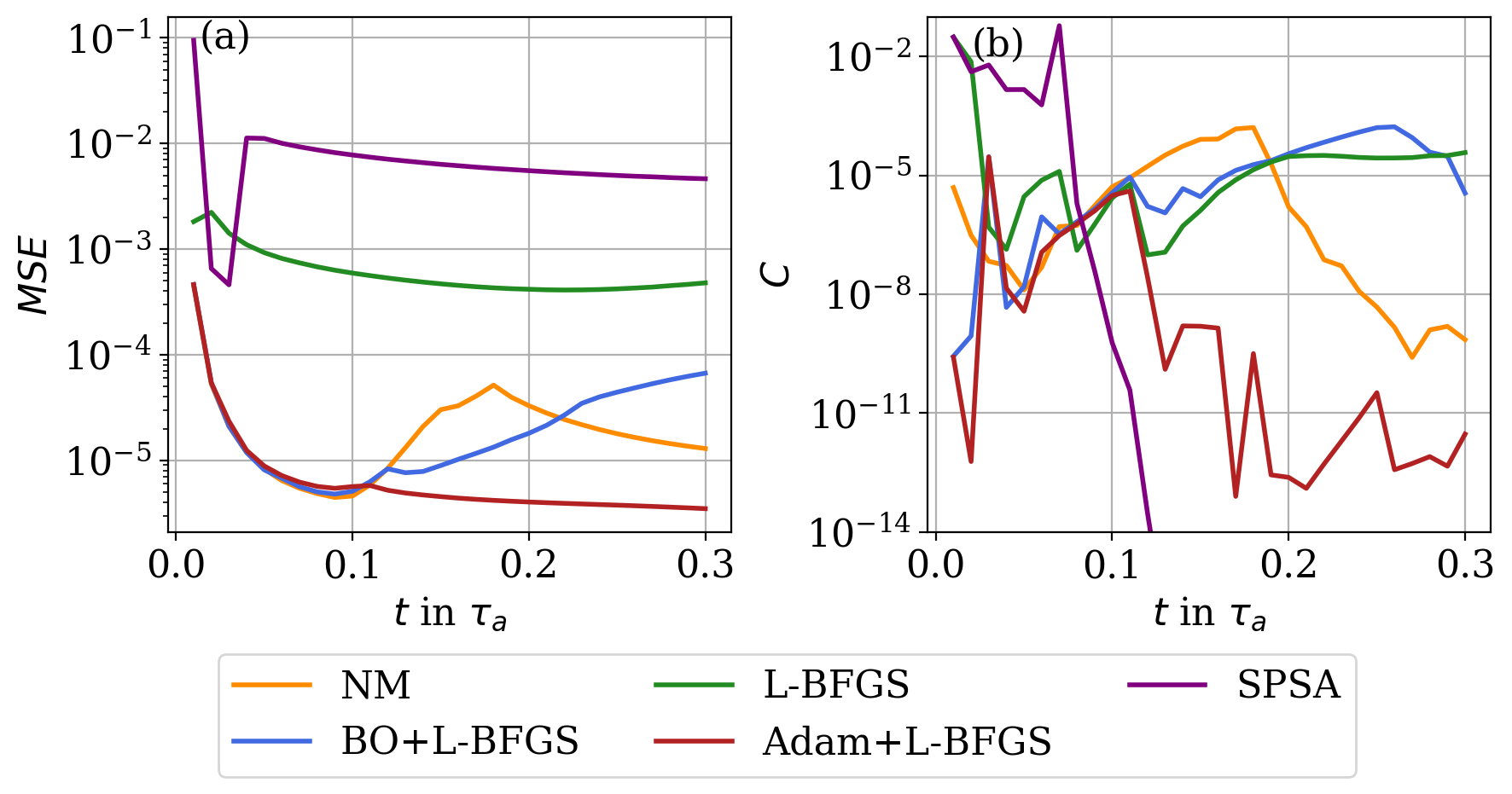}
    \caption{Comparison of investigated optimization methods with (a) the mean squared error (MSE) and (b) the cost function over time. Overall the combined optimization with Adam and L-BFGS algorithm shows the best performance for the considered test case with $N=16$, $D=1$, $U=10$ and $\tau=0.001$.}
    \label{fig:Comp_Optimizer}
\end{figure*}

In conclusion, the Nelder-Mead algorithm can be recommended in cases of low parameter spaces ($N\leq 16$) due to its accuracy and the computation time. If the system is increased it is advisable to chose the combination of Bayesian optimization and L-BFGS-B algorithm. The comparison of the mean squared errors and the cost functions of the VQA with the presented optimization algorithms are shown in Fig. \ref{fig:Comp_Optimizer}. Furthermore, an overview of the used optimization algorithms, the corresponding methods, accuracy and computation efforts is presented in table \ref{tab:optimizer}.

%-----------------------------------------
%-----------------------------------------
%-----------------------------------------
\begin{table}[ht]
\centering
% To place a caption above a table
\begin{tabular}[h]{l c c c}
\hline
Optimizer & Method  & Accuracy & Computation time \\
\hline

NM & geometric & $\sim 10^{-5}$ & $0.75h$ \\

L-BFGS-B & gradient-based & $\sim 10^{-4}$ & $0.75h$ \\

BO & approximation & $\sim 10^{-5}$ & $2.25h$ \\
 \& L-BFGS-B & gradient-based &  &  \\

Adam  & gradient-based & $\sim 10^{-5}$ & $7.5h$  \\
\& L-BFGS-B & gradient-based &  &   \\

SPSA & stochastic  & $\sim 10^{-3}$ &  $6.25h$\\
 & approximation  &  &  \\ 
 \hline
\end{tabular}
\caption{Comparison of classical optimization methods for VQA. The accuracy is given as the magnitude of the MSE, see eq. \eqref{eq:MSE} and the computation time is the mean value for the computation of one time step.}
\label{tab:optimizer}
\end{table}
%-----------------------------------------

\bibliographystyle{elsarticle-num-names} 
\bibliography{references}

\begin{thebibliography}{53}
\expandafter\ifx\csname natexlab\endcsname\relax\def\natexlab#1{#1}\fi
\providecommand{\url}[1]{\texttt{#1}}
\providecommand{\href}[2]{#2}
\providecommand{\path}[1]{#1}
\providecommand{\DOIprefix}{doi:}
\providecommand{\ArXivprefix}{arXiv:}
\providecommand{\URLprefix}{URL: }
\providecommand{\Pubmedprefix}{pmid:}
\providecommand{\doi}[1]{\href{http://dx.doi.org/#1}{\path{#1}}}
\providecommand{\Pubmed}[1]{\href{pmid:#1}{\path{#1}}}
\providecommand{\bibinfo}[2]{#2}
\ifx\xfnm\relax \def\xfnm[#1]{\unskip,\space#1}\fi
%Type = Article
\bibitem[{Preskill(2018)}]{Preskill2018}
\bibinfo{author}{J.~Preskill},
\newblock \bibinfo{title}{{Quantum computing in the NISQ era and beyond}},
\newblock \bibinfo{journal}{Quantum} \bibinfo{volume}{2} (\bibinfo{year}{2018})
  \bibinfo{pages}{79}.
%Type = Article
\bibitem[{Deutsch(2020)}]{Deutsch2020}
\bibinfo{author}{I.~H. Deutsch},
\newblock \bibinfo{title}{{Harnessing the power of the second quantum
  revolution}},
\newblock \bibinfo{journal}{PRX Quantum} \bibinfo{volume}{1}
  (\bibinfo{year}{2020}) \bibinfo{pages}{020101}.
%Type = Book
\bibitem[{Nielsen and Chuang(2011)}]{Nielsen2010}
\bibinfo{author}{M.~Nielsen}, \bibinfo{author}{I.~Chuang},
  \bibinfo{title}{Quantum Computation and Quantum Information: 10th Anniversary
  Edition}, \bibinfo{publisher}{Cambridge University Press},
  \bibinfo{year}{2011}.
%Type = Article
\bibitem[{Shor(1997)}]{Shor1997}
\bibinfo{author}{P.~W. Shor},
\newblock \bibinfo{title}{Polynomial-time algorithms for prime factorization
  and discrete logarithms on a quantum computer},
\newblock \bibinfo{journal}{SIAM J. Comput.} \bibinfo{volume}{26}
  (\bibinfo{year}{1997}) \bibinfo{pages}{1484--1509}.
%Type = Article
\bibitem[{Grover(1997)}]{Grover1997}
\bibinfo{author}{L.~K. Grover},
\newblock \bibinfo{title}{Quantum mechanics helps in searching for a needle in
  a haystack},
\newblock \bibinfo{journal}{Phys. Rev. Lett.} \bibinfo{volume}{79}
  (\bibinfo{year}{1997}) \bibinfo{pages}{325--328}.
%Type = Article
\bibitem[{Deng et~al.(2023)Deng, Gu, Liu, Gong, Su, Zhang, Tang, Jia, Xu, Chen,
  Qin, Peng, Yan, Hu, Huang, Li, Li, Chen, Jiang, Gan, Yang, You, Li, Zhong,
  Wang, Liu, Renema, Lu, and Pan}]{Deng2023}
\bibinfo{author}{Y.-H. Deng}, \bibinfo{author}{Y.-C. Gu},
  \bibinfo{author}{H.-L. Liu}, \bibinfo{author}{S.-Q. Gong},
  \bibinfo{author}{H.~Su}, \bibinfo{author}{Z.-J. Zhang},
  \bibinfo{author}{H.-Y. Tang}, \bibinfo{author}{M.-H. Jia},
  \bibinfo{author}{J.-M. Xu}, \bibinfo{author}{M.-C. Chen},
  \bibinfo{author}{J.~Qin}, \bibinfo{author}{L.-C. Peng},
  \bibinfo{author}{J.~Yan}, \bibinfo{author}{Y.~Hu},
  \bibinfo{author}{J.~Huang}, \bibinfo{author}{H.~Li}, \bibinfo{author}{Y.~Li},
  \bibinfo{author}{Y.~Chen}, \bibinfo{author}{X.~Jiang},
  \bibinfo{author}{L.~Gan}, \bibinfo{author}{G.~Yang},
  \bibinfo{author}{L.~You}, \bibinfo{author}{L.~Li}, \bibinfo{author}{H.-S.
  Zhong}, \bibinfo{author}{H.~Wang}, \bibinfo{author}{N.-L. Liu},
  \bibinfo{author}{J.~J. Renema}, \bibinfo{author}{C.-Y. Lu},
  \bibinfo{author}{J.-W. Pan},
\newblock \bibinfo{title}{Gaussian boson sampling with
  pseudo-photon-number-resolving detectors and quantum computational
  advantage},
\newblock \bibinfo{journal}{Phys. Rev. Lett.} \bibinfo{volume}{131}
  (\bibinfo{year}{2023}) \bibinfo{pages}{{150601}}.
%Type = Misc
\bibitem[{Choi et~al.(2023)Choi, Moses, and Thompson}]{Choi2023}
\bibinfo{author}{S.~Choi}, \bibinfo{author}{W.~S. Moses},
  \bibinfo{author}{N.~Thompson}, \bibinfo{title}{{The Quantum Tortoise and the
  Classical Hare: A simple framework for understanding which problems quantum
  computing will accelerate (and which it won’t)}}, \bibinfo{year}{2023}.
  \href{http://arxiv.org/abs/2310.15505}{{\tt arXiv:2310.15505}}.
%Type = Article
\bibitem[{Moin and Mahesh(1998)}]{Moin1998}
\bibinfo{author}{P.~Moin}, \bibinfo{author}{K.~Mahesh},
\newblock \bibinfo{title}{{Direct Numerical Simulation: A tool in turbulence
  research}},
\newblock \bibinfo{journal}{Annu. Rev. Fluid Mech.} \bibinfo{volume}{30}
  (\bibinfo{year}{1998}) \bibinfo{pages}{539--578}.
%Type = Article
\bibitem[{Iyer et~al.(2019)Iyer, Schumacher, Sreenivasan, and Yeung}]{Iyer2019}
\bibinfo{author}{K.~P. Iyer}, \bibinfo{author}{J.~Schumacher},
  \bibinfo{author}{K.~R. Sreenivasan}, \bibinfo{author}{P.~K. Yeung},
\newblock \bibinfo{title}{Scaling of locally averaged energy dissipation and
  enstrophy density in isotropic turbulence},
\newblock \bibinfo{journal}{New J. Phys.} \bibinfo{volume}{21}
  (\bibinfo{year}{2019}) \bibinfo{pages}{033016}.
%Type = Article
\bibitem[{Buaria et~al.(2019)Buaria, Pumir, Bodenschatz, and
  Yeung}]{Buaria2019}
\bibinfo{author}{D.~Buaria}, \bibinfo{author}{A.~Pumir},
  \bibinfo{author}{E.~Bodenschatz}, \bibinfo{author}{P.~K. Yeung},
\newblock \bibinfo{title}{Extreme velocity gradients in turbulent flows},
\newblock \bibinfo{journal}{New J. Phys.} \bibinfo{volume}{21}
  (\bibinfo{year}{2019}) \bibinfo{pages}{043004}.
%Type = Article
\bibitem[{Gourianov et~al.(2022)Gourianov, Lubasch, Dolgov, van~den Berg,
  Babaee, Givi, Kiffner, and Jaksch}]{Gourianov2022}
\bibinfo{author}{N.~Gourianov}, \bibinfo{author}{M.~Lubasch},
  \bibinfo{author}{S.~Dolgov}, \bibinfo{author}{Q.~Y. van~den Berg},
  \bibinfo{author}{H.~Babaee}, \bibinfo{author}{P.~Givi},
  \bibinfo{author}{M.~Kiffner}, \bibinfo{author}{D.~Jaksch},
\newblock \bibinfo{title}{A quantum-inspired approach to exploit turbulence
  structures},
\newblock \bibinfo{journal}{Nat. Comput. Sci.} \bibinfo{volume}{2}
  (\bibinfo{year}{2022}) \bibinfo{pages}{30--37}.
%Type = Article
\bibitem[{Meng and Yang(2023)}]{Meng2023}
\bibinfo{author}{Z.~Meng}, \bibinfo{author}{Y.~Yang},
\newblock \bibinfo{title}{{Quantum computing of fluid dynamics using the
  hydrodynamic Schr\"odinger equation}},
\newblock \bibinfo{journal}{Phys. Rev. Research} \bibinfo{volume}{5}
  (\bibinfo{year}{2023}) \bibinfo{pages}{033182}.
%Type = Article
\bibitem[{Jin et~al.(2023)Jin, Liu, and Yu}]{Jin2023}
\bibinfo{author}{S.~Jin}, \bibinfo{author}{N.~Liu}, \bibinfo{author}{Y.~Yu},
\newblock \bibinfo{title}{{Quantum simulation of partial differential
  equations: Applications and detailed analysis}},
\newblock \bibinfo{journal}{Phys. Rev. A} \bibinfo{volume}{108}
  (\bibinfo{year}{2023}) \bibinfo{pages}{032603}.
%Type = Misc
\bibitem[{Succi and Tiribocchi(2023)}]{Succi2023}
\bibinfo{author}{S.~Succi}, \bibinfo{author}{A.~Tiribocchi},
  \bibinfo{title}{{Navier-Stokes-Schrödinger equation for dissipative
  fluids}}, \bibinfo{year}{2023}. \href{http://arxiv.org/abs/2308.05879}{{\tt
  arXiv:2308.05879}}.
%Type = Article
\bibitem[{Pfeffer et~al.(2022)Pfeffer, Heyder, and Schumacher}]{Pfeffer2022}
\bibinfo{author}{P.~Pfeffer}, \bibinfo{author}{F.~Heyder},
  \bibinfo{author}{J.~Schumacher},
\newblock \bibinfo{title}{{Hybrid quantum-classical reservoir computing of
  thermal convection flow}},
\newblock \bibinfo{journal}{Phys. Rev. Research} \bibinfo{volume}{4}
  (\bibinfo{year}{2022}) \bibinfo{pages}{033176}.
%Type = Article
\bibitem[{Pfeffer et~al.(2023)Pfeffer, Heyder, and Schumacher}]{Pfeffer2023}
\bibinfo{author}{P.~Pfeffer}, \bibinfo{author}{F.~Heyder},
  \bibinfo{author}{J.~Schumacher},
\newblock \bibinfo{title}{Reduced-order modeling of two-dimensional turbulent
  {R}ayleigh-{B}{\'{e}}nard flow by hybrid quantum-classical reservoir
  computing},
\newblock \bibinfo{journal}{Phys. Rev. Research} \bibinfo{volume}{5}
  (\bibinfo{year}{2023}) \bibinfo{pages}{043242}.
%Type = Article
\bibitem[{Bharadwaj and Sreenivasan(2020)}]{Bharadwaj2020}
\bibinfo{author}{S.~S. Bharadwaj}, \bibinfo{author}{K.~R. Sreenivasan},
\newblock \bibinfo{title}{Quantum computation of fluid dynamics},
\newblock \bibinfo{journal}{Indian Academy of Sciences Conference Series}
  \bibinfo{volume}{3} (\bibinfo{year}{2020}) \bibinfo{pages}{77--96}.
%Type = Article
\bibitem[{Bharadwaj and Sreenivasan(2023)}]{Bharadwaj2023}
\bibinfo{author}{S.~S. Bharadwaj}, \bibinfo{author}{K.~R. Sreenivasan},
\newblock \bibinfo{title}{Hybrid quantum algorithms for flow problems},
\newblock \bibinfo{journal}{Proc. Natl. Acad. Sci. USA} \bibinfo{volume}{120}
  (\bibinfo{year}{2023}) \bibinfo{pages}{{e2311014120}}.
%Type = Article
\bibitem[{Bharadwaj et~al.(2023)Bharadwaj, Nadiga, Eidenbenz, and
  Sreenivasan}]{Bharadwaj2023(2)}
\bibinfo{author}{S.~S. Bharadwaj}, \bibinfo{author}{B.~Nadiga},
  \bibinfo{author}{S.~Eidenbenz}, \bibinfo{author}{K.~R. Sreenivasan},
\newblock \bibinfo{title}{Quantum computing of nonlinear flow problems with a
  homotopy analysis algorithm},
\newblock \bibinfo{journal}{Bull. Am. Phys. Soc.} \bibinfo{volume}{ZC17}
  (\bibinfo{year}{2023}) \bibinfo{pages}{002}.
%Type = Article
\bibitem[{Gaitan(2021)}]{Gaitan2020}
\bibinfo{author}{F.~Gaitan},
\newblock \bibinfo{title}{{Finding flows in a Navier-Stokes fluid through
  quantum computing}},
\newblock \bibinfo{journal}{npj Quantum Inf.} \bibinfo{volume}{6}
  (\bibinfo{year}{2021}) \bibinfo{pages}{{61}}.
%Type = Article
\bibitem[{Lubasch et~al.(2020)Lubasch, Joo, Moinier, Kiffner, and
  Jaksch}]{Lubasch2020}
\bibinfo{author}{M.~Lubasch}, \bibinfo{author}{J.~Joo},
  \bibinfo{author}{P.~Moinier}, \bibinfo{author}{M.~Kiffner},
  \bibinfo{author}{D.~Jaksch},
\newblock \bibinfo{title}{Variational quantum algorithms for nonlinear
  problems},
\newblock \bibinfo{journal}{Phys. Rev. A} \bibinfo{volume}{101}
  (\bibinfo{year}{2020}) \bibinfo{pages}{010301}.
  \DOIprefix\doi{10.1103/PhysRevA.101.010301}.
%Type = Article
\bibitem[{Pool et~al.(2022)Pool, Somoza, Lubasch, and Horstmann}]{Pool2022}
\bibinfo{author}{A.~J. Pool}, \bibinfo{author}{A.~D. Somoza},
  \bibinfo{author}{M.~Lubasch}, \bibinfo{author}{B.~Horstmann},
\newblock \bibinfo{title}{{Solving partial differential equations using a
  quantum computer}},
\newblock \bibinfo{journal}{2022 IEEE International Conference on Quantum
  Computing and Engineering}  (\bibinfo{year}{2022}) \bibinfo{pages}{864--866}.
%Type = Article
\bibitem[{Demirdjian et~al.(2020)Demirdjian, Gunlycke, Reynolds, Doyle, and
  Tafur}]{Demirdjian2022}
\bibinfo{author}{R.~Demirdjian}, \bibinfo{author}{D.~Gunlycke},
  \bibinfo{author}{C.~A. Reynolds}, \bibinfo{author}{J.~D. Doyle},
  \bibinfo{author}{S.~Tafur},
\newblock \bibinfo{title}{Variational quantum solutions to the
  advection–diffusion equation for applications in fluid dynamics},
\newblock \bibinfo{journal}{Quantum Inf. Process.} \bibinfo{volume}{21}
  (\bibinfo{year}{2020}) \bibinfo{pages}{322}.
%Type = Article
\bibitem[{Leong et~al.(2022)Leong, Ewe, and Koh}]{Leong2022}
\bibinfo{author}{F.~Y. Leong}, \bibinfo{author}{W.-B. Ewe},
  \bibinfo{author}{D.~E. Koh},
\newblock \bibinfo{title}{Variational quantum evolution equation solver},
\newblock \bibinfo{journal}{Sci. Rep.} \bibinfo{volume}{12}
  (\bibinfo{year}{2022}) \bibinfo{pages}{{10817}}.
%Type = Article
\bibitem[{Leong et~al.(2023)Leong, Koh, Ewe, and Kong}]{Leong2023}
\bibinfo{author}{F.~Y. Leong}, \bibinfo{author}{D.~E. Koh},
  \bibinfo{author}{W.-B. Ewe}, \bibinfo{author}{J.~F. Kong},
\newblock \bibinfo{title}{Variational quantum simulation of partial
  differential equations: Applications in colloidal transport},
\newblock \bibinfo{journal}{Int. J. Numer. Method H.} \bibinfo{volume}{33}
  (\bibinfo{year}{2023}) \bibinfo{pages}{3669--3690}.
%Type = Article
\bibitem[{Todorova and {de Steijl}(2020)}]{Todorova2020}
\bibinfo{author}{B.~N. Todorova}, \bibinfo{author}{R.~{de Steijl}},
\newblock \bibinfo{title}{{Quantum algorithm for the collisionless Boltzmann
  equation}},
\newblock \bibinfo{journal}{J. Comp. Phys.} \bibinfo{volume}{409}
  (\bibinfo{year}{2020}) \bibinfo{pages}{109347}.
%Type = Article
\bibitem[{Budinski(2021)}]{Budinski2021}
\bibinfo{author}{L.~Budinski},
\newblock \bibinfo{title}{{Quantum algorithm for the advection–diffusion
  equation simulated with the Lattice Boltzmann method}},
\newblock \bibinfo{journal}{Quantum Inf. Process.} \bibinfo{volume}{20}
  (\bibinfo{year}{2021}) \bibinfo{pages}{57}.
%Type = Article
\bibitem[{Succi et~al.(2023)Succi, Itani, Sreenivasan, and Steijl}]{Succi2023a}
\bibinfo{author}{S.~Succi}, \bibinfo{author}{W.~Itani}, \bibinfo{author}{K.~R.
  Sreenivasan}, \bibinfo{author}{R.~Steijl},
\newblock \bibinfo{title}{Quantum computing for fluids: Where do we stand?},
\newblock \bibinfo{journal}{Europhys. Lett.} \bibinfo{volume}{144}
  (\bibinfo{year}{2023}) \bibinfo{pages}{{10001}}.
%Type = Article
\bibitem[{Harrow et~al.(2009)Harrow, Hassidim, and Lloyd}]{Harrow2009}
\bibinfo{author}{A.~H. Harrow}, \bibinfo{author}{A.~Hassidim},
  \bibinfo{author}{S.~Lloyd},
\newblock \bibinfo{title}{Quantum algorithm for linear systems of equations},
\newblock \bibinfo{journal}{Phys. Rev. Lett.} \bibinfo{volume}{103}
  (\bibinfo{year}{2009}) \bibinfo{pages}{150502}.
  \DOIprefix\doi{10.1103/PhysRevLett.103.150502}.
%Type = Article
\bibitem[{Aaronson(2015)}]{Aaronson2015}
\bibinfo{author}{S.~Aaronson},
\newblock \bibinfo{title}{{Read the fine print}},
\newblock \bibinfo{journal}{Nature Physics} \bibinfo{volume}{11}
  (\bibinfo{year}{2015}) \bibinfo{pages}{291--293}.
%Type = Article
\bibitem[{Montanaro and Pallister(2016)}]{Montanaro2016}
\bibinfo{author}{A.~Montanaro}, \bibinfo{author}{S.~Pallister},
\newblock \bibinfo{title}{Quantum algorithms and the finite element method},
\newblock \bibinfo{journal}{Phys. Rev. A} \bibinfo{volume}{93}
  (\bibinfo{year}{2016}) \bibinfo{pages}{032324}.
%Type = Article
\bibitem[{Guseynov et~al.(2023)Guseynov, Zhukov, Pogosov, and
  Lebedev}]{Guseynov2023}
\bibinfo{author}{N.~M. Guseynov}, \bibinfo{author}{A.~A. Zhukov},
  \bibinfo{author}{W.~V. Pogosov}, \bibinfo{author}{A.~V. Lebedev},
\newblock \bibinfo{title}{Depth analysis of variational quantum algorithms for
  the heat equation},
\newblock \bibinfo{journal}{Phys. Rev. A} \bibinfo{volume}{107}
  (\bibinfo{year}{2023}) \bibinfo{pages}{052422}.
  \DOIprefix\doi{10.1103/PhysRevA.107.052422}.
%Type = Article
\bibitem[{Liu et~al.(2023)Liu, Chen, Shu, Chew, Khoo, Zhao, and Cui}]{Liu2023}
\bibinfo{author}{Y.~Y. Liu}, \bibinfo{author}{Z.~Chen},
  \bibinfo{author}{C.~Shu}, \bibinfo{author}{S.~C. Chew},
  \bibinfo{author}{B.~C. Khoo}, \bibinfo{author}{X.~Zhao},
  \bibinfo{author}{Y.~D. Cui},
\newblock \bibinfo{title}{{Application of a variational hybrid
  quantum-classical algorithm to heat conduction equation and analysis of time
  complexity}},
\newblock \bibinfo{journal}{Phys. Fluids} \bibinfo{volume}{34}
  (\bibinfo{year}{2023}) \bibinfo{pages}{117121}.
%Type = Article
\bibitem[{Qis(2023)}]{Qiskit}
\bibinfo{title}{Qiskit version 0.23.2}  (\bibinfo{year}{2023}).
%Type = Article
\bibitem[{Childs et~al.(2017)Childs, Kothari, and Somma}]{Childs2017}
\bibinfo{author}{A.~M. Childs}, \bibinfo{author}{R.~Kothari},
  \bibinfo{author}{R.~D. Somma},
\newblock \bibinfo{title}{Quantum algorithm for systems of linear equations
  with exponentially improved dependence on precision},
\newblock \bibinfo{journal}{SIAM J. Comput.} \bibinfo{volume}{46}
  (\bibinfo{year}{2017}) \bibinfo{pages}{1920--1950}.
%Type = Article
\bibitem[{Childs et~al.(2021)Childs, Liu, and Ostrander}]{Childs2021}
\bibinfo{author}{A.~M. Childs}, \bibinfo{author}{J.-P. Liu},
  \bibinfo{author}{A.~Ostrander},
\newblock \bibinfo{title}{High-precision quantum algorithms for partial
  differential equations},
\newblock \bibinfo{journal}{{Quantum}} \bibinfo{volume}{5}
  (\bibinfo{year}{2021}) \bibinfo{pages}{574}.
%Type = Article
\bibitem[{Liu et~al.(2021)Liu, Kolden, Krovi, Loureiro, Trivisa, and
  Childs}]{Liu2021}
\bibinfo{author}{J.-P. Liu}, \bibinfo{author}{H.~{\O}. Kolden},
  \bibinfo{author}{H.~K. Krovi}, \bibinfo{author}{N.~F. Loureiro},
  \bibinfo{author}{K.~Trivisa}, \bibinfo{author}{A.~M. Childs},
\newblock \bibinfo{title}{Efficient quantum algorithm for dissipative nonlinear
  differential equations},
\newblock \bibinfo{journal}{Proc. Natl. Acad. Sci. USA} \bibinfo{volume}{118}
  (\bibinfo{year}{2021}) \bibinfo{pages}{e2026805118}.
%Type = Article
\bibitem[{Cerezo et~al.(2021)Cerezo, Arrasmith, Babbush, Benjamin, Endo, Fujii,
  McClean, Mitarai, Yuan, Cincio, and Coles}]{Cerezo2021}
\bibinfo{author}{M.~Cerezo}, \bibinfo{author}{A.~Arrasmith},
  \bibinfo{author}{R.~Babbush}, \bibinfo{author}{S.~C. Benjamin},
  \bibinfo{author}{S.~Endo}, \bibinfo{author}{K.~Fujii}, \bibinfo{author}{J.~R.
  McClean}, \bibinfo{author}{K.~Mitarai}, \bibinfo{author}{X.~Yuan},
  \bibinfo{author}{L.~Cincio}, \bibinfo{author}{P.~J. Coles},
\newblock \bibinfo{title}{Variational quantum algorithms},
\newblock \bibinfo{journal}{Nat. Rev. Phys.} \bibinfo{volume}{3}
  (\bibinfo{year}{2021}) \bibinfo{pages}{625--644}.
%Type = Article
\bibitem[{Nelder and Mead(1965)}]{Nelder1965}
\bibinfo{author}{J.~Nelder}, \bibinfo{author}{R.~Mead},
\newblock \bibinfo{title}{A simplex method for function minimization},
\newblock \bibinfo{journal}{The Computer Journal} \bibinfo{volume}{7}
  (\bibinfo{year}{1965}) \bibinfo{pages}{308--313}.
  \DOIprefix\doi{10.1093/comjnl/7.4.308}.
%Type = Article
\bibitem[{Barratt et~al.(2021)Barratt, Dborin, Bal, Stojevic, Pollmann, and
  Green}]{Barratt2021}
\bibinfo{author}{F.~Barratt}, \bibinfo{author}{J.~Dborin},
  \bibinfo{author}{M.~Bal}, \bibinfo{author}{V.~Stojevic},
  \bibinfo{author}{F.~Pollmann}, \bibinfo{author}{A.~G. Green},
\newblock \bibinfo{title}{{Parallel quantum simulation of large systems on
  small NISQ computers}},
\newblock \bibinfo{journal}{npj Quantum Inf.} \bibinfo{volume}{7}
  (\bibinfo{year}{2021}) \bibinfo{pages}{{79}}.
%Type = Article
\bibitem[{Shaffer et~al.(2023)Shaffer, Kocia, and Sarovar}]{Shaffer2023}
\bibinfo{author}{R.~Shaffer}, \bibinfo{author}{L.~Kocia},
  \bibinfo{author}{M.~Sarovar},
\newblock \bibinfo{title}{{Surrogate-based optimization for variational quantum
  algorithms}},
\newblock \bibinfo{journal}{Phys. Rev. A} \bibinfo{volume}{107}
  (\bibinfo{year}{2023}) \bibinfo{pages}{032415}.
%Type = Article
\bibitem[{Dudley et~al.(2019)Dudley, Genty, Mussot, Chabchoub, and
  Dias}]{Dudley2019}
\bibinfo{author}{J.~M. Dudley}, \bibinfo{author}{G.~Genty},
  \bibinfo{author}{A.~Mussot}, \bibinfo{author}{A.~Chabchoub},
  \bibinfo{author}{F.~Dias},
\newblock \bibinfo{title}{Rogue waves and analogies in optics and
  oceanography},
\newblock \bibinfo{journal}{Nat. Rev. Phys.} \bibinfo{volume}{1}
  (\bibinfo{year}{2019}) \bibinfo{pages}{675--689}.
%Type = Article
\bibitem[{Brassard et~al.(2002)Brassard, Hoyer, Mosca, and Tapp}]{Brassard2002}
\bibinfo{author}{G.~Brassard}, \bibinfo{author}{P.~Hoyer},
  \bibinfo{author}{M.~Mosca}, \bibinfo{author}{A.~Tapp},
\newblock \bibinfo{title}{Quantum amplitude amplification and estimation},
\newblock \bibinfo{journal}{Contemp. Math.} \bibinfo{volume}{305}
  (\bibinfo{year}{2002}) \bibinfo{pages}{53--74}.
%Type = Article
\bibitem[{Engel et~al.(2021)Engel, Smith, and Parker}]{Engel2021}
\bibinfo{author}{A.~Engel}, \bibinfo{author}{G.~Smith}, \bibinfo{author}{S.~E.
  Parker},
\newblock \bibinfo{title}{Linear embedding of nonlinear dynamical systems and
  prospects for efficient quantum algorithms},
\newblock \bibinfo{journal}{Phys. Plasmas} \bibinfo{volume}{28}
  (\bibinfo{year}{2021}) \bibinfo{pages}{062305}.
%Type = Article
\bibitem[{Joseph(2020)}]{Joseph2020}
\bibinfo{author}{I.~Joseph},
\newblock \bibinfo{title}{{Koopman–von Neumann approach to quantum simulation
  of nonlinear classical dynamics}},
\newblock \bibinfo{journal}{Phys. Rev. Research} \bibinfo{volume}{2}
  (\bibinfo{year}{2020}) \bibinfo{pages}{043102}.
%Type = Article
\bibitem[{Giannakis et~al.(2022)Giannakis, Ourmazd, Pfeffer, Schumacher, and
  Slawinska}]{Giannakis2022}
\bibinfo{author}{D.~Giannakis}, \bibinfo{author}{A.~Ourmazd},
  \bibinfo{author}{P.~Pfeffer}, \bibinfo{author}{J.~Schumacher},
  \bibinfo{author}{J.~Slawinska},
\newblock \bibinfo{title}{{Embedding classical dynamics in a quantum
  computer}},
\newblock \bibinfo{journal}{Phys. Rev. A} \bibinfo{volume}{105}
  (\bibinfo{year}{2022}) \bibinfo{pages}{{052404}}.
%Type = Misc
\bibitem[{Lin et~al.(2022)Lin, Lowrie, Aslangil, Suba{\c{s}}i, and
  Sornborger}]{Lin2022}
\bibinfo{author}{Y.~T. Lin}, \bibinfo{author}{R.~B. Lowrie},
  \bibinfo{author}{D.~Aslangil}, \bibinfo{author}{Y.~Suba{\c{s}}i},
  \bibinfo{author}{A.~T. Sornborger}, \bibinfo{title}{{Koopman von Neumann
  mechanics and the Koopman representation: A perspective on solving nonlinear
  dynamical systems with quantum computers}}, \bibinfo{year}{2022}.
  \href{http://arxiv.org/abs/2202.02188}{{\tt arXiv:2202.02188}}.
%Type = Article
\bibitem[{Cross et~al.(2019)Cross, Bishop, Sheldon, Nation, and
  Gambetta}]{Cross2019}
\bibinfo{author}{A.~W. Cross}, \bibinfo{author}{L.~S. Bishop},
  \bibinfo{author}{S.~Sheldon}, \bibinfo{author}{P.~D. Nation},
  \bibinfo{author}{J.~M. Gambetta},
\newblock \bibinfo{title}{Validating quantum computers using randomized model
  circuits},
\newblock \bibinfo{journal}{Phys. Rev. A} \bibinfo{volume}{100}
  (\bibinfo{year}{2019}) \bibinfo{pages}{032328}.
%Type = Article
\bibitem[{Uvarov and Biamonte(2021)}]{Uvarov2021}
\bibinfo{author}{A.~V. Uvarov}, \bibinfo{author}{J.~D. Biamonte},
\newblock \bibinfo{title}{On barren plateaus and cost function locality in
  variational quantum algorithms},
\newblock \bibinfo{journal}{Journal of Physics A: Mathematical and Theoretical}
  \bibinfo{volume}{54} (\bibinfo{year}{2021}).
  \DOIprefix\doi{10.1088/1751-8121/abfac7}.
%Type = Article
\bibitem[{Liu and Nocedal(1989)}]{Liu1989}
\bibinfo{author}{D.~Liu}, \bibinfo{author}{J.~Nocedal},
\newblock \bibinfo{title}{{On the limited memory BFGS method for large scale
  optimization}},
\newblock \bibinfo{journal}{Mathematical Programming} \bibinfo{volume}{45}
  (\bibinfo{year}{1989}) \bibinfo{pages}{503–528}.
  \DOIprefix\doi{10.1007/BF01589116}.
%Type = Article
\bibitem[{Mockus et~al.(1978)Mockus, Tiesis, and Zilinskas}]{Mockus1978}
\bibinfo{author}{J.~Mockus}, \bibinfo{author}{V.~Tiesis},
  \bibinfo{author}{A.~Zilinskas},
\newblock \bibinfo{title}{The application of {B}ayesian methods for seeking the
  extremum},
\newblock \bibinfo{journal}{Towards Global Optimization} \bibinfo{volume}{2}
  (\bibinfo{year}{1978}) \bibinfo{pages}{117--129}.
%Type = Article
\bibitem[{Kingma and Ba(2015)}]{Kingma2014}
\bibinfo{author}{D.~K. Kingma}, \bibinfo{author}{J.~L. Ba},
\newblock \bibinfo{title}{Adam: A method for stochastic optimization},
\newblock \bibinfo{journal}{3rd International Conference for Learning
  Representations}  (\bibinfo{year}{2015}).
  \DOIprefix\doi{10.48550/arXiv.1412.6980}.
%Type = Article
\bibitem[{Spall(1998)}]{Spall1998}
\bibinfo{author}{J.~C. Spall},
\newblock \bibinfo{title}{An overview of the simultaneous perturbation method
  for efficient optimization},
\newblock \bibinfo{journal}{Johns Hopkins Apl Technical Digest}
  \bibinfo{volume}{19} (\bibinfo{year}{1998}).

\end{thebibliography}

\end{document}